\documentclass[usenatbib]{mnras}

\usepackage{multirow}
\usepackage{rotating}
\usepackage{colortbl}
\usepackage{color}
\usepackage[fleqn]{amsmath}
\usepackage{amssymb}
\usepackage{amsfonts}
\usepackage{verbatim}
\usepackage{scalefnt}
\usepackage[percent]{overpic}
\usepackage[T1]{fontenc} 
\usepackage{aecompl} 
\usepackage{ulem}
\usepackage{cleveref}
\usepackage{orcidlink}
\usepackage{xspace}

\def\lsim{\mathrel{\rlap{\lower 3pt \hbox{$\sim$}} \raise 2.0pt \hbox{$<$}}}
\def\gsim{\mathrel{\rlap{\lower 3pt \hbox{$\sim$}} \raise 2.0pt \hbox{$>$}}}

\newcommand{\comments}[1]{} 
\newcommand\T{\rule{0pt}{2.6ex}}       
\newcommand\B{\rule[-1.2ex]{0pt}{0pt}} 
\defcitealias{Bortolas_et_al_2022}{B22}

\newcommand{\soutPC}{\bgroup\markoverwith{\textcolor{cyan}{\rule[0.5ex]{2pt}{1pt}}}\ULon}

\newcommand{\soutTZ}{\bgroup\markoverwith{\textcolor{green}{\rule[0.5ex]{2pt}{1pt}}}\ULon}

\newcommand{\seb}{\bgroup\markoverwith{\textcolor{magenta}{\rule[0.5ex]{2pt}{1pt}}}\ULon}

\newcommand{\soutrev}{\bgroup\markoverwith{\textcolor{red}{\rule[0.5ex]{2pt}{1pt}}}\ULon}

\newcommand{\td}{\ensuremath{\tau_{\rm d}}\xspace}
\newcommand{\zinj}{\ensuremath{z_{\rm inj}}\xspace}
\newcommand{\zbinary}{\ensuremath{z_{\rm binary}}\xspace}
\newcommand{\kzz}{\ensuremath{K(\zbinary | \zinj)}\xspace}

\defcitealias{Navarro_et_al_1996}{NFW}

\title[Emulating MBH dynamics with normalizing flows]{Emulating the complex galactic-scale orbital dynamics of LISA massive black hole pairs with normalizing flows}

\author[P.~R.~Capelo et al.]{Pedro R. Capelo$^{\orcidlink{0000-0002-1786-963X}}$,$^{1}$\thanks{\!\!E-mails: \!\href{pcapelo@physik.uzh.ch}{pcapelo@physik.uzh.ch}; \!\href{carlos.moreno.martinez@cern.ch}{carlos.morenomartinez@unige.ch}}\label{emails}\thanks{\!\!These authors contributed equally to this work.}\label{contrib}
Carlos Moreno Martinez$^{\orcidlink{0000-0002-5719-7655}}$,$^{2}$\hyperref[emails]{\footnotemark[1]}\hyperref[contrib]{\footnotemark[2]}
Nodens Koren,$^{3}$
Tommaso Zana$^{\orcidlink{0000-0003-4244-8527}}$,$^{4,5,6}$\newauthor
Elisa Bortolas$^{\orcidlink{0000-0001-9458-821X}}$,$^{7}$
Matteo Bonetti$^{\orcidlink{0000-0001-7889-6810}}$,$^{8,9,10}$
Janis Fluri$^{\orcidlink{0000-0002-4813-4935}}$,$^{3}$
Thomas Hofmann$^{3}$
and Lucio Mayer$^{\orcidlink{0000-0002-7078-2074}}$$^{1}$
\\
$^{1}$Department of Astrophysics, University of Zurich, Winterthurerstrasse 190, CH-8057 Z{\"u}rich, Switzerland\\
$^{2}$D\'{e}partement de Physique Nucl\'{e}aire et Corpusculaire, Universit\'{e} de Gen\`{e}ve, 24, quai Ernest-Ansermet, CH-1205 Gen\`{e}ve, Switzerland\\
$^{3}$Department of Computer Science, ETH Zurich, Universit{\"a}tstrasse 6, CH-8092 Z{\"u}rich, Switzerland\\
$^{4}$Dipartimento di Fisica, Sapienza, Universit\`{a} di Roma, Piazzale Aldo Moro 5, IT-00185 Roma, Italy\\
$^{5}$INAF, Osservatorio Astronomico di Roma, Via di Frascati 33, IT-00078 Monte Porzio Catone, Italy\\
$^{6}$INFN, Sezione di Roma I, Piazzale Aldo Moro 2, IT-00185 Roma, Italy\\
$^{7}$INAF, Osservatorio Astronomico di Padova, Vicolo dell'Osservatorio 5, IT-35122 Padova, Italy\\
$^{8}$Universit\`{a} degli Studi di Milano-Bicocca, Piazza della Scienza 3, IT-20126 Milano, Italy\\
$^{9}$INFN, Sezione di Milano-Bicocca, Piazza della Scienza 3, IT-20126 Milano, Italy\\
$^{10}$INAF, Osservatorio Astronomico di Brera, via Brera 20, IT-20121 Milano, Italy
}

\date{Accepted XXX. Received YYY; in original form ZZZ}

\pubyear{2026}

\begin{document}

\label{firstpage}

\pagerange{\pageref{firstpage}--\pageref{lastpage}}

\maketitle


\begin{abstract}
Massive black hole (MBH) pairs, formed in galaxy mergers, may coalesce in a burst of gravitational waves (GWs). Estimating the coalescence time-scales and rates is a long-standing astrophysical problem, essential to inform predictions for future GW detectors such as the Laser Interferometer Space Antenna, but remains challenging: MBH orbital decay in realistic galactic environments is complex and stochastic, and non-axisymmetric structures such as stellar bars can perturb MBH pair dynamics, delaying or accelerating binary formation and undermining the assumption that dynamical friction alone sets the inspiral duration. Capturing this evolution requires simulations too expensive to run at population scale. Here we present an artificial intelligence framework that emulates the galactic-scale orbital decay of an inspiralling MBH using conditional normalizing flows trained on a large suite of semi-analytical orbital integrations. Our model captures the evolution of secondary MBHs orbiting within multi-component galactic merger remnants featuring rotating stellar discs and bars, across a broad range of MBH masses, orbital configurations, and bar properties. The trained emulator reproduces the simulations' decay-time distributions while reducing computational cost by orders of magnitude. For the first time, we apply this model to galaxy populations drawn from a cosmological simulation, exploiting morphological information on barred and non-barred galaxies to infer MBH binary formation time-scales across cosmic time. Our results show that stellar bars can alter the distribution of MBH binary formation times. More broadly, this demonstrates how simulation-based, surrogate machine learning emulators can unlock a class of astrophysical problems where the physics is well understood system-by-system but intractable at scale.
\end{abstract}

\begin{keywords}
black hole physics -- methods: numerical -- stars: kinematics and dynamics -- galaxies: kinematics and dynamics -- galaxies: structure
\end{keywords}


\section{Introduction}\label{sec:introduction}

Massive black holes (MBHs) are postulated to reside at the centre of most massive galaxies \citep[e.g.][]{Kormendy_Ho_2013}. When these galaxies merge with each other, conforming to the hierarchical paradigm of galaxy formation and evolution \citep[e.g.][]{Blumenthal_et_al_1984}, their central MBHs are thought to grow via gas accretion -- due to gas inflows caused by merger-induced gravitational \citep[e.g.][]{Hopkins_Quataert_2010,Capelo_et_al_2015} and hydrodynamical \citep[e.g.][]{Capelo_Dotti_2017,Blumenthal_Barnes_2018} torques -- and are also conjectured to eventually coalesce with each other \citep[][]{Begelman_et_al_1980}, releasing gravitational waves \citep[GWs; ][]{Einstein_1916} in the process.

Space-borne missions such as the Laser Interferometer Space Antenna \citep[LISA;][]{AmaroSeoane_et_al_2023,Colpi_et_al_2024}, TianQin \citep[][]{Luo_et_al_2016,Li_et_al_2025}, and Taiji \citep[][]{Ruan_et_al_2020,Luo_et_al_2021} should be able to detect such events in the mHz GW band for a large span of MBH masses ($M_{\bullet} \sim 10^3$--$10^7$~M$_{\sun}$) and mass ratios, and a wide range of redshifts, recording their GW signals in the late inspiral, coalescence, and ringdown phases and providing information on many MBH properties, such as the total mass of the two MBHs, their mass ratio, their spins, and their location \citep[][]{AmaroSeoane_et_al_2023}.

Moreover, pulsar timing arrays have shown evidence for a GW background in the nHz band \citep[e.g.][]{Agazie_et_al_2023a}, which could be caused by the superposition of many coalescence events between very massive MBHs \citep[$M_{\bullet} \gtrsim 10^8$~M$_{\sun}$;][]{Agazie_et_al_2023b}, and may soon detect GW signals from individual sources, thanks to the upcoming Square Kilometre Array Observatory's ability to track several pulsars with improved timing precision \citep[][]{Shannon_et_al_2025}. These high-mass individual sources may also be subjected to hydrodynamical torques from the surrounding gas, causing perturbations that produce higher-frequency GWs, thus detectable in the mHz (i.e. LISA) band, and potentially providing information on the surroundings of these systems \citep[][]{Zwick_et_al_2022,Capelo_et_al_2026}.

Concurrently, accretion of the surrounding gas on to these MBHs -- especially the relatively low-mass ones, which tend to reside in more gas-rich systems -- potentially yields electromagnetic (EM) counterparts of these events, for a wide range of wavelengths \citep[e.g.][]{DeRosa_et_al_2019,Bogdanovic_et_al_2022,Capelo_et_al_2023,Krause_et_al_2025,Chakraborty_et_al_2025}. For example, in the past few decades, a plethora of dual active galactic nuclei (AGN) have been observed at separations ranging between 7~pc and hundreds of kpc, and detected from radio to X-rays \citep[see, e.g.][]{DeRosa_et_al_2019,Pfeifle_et_al_2025}. These dual AGN are surmised to be the precursors of GW events: accreting MBH pairs \citep[e.g.][]{Koss_et_al_2012,Capelo_et_al_2017,Battistini_et_al_2026} on their way to form a binary and ultimately coalesce, with EM emission occurring before, during, and after the MBH coalescence \citep[e.g.][]{Bogdanovic_et_al_2022}.

The possibility of using both GW detections and EM observations at various bands and wavelengths ushers in the definite possibility of both multiwavelength and multimessenger studies of MBHs and their host galaxies, in the context of understanding how both types of astrophysical objects assemble and (co-)evolve \citep[e.g.][]{Capelo_et_al_2026}. In particular, the rate of GW detections, along with the characterization of such events, will be crucial for our understanding of the underlying population of MBHs, including how they formed and evolved \citep[e.g.][]{AmaroSeoane_et_al_2023,Sato-Polito_et_al_2025}. To achieve this ambitious goal, we must improve our comprehension of the physical processes concerning the formation, growth, and dynamics of MBHs, in order to correctly interpret the data, both that coming from EM surveys and that from GW events, which are inextricably linked to the underlying MBH coalescence rate \citep[][]{Truant_et_al_2026,Izquierdo-Villalba_et_al_2026}.

Coalescence rates depend on many factors, including the mass of the MBH, the galactic merger rate, the MBH occupation fraction, and, ultimately, MBH dynamics. MBHs first approach each other as entities embedded in their own hosts: as such, the initial evolution is that of the individual host galaxies. Once a galactic remnant has formed, as a consequence of a galactic interaction, the two MBHs migrate towards the centre due to dynamical friction (DF), mostly against gas and stars \citep[e.g.][]{Khan_et_al_2011,Khan_et_al_2018,Gualandris_et_al_2022,Mannerkoski_et_al_2023,Fastidio_et_al_2025}, either as `naked' entities or as part of larger systems \citep[e.g. inside stellar cusps;][]{VanWassenhove_et_al_2014}. Eventually, at smaller scales, DF ceases to be the dominant physical process and is supplanted by stellar and, depending on the environment, gas-driven interactions, which promote the orbital decay down to the GW-driven inspiral and MBH coalescence \citep[][]{Begelman_et_al_1980}.

In this work, we focus on the orbital evolution of LISA MBHs within a galaxy merger remnant, at spatial scales above 10~pc separations \citep[i.e. the typical distance at which an MBH pair of $10^6$--$10^7$~M$_{\sun}$ becomes a binary; e.g.][]{Merritt_2013,Pfister_et_al_2017}, where DF is still the dominant mechanism. DF can be understood as the drag force arising from a global response of the host to the perturbation induced by the MBH \citep[e.g.][]{White_1983,Tremaine_Weinberg_1984,Weinberg_1986,Weinberg_1989,Tamfal_et_al_2021}, which causes the sinking of the MBH towards the host's centre. A more common interpretation \citep[][]{Chandrasekhar_1943} is that of a local drag induced by an overdensity of star \citep[and gas;][]{Ostriker_1999} trailing the orbiting MBH and thus resulting in a deceleration of the perturber. In either case, the consequence is that gravitational interactions between the MBH and its host usually cause the perturber to sink towards the centre.\footnote{In some cases, the perturber can experience negative DF \citep[e.g.][]{Park_Bogdanovic_2017,Park_Bogdanovic_2019} and/or dynamical stalling/buoyancy \citep[e.g.][]{Tamfal_et_al_2018,Dattathri_et_al_2025}. Since in this work we do not model MBH feedback nor dark matter (DM) cores, we do not consider these scenarios.}

The rate of this decay, however, strongly depends on the mass of the perturber, its orbit, and the inner structure of the host, all quantities that are difficult to resolve in cosmological simulations, given their relatively coarse spatial, mass, and time resolution. For this reason, many approximations have been implemented in the literature, ranging from the simple repositioning of the MBH to a local minimum of the gravitational potential \citep[e.g.][]{Sijacki_et_al_2015}, to a boosted dynamical mass \citep[e.g.][]{Debuhr_et_al_2011}, to more sophisticated sub-grid models that mimick the effects of DF \citep[e.g.][]{Tremmel_et_al_2015,Pfister_et_al_2019,Bird_et_al_2022,Damiano_et_al_2024}, or a combination of these methods \citep[e.g.][]{Chen_et_al_2022}. All these practices present limitations, as they rely on simplified assumptions (e.g. in the case of the DF sub-grid models), and often lead to unphysical results \citep[e.g. when using the repositioning scheme; see][]{Buttigieg_et_al_2025}.

Another approach is to combine cosmological simulations with semi-analytical models (SAMs) or isolated $N$-body simulations in post-processing, to model the dynamics of the MBHs below the resolution limit, assuming a given host's structure, informed by the resolved galactic properties \citep[e.g.][]{Salcido_et_al_2016,DeGraf_Sijacki_2020,Katz_et_al_2020,Volonteri_et_al_2020,Chen_et_al_2022,Francois_et_al_2026}. Recently, \citet{Li_et_al_2022} evaluated the decay time distribution of LISA MBHs using a cosmological simulation and a SAM developed by \citet{Li_et_al_2020a,Li_et_al_2020b}, finding that DF is the most important mechanism determining the coalescence rate of the majority of MBH pairs that coalesce within a Hubble time.

All the above works rely on relatively simple models, based on simplified galactic structures bearing some level of symmetry (e.g. spherical or axi-symmetric). However, recent works have shown that the simple model of DF that has been used for decades is not able to explain the complexity of the orbital decay of MBHs occurring in complex, non-symmetric systems. A short summary review of the different galactic and sub-galactic scale effects that can significantly alter the orbital decay, both delaying and accelerating it depending on the specific configuration, can be found in \citeauthor{Krause_et_al_2025} (\citeyear{Krause_et_al_2025}; see also \citealt{AmaroSeoane_et_al_2023}). Such effects range from the most dramatic perturbations on kpc scales induced by giant molecular clouds and massive star clusters at high redshift to bars and spiral structures, in both gas and stars, ubiquitous at both high and low redshift \citep[][]{Tamburello_et_al_2017,Bortolas_et_al_2020}. In addition to changing the orbital decay time-scale of the MBH pair, they can eject lighter MBHs out of galactic discs, thus completely suppressing the formation of an MBH binary. This has led to the new notion of the ``last kiloparsec problem'' \citep[][]{AmaroSeoane_et_al_2023}, which highlights the difficulty of predicting binary formation time-scales in a realistic galactic potential. We emphasize that this is completely different from the ``last parsec problem'' \citep[][]{Milosavljevic_Merritt_2001}, by now completely overcome in modern simulations \citep[e.g.][]{Berczik_et_al_2006,Vasiliev_et_al_2015,Gualandris_et_al_2017}, which was caused by the oversimplified galactic stellar potential models lacking elongated stellar orbits. The main outcome of all these different types of perturbations is thus to render the MBH decay and binary formation process stochastic by a significant degree.

While the clumpiness of the gaseous and stellar phase of galactic discs, likely hosts of LISA MBHs, is an issue mainly at $z > 1$, since at lower redshift the typical masses of clouds and clusters become lower than those of MBHs \citep[][]{Shibuya_et_al_2016}, non-axisymmetric global distortions such as bars are present in many galaxies and at all redshifts \citep[e.g.][]{LeConte_et_al_2026}. Our own Milky Way possesses a bar encompassing a significant fraction of the disc stellar mass \citep[][]{Portail_et_al_2017}. It is thus sensible to start addressing the effect of bars on the MBH pair dynamics, given their general relevance. \citeauthor{Bortolas_et_al_2022} (\citeyear{Bortolas_et_al_2022}; hereafter \citetalias{Bortolas_et_al_2022}), building upon the work of \citet{Bonetti_et_al_2020,Bonetti_et_al_2021}, have studied the effect of an analytical bar potential on the decay of a lighter secondary MBH with mass 0.05--$1 \times 10^8$~M$_{\sun}$, with the (unmodelled) primary MBH set at the centre of the galaxy, exploring a large parameter space varying the orbital parameters of such MBH (eccentricity, inclination, phase of the orbit, etc.). The outcome was indeed stochastic, with only limited islands of regularity for small subsets of the parameter space, yet it provided a physical understanding of the results. For example, the stalling of the secondary MBH decay was observed as a result in orbit trapping at resonances between the bar pattern speed and the orbit of the MBH, which requires coplanar orbits and is sensitive also to eccentricity and on whether the MBH co-rotates or counter-rotates with the bar. However, this simple toy-model considered only one specific realization for the bar parameters.

In this work, we introduce a new SAM that includes an MBH orbiting a non-trivial galactic structure, in which not only the MBH quantities are varied (as it was the case in \citetalias{Bortolas_et_al_2022}), but also the mass, length, and orbital frequency of a stellar bar are modified. In addition to the enhanced complexity of the new SAM, we also explore the acceleration of the computation of the MBH binary formation time by means of machine learning (ML) techniques. In recent years, ML-based surrogate models have become increasingly common in astrophysics as alternatives to tackle computationally expensive simulations. Once trained on a representative sample of simulated data, they can reproduce the underlying physics processes while reducing computational cost by orders of magnitude. Such approaches have been applied to a broad range of astrophysical problems, including cosmological simulations, GW data analysis, and population studies \citep[e.g.][]{Shallue_et_al_2018,Charnock_et_al_2018,Ntampaka_et_al_2019,Dax_et_al_2021,Villaescusa_Navarro_et_al_2022}. Here, we construct a probabilistic emulator based on conditional normalizing flows \citep[NFs;][]{JimenezRezende_Mohamed_2015,Papamakarios_et_al_2019} as a surrogate model for the MBH orbital decays induced by DF. The model is trained on a large dataset generated with the enhanced SAM to learn the conditional probability distribution of MBH binary formation time as a function of the relevant properties of the host galaxy and the secondary MBH. Unlike deterministic surrogate models, this approach is able to capture the full distribution of possible outcomes and reproduce the stochastic behaviour encoded in the simulations used to produce the training data. Once trained, the emulator provides fast sampling of decay time distributions with a computational cost orders of magnitude lower than that of the original SAM, making it possible to evaluate large cosmological galaxy populations while retaining the information encoded with finer resolution in the galactic-scale dynamic simulations.

The paper is organized as follows: in Section~\ref{sec:methods}, we describe the new SAM, ML, and cosmological framework of our work; in Section~\ref{sec:results}, we present the results; we summarize and conclude in Section~\ref{sec:conclusions}.


\begin{table*}
  \centering
  \caption{Galaxy structural parameters for our $N_{\rm SAM}$ simulations. All models have the same total stellar mass $M_{\star} = M_{\rm bulge} + M_{\rm disc} + M_{\rm bar} = 5.3 \times 10^{10}$~M$_{\sun}$ and the same virial mass $M_{\rm vir} + M_{\star}$ (the stellar mass beyond the virial radius being negligible). The quantities that never vary are: $M_{\rm vir}$, $r_{\rm DM}$, $r_{\rm vir}$ (and, consequently, $c_{\rm vir}$), $a_{\rm bulge}$, $\gamma$, $R_{\rm disc}$, $z_{\rm disc}$, $b_{\rm bar}$, and $c_{\rm bar}$. The bar length ($a_{\rm bar}$) and mass ($M_{\rm bar}$) are taken from a uniform and a log-uniform distribution, respectively, using a \citeauthor{Sobol_1967} sequence. The bar orbital frequency is given by Equation~\eqref{eq:omega}, and the disc and bulge mass are given by Equations~\eqref{eq:discmass} and \eqref{eq:bulgemass}, respectively. Not included in this table is the case with $M_{\rm bar} = 0$, $M_{\rm disc} = 4.7748 \times 10^{10}$~M$_{\sun}$, and $M_{\rm bulge} = 5.252 \times 10^9$~M$_{\sun}$, for which we re-utilized $N_{\rm SAM,B22}$ simulations of \citetalias{Bortolas_et_al_2022}.}
  \label{tab:model}
  \begin{center}
  \tabcolsep=0.11cm
    \begin{tabular}{ccccc}
        \hline
        \hline
        Component & Model & Mass [M$_{\sun}$] & Length [kpc] & Others \T \B \\
        \hline
        \hline
        Halo & \citetalias{Navarro_et_al_1996} (Eq.~\ref{eq:NFW}) & $M_{\rm vir} = 8 \times 10^{11}$ & $r_{\rm DM} = 16$, $r_{\rm vir} = 245$  & $c_{\rm vir} = 15.3125$ \T \B \\
        Bulge & \citeauthor{Dehnen_1993} (Eq.~\ref{eq:bulge}) & $M_{\rm bulge} = [4.16$--$6.05] \times 10^9$ & $a_{\rm bulge} = 0.7$ & $\gamma = 1$ \citep[][]{Hernquist_1990} \T \B \\
        Disc & Exponential (Eq.~\ref{eq:disc}) & $M_{\rm disc} = [1.809$--$4.488] \times 10^{10}$ & $(R_{\rm disc}, z_{\rm disc}) = (3, 0.3)$ & -- \T \B \\
        Bar & Softened needle (Eqs~\ref{eq:bar1}--\ref{eq:bar2}) & $M_{\rm bar} = [0.3$--$3.0] \times 10^{10}$ & $(a,b,c)_{\rm bar}$ = ([1--9], $z_{\rm disc}$, $z_{\rm disc}$) & $\omega_{\rm bar} = 23$--122~km~s$^{-1}$~kpc$^{-1}$ \T \B \\
         \hline
         \hline
    \end{tabular}
  \end{center}
\end{table*}

\section{Methods}\label{sec:methods}

In this section, we describe the SAM employed to construct the training set (Section~\ref{sec:SAM}), the ML surrogate model (Section~\ref{sec:cnf_model}), the cosmological simulation we exploited (Section~\ref{sec:TNG50}), and how we sampled between the SAM and cosmological data (Section~\ref{sec:sampling}).

\subsection{The semi-analytical model}\label{sec:SAM}

We run $N_{\rm SAM} = 10^5$ SAM simulations, employing an improved version of the model described by \citetalias{Bortolas_et_al_2022}, in which we integrate the evolution of an MBH orbiting a multi-component, non-axisymmetric galaxy. This galaxy is deemed to be the remnant of a merger between two galaxies, and the perturber is assumed to be the central MBH of the secondary galaxy. In some instances, this secondary MBH may not be ``naked'', but it could be surrounded by a stellar cusp, which effectively increases the mass of the perturber and affects its dynamics \citep[e.g.][]{VanWassenhove_et_al_2014}. In such cases, the mass of the perturber in our model is assumed to be that of the MBH and the cusp combined, but we caution that we do not allow for any perturber's mass change. The central MBH of the primary galaxy is conjectured to be already at the centre of the remnant galaxy (since DF is more efficient for heavier MBHs)\footnote{This assumption might not hold in the case of a major galactic merger, wherein the masses of the two MBHs are comparable. However, minor mergers are vastly more common than major mergers \citep[e.g.][]{Fakhouri_et_al_2010}}. and never move,\footnote{MBHs are not exactly pinned to the centre, as they may move due to, e.g. gravitational Brownian motion, but it has been shown that this effect is modest \citep[e.g.][]{Merritt_et_al_2007}.} and it is not modelled. The number of simulations run, $N_{\rm SAM}$, was chosen as a compromise between the computational cost needed to generate them and the demand for a sample large enough for the training of a surrogate model (Section~\ref{sec:cnf_model}).

For each simulation, we vary the properties of the galaxy and of the MBH, as described in Sections~\ref{sec:galactic_model} and \ref{sec:black_holes}. All free parameters are allowed to vary within their respective ranges and are sampled uniformly or log-uniformly, depending on the quantity. Since the output of these SAMs is the training set for our ML surrogate model (see Section~\ref{sec:cnf_model}), the samples should be distributed independently on each dimension (to avoid too many repeated values), as uniform as possible (to avoid clustering), and easy to extend. For this reason, we avoided regular grids and independent and identically distributed random variables, and opted instead for the \citet{Sobol_1967} sequence, a deterministic low-discrepancy sequence of quasi-random points assured to evenly cover the entire distribution, recently used in works with astrophysical simulations \citep[e.g.][]{Kacprzak_et_al_2023,Bairagi_et_al_2025}.

The motion of the MBH is integrated until the distance from the galactic centre reaches 10~pc (when we assume the MBH binary has formed\footnote{An MBH binary forms when the separation between the two MBHs is equal to the influence radius of the more massive MBH \citep[e.g.][]{Merritt_2013}, which ranges between $\sim$1 and $\sim$20~pc over more than three orders of magnitude of MBH mass \citep[e.g.][]{Capelo_et_al_2023}. Moreover, the decay time of the last few tens of pc is very short compared to the total decay time \citep[see, e.g.][]{Khan_et_al_2018}. Hence, we simplified the computations by fixing the threshold to 10~pc.} and we record the binary formation time $\tau_{\rm d}$) or when the integration time reaches $2 \, t_{\rm H}$, where $t_{\rm H} = 13.8$~Gyr is the Hubble time (assuming the \citealt{Planck_2016} cosmological parameters; see Section~\ref{sec:TNG50}), whichever occurs first.

\subsubsection{The galactic model}\label{sec:galactic_model}

\begin{figure*}
\includegraphics[width=0.99\textwidth]{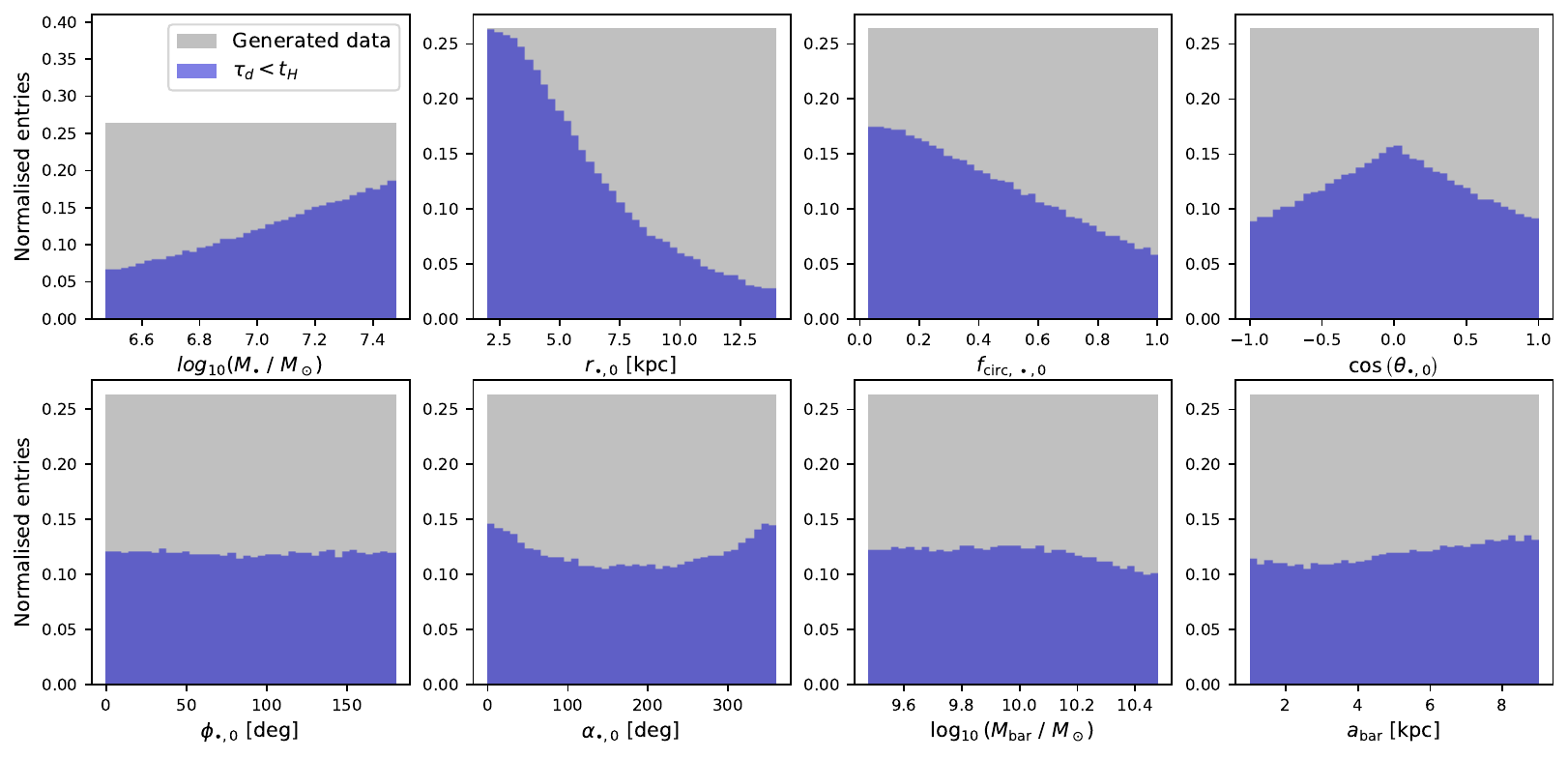}
\caption{Initial distribution of the eight free parameters of our SAM, considering all $N_{\rm SAM}$ simulations (gray), shown together with the distribution of the same quantities, but considering only the 45239 simulations yielding $\tau_{\rm d} < t_{\rm H}$ (blue). All distributions use 40 equally spaced bins and are normalized to the initial number of generated simulations. Orbital decay is promoted when MBHs are initially relatively massive ($M_{\bullet}$), close to the centre ($r_{\bullet,0}$), in radial orbits ($f_{\rm circ,\bullet,0}$), in coplanar orbits ($\theta_{\bullet,0}$), or co-rotating with respect to the bar and disc ($\alpha_{\bullet,0}$); it does not depend on the initial position of the MBH with respect to the bar ($\phi_{\bullet,0}$); and it lightly depends on the bar properties ($M_{\rm bar}$ and $a_{\rm bar}$).}
\label{fig:SAM}
\end{figure*}

Each galaxy is composed of a spherical DM halo and a stellar component, modelled with a spherical stellar bulge and an axisymmetric stellar disc \citep[][]{Bonetti_et_al_2020,Bonetti_et_al_2021}, and a stellar bar (\citetalias{Bortolas_et_al_2022}).

The DM halo is described by a \citeauthor{Navarro_et_al_1996} (\citeyear{Navarro_et_al_1996}; hereafter \citetalias{Navarro_et_al_1996}) profile of density

\begin{equation}
\rho_{\rm DM}  = \frac{M_{\rm DM}}{4 \pi r_{\rm DM}^3} \frac{r_{\rm DM}}{r (1+r/r_{\rm DM})^2}\,,
\label{eq:NFW}
\end{equation}

\noindent where $r$ is the spherical radius, $r_{\rm DM}$ is the DM scale radius, and $M_{\rm DM} = M_{\rm vir} / [\ln (1+c_{\rm vir}) - c_{\rm vir}/(1+c_{\rm vir})]$, with $c_{\rm vir} = r_{\rm vir}/r_{\rm DM}$, $r_{\rm vir}$, and $M_{\rm vir}$ being the DM concentration, virial radius, and virial DM mass, respectively.

The stellar bulge is described by a \citeauthor{Dehnen_1993} (\citeyear{Dehnen_1993}; see also \citealt{Tremaine_et_al_1994}) profile of density

\begin{equation}
\rho_{\rm bulge}  = \frac{(3 - \gamma) M_{\rm bulge}}{4 \pi} \frac{a_{\rm bulge}}{r^{\gamma}(r+a_{\rm bulge})^{3 - \gamma}}\,,
\label{eq:bulge}
\end{equation}

\noindent where $M_{\rm bulge}$ is the (total, integrated to infinity) bulge mass, $a_{\rm bulge}$ is the bulge scale radius, and $\gamma$ is the \citeauthor{Dehnen_1993} index.

The stellar disc is described by a three-dimensional exponential profile \citep[][]{Spitzer_1942,Binney_Tremaine_2008},

\begin{equation}
\rho_{\rm disc}  = \frac{M_{\rm disc}}{4 \pi R_{\rm disc}^2 z_{\rm disc}} \exp \left(-\frac{R}{R_{\rm disc}}\right) {\rm sech}^2 \left( \frac{z}{z_{\rm disc}}\right),
\label{eq:disc}
\end{equation}

\noindent where $M_{\rm disc}$ is the (total, integrated to infinity) disc mass, $R_{\rm disc}$ is the disc scale length, $z_{\rm disc}$ is the disc scale height, and $R$ and $z$ are the cylindric radius and height direction, respectively.

We also include a stellar bar, modelled with a ``softened needle'' profile \citep[][]{Long_Murali_1992}, whose potential is

\begin{equation}
\phi_{\rm bar} (x, y, z) = \frac{G M_{\rm bar}}{2 a_{\rm bar}} \ln \left( \frac{x - a_{\rm bar} + T_-}{x + a_{\rm bar} + T_+} \right),
\label{eq:bar1}
\end{equation}

\noindent with

\begin{equation}
T_{\pm} = \sqrt{(a_{\rm bar}  \pm x)^2 + y^2 + \left[ b_{\rm bar} +(c_{\rm bar}^2 + z^2)^{1/2} \right]^2}\,,
\label{eq:bar2}
\end{equation}

\noindent where $M_{\rm bar}$ is the total bar mass, $a_{\rm bar}$ is the bar scale length in the $x$ direction (hereafter the bar major scale length), $b_{\rm bar}$ and $c_{\rm bar}$ are the bar scale lengths in the $y$ and $z$ directions, respectively (hereafter the bar minor scale lengths), and $G$ is the gravitational constant. The bar is assumed to initially have its major axis along the $x$ axis and to rotate within the $x$-$y$ plane (counterclockwise when viewed from the $z > 0$ half-space) with an orbital frequency, $\omega_{\rm bar}$.

The values of the parameters of our model are listed in Table~\ref{tab:model}. The DM halo parameters -- $M_{\rm vir}$, $r_{\rm DM}$, and $r_{\rm vir}$ -- are always the same for all $N_{\rm SAM}$ simulations and are equal to those quoted in \citetalias{Bortolas_et_al_2022}. The stellar parameters, on the other hand, do change, since we let the mass, major scale length (hereafter, bar length), and orbital frequency of the bar vary for each of the runs. The bar length can vary between 1 and 9~kpc (uniformly sampled), whereas the bar mass can range between $3 \times 10^9$ and $3 \times 10^{10}$~M$_{\sun}$ (log-uniformly sampled). We note that we let $M_{\rm bar}$ and $a_{\rm bar}$ vary independently from each other. This means that there will be a region of the parameter space with small $M_{\rm bar}$ and large $a_{\rm bar}$. In this region, the effect of the bar will be negligible. The stellar bar orbital frequency also varies, and we apply a simple relation with the bar length, by assuming that the bar length is equal to the corotation radius of the galaxy \citep[][]{Sellwood_Wilkinson_1993}, so that the bar frequency is equal to that of the disc. We fit (between 1 and 9~kpc) the galactic rotation curve\footnote{Each of the galactic models we adopt is associated with a different rotation curve. However, in order to minimise the computational burden, we always use the same velocity curve -- computed by \citetalias{Bortolas_et_al_2022} for a barred galaxy (with a bar of mass $3 \times 10^9$~M$_{\sun}$, orbital frequency 40~km~s$^{-1}$~kpc$^{-1}$, and scale lengths 5, 2, and 0.3~kpc) -- as it was shown by \citetalias{Bortolas_et_al_2022} that the velocity curves with and without a bar are very similar (as long as the total stellar mass is the same). We also note that Equation~\eqref{eq:omega} does indeed recover the value of \citetalias{Bortolas_et_al_2022} \citep[40~km~s$^{-1}$~kpc$^{-1}$ at 5~kpc; see also][]{Portail_et_al_2017}.} employed for the DF computations and use

\begin{equation}
\omega_{\rm bar} = \frac{0.21\, a_{\rm bar}^3 - 5.56\, a_{\rm bar}^2 + 47.13\, a_{\rm bar} + 80.15}{a_{\rm bar}}\,.
\label{eq:omega}
\end{equation}

All other parameters -- the stellar bulge and disc scale lengths, the \citeauthor{Dehnen_1993} index, and the stellar bar minor scale lengths -- are always the same for all $N_{\rm SAM}$ SAM simulations. Also, they are equal to those in \citetalias{Bortolas_et_al_2022}, except for the bar minor scale lengths, which are here both equal to the disc scale height (0.3~kpc), since bars originate from gravitational instabilities within the disc, thus inheriting its vertical profile.

When varying the mass of the stellar bar, we require that the total stellar mass is constant and equal to the value given in \citetalias{Bortolas_et_al_2022} ($5.3 \times 10^{10}$~M$_{\sun}$). Thus, we concurrently vary also the mass\footnote{When varying the masses, we keep the scale radii ($R_{\rm disc}$, $z_{\rm disc}$, and $a_{\rm bulge}$) fixed for simplicity, since the variation in mass is not too large.} of the stellar disc as

\begin{equation}
M_{\rm disc} = M_{\rm disc,D} - (M_{\rm bar} - M_{\rm bar,D}) \left(1 - 0.1\frac{a_{\rm bulge}}{a_{\rm bar}}\right)
\label{eq:discmass}
\end{equation}

\noindent and that of the stellar bulge as

\begin{equation}
M_{\rm bulge} = M_{\rm bulge,D} - (M_{\rm bar} - M_{\rm bar,D}) \left(0.1\frac{a_{\rm bulge}}{a_{\rm bar}}\right),
\label{eq:bulgemass}
\end{equation}

\noindent where $M_{\rm disc,D} = 3 \times 10^{10}$~M$_{\sun}$, $M_{\rm bulge,D} = 5 \times 10^9$~M$_{\sun}$, and $M_{\rm bar,D} = 1.8 \times 10^{10}$~M$_{\sun}$, following a similar recipe to that of \citetalias{Bortolas_et_al_2022}. The bar stellar mass fraction, $f_{\rm bar} = M_{\rm bar}/M_{\star}$, varies therefore between 0.0566 and 0.566.

For our study, we also considered the non-barred model by \citetalias{Bortolas_et_al_2022}, wherein $M_{\rm bar} = 0$, $M_{\rm disc} = 4.7748 \times 10^{10}$~M$_{\sun}$, and $M_{\rm bulge} = 5.252 \times 10^9$~M$_{\sun}$. The disc and bulge mass obey Equations~\ref{eq:discmass}--\ref{eq:bulgemass} (with $a_{\rm bar} = 5$~kpc, the bar length in \citetalias{Bortolas_et_al_2022}), so that the total stellar mass is the same as in the barred case. For this particular case, however, we did not re-run our SAM but re-utilized the results by \citetalias{Bortolas_et_al_2022}, who performed $N_{\rm SAM,B22} = 10075$ non-barred runs,\footnote{\citetalias{Bortolas_et_al_2022} actually performed $3 \times 10^4$ runs, using an MBH mass range between $10^5$ and $10^8$~M$_{\sun}$. Since our SAM uses a range $M_{\bullet} \in [3 \times 10^6,3 \times 10^7]$~M$_{\sun}$ (see Section~\ref{sec:black_holes}), we restricted their results to the same range. We additionally note that their integration stopped at 10~pc or $t_{\rm H}$, rather than at 10~pc or $2\,t_{\rm H}$, but this has negligible consequences on the extrapolation of the decay-time curve (see Section~\ref{sec:cnf_model}).} randomly sampling the MBH quantities (described in Section~\ref{sec:black_holes}).

\begin{figure*}
\includegraphics[width=0.99\textwidth]{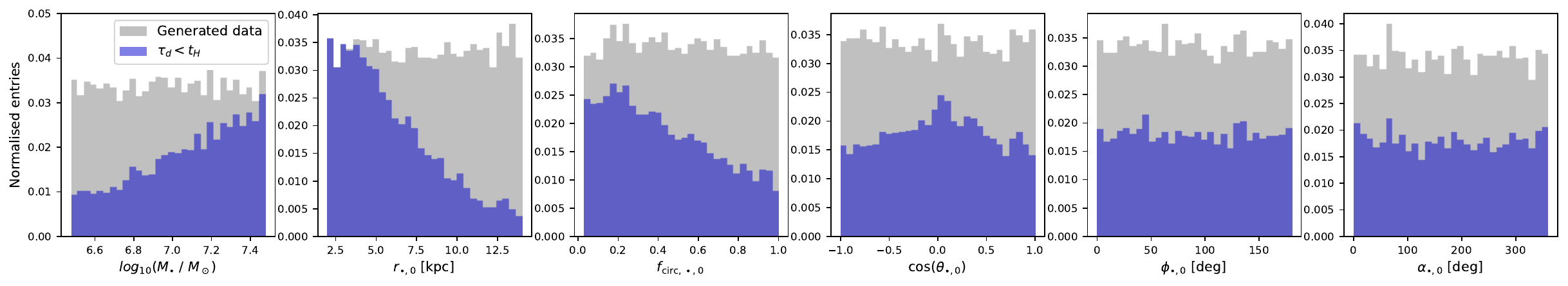}
\includegraphics[width=0.99\textwidth]{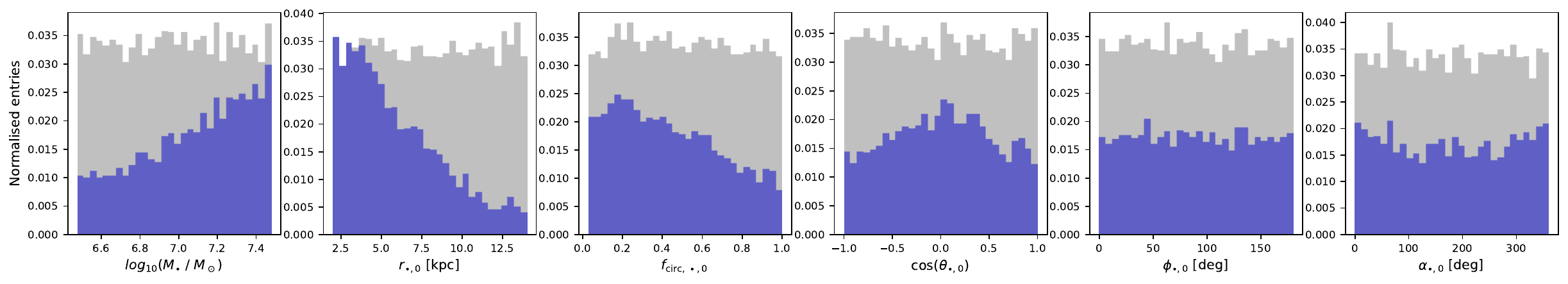}
\caption{Initial distribution of the six free parameters used in \citetalias{Bortolas_et_al_2022} for $N_{\rm SAM,B22}$ non-barred (top panels; part of our training set) and barred (bottom panels; shown here simply for comparison with Figure~\ref{fig:SAM}) galaxies, shown together with the distribution of the same quantities, but considering only the 5372 non-barred (5150 barred) simulations yielding $\tau_{\rm d} < t_{\rm H}$ (blue). All distributions use 40 equally spaced bins and are normalized to the initial number of generated simulations. The granularity of the plots, compared to that of Figure~\ref{fig:SAM}, is due to the sampling method (random instead of \citeauthor{Sobol_1967}). The orbital decay's qualitative dependence on the MBH quantities is the same as in Figure~\ref{fig:SAM}.}
\label{fig:SAM_B22}
\end{figure*}

\subsubsection{The massive black holes}\label{sec:black_holes}

For each SAM run, not only the galaxy's (stellar) structure varies, but so do the MBH's properties.

The MBH parameters that vary are

\begin{itemize}

\item[--] the MBH mass, $M_{\bullet}$, whose value ranges between $3 \times 10^6$~M$_{\sun}$ and $3 \times 10^7$~M$_{\sun}$ (thus capturing a wide range of galactic merger mass ratios). The upper limit's choice is driven by the fact that we wish to apply the results of the SAM to a subset of simulated galaxies from the \textsc{tng50} simulation (see Section~\ref{sec:TNG50}), which we assume are the merger remnants within which the MBHs orbit. Since we will select galaxies with a minimum central stellar mass of $3 \times 10^9$~M$_{\sun}$, we set the maximum MBH mass to one per cent of that value. This ensures that the ratio of the mass of the orbiting MBH to that of its merging host galaxy is consistent with observations \citep[e.g.][]{Zhang_et_al_2023}. Near the upper limit of the MBH mass range, this is valid if the merger mass ratio is $\gtrsim 0.1$ or, in the case of more minor mergers, if the perturber can be thought as an MBH surrounded by a stellar component \citep[e.g.][]{VanWassenhove_et_al_2014}. The lower limit is given instead by the fact that most low-mass MBHs would not reach the centre of the remnant within a Hubble time (see Figure~\ref{fig:SAM}).

\item[--] the initial position of the MBH, described by the distance from the galactic centre, $r_{\bullet,0} \in 2$--14~kpc; the azimuthal angle, $\phi_{\bullet,0} \in [0,180^{\circ})$,\footnote{The range should be $[0,360^{\circ})$, but for symmetry reasons it can be halved.} with 0 meaning that the MBH sits along the longest principal axis of the bar at time 0; and the angle between the positive polar axis and the MBH initial position vector, $\theta_{\bullet,0} \in [0$,$180^{\circ}]$, with $90^{\circ}$ meaning that the MBH sits on the $x$-$y$ plane.

\item[--] the initial velocity of the MBH, ${\bf v}_{\bullet,0}$, described by $f_{\rm circ,\bullet,0} \in [0.03,1]$ (we avoid extremely low values, as they effectively imply an MBH with initial null velocity), the ratio between the tangential velocity and the circular velocity at $r_{\bullet,0}$ (computed in the plane), with a small $f_{\rm circ,\bullet,0}$ indicating a very radial (plunging) orbit, since the MBH has very little angular momentum for its radius, and a large $f_{\rm circ,\bullet,0}$ implying a very circular orbit; and by the angle between the initial velocity vector (always set perpendicular to the position vector; hence, we only need one angle to determine the velocity's direction) and the disc's $x$--$y$ plane, $\alpha_{\bullet,0} \in [0,360^{\circ})$, with 0 (prograde with respect to the bar and disc rotation, except when $\theta_{\bullet,0} = 0$ or $180^{\circ}$) and $180^{\circ}$ (retrograde) meaning that the velocity vector lies parallel to the $x$-$y$ plane. This setup implies that the initial radial velocity is always zero: in other words, we are assuming that we start modelling the MBH's orbit when it is at apocentre. By sampling all possible values of positions and velocities, we implicitly sample all possible orientations between the bar and the MBH orbital plane.

\end{itemize}

The MBH quantities above are sampled using \citeauthor{Sobol_1967} sequences,\footnote{Except in the non-barred case, wherein the quantities are randomly sampled, since we re-utilized the results by \citetalias{Bortolas_et_al_2022}.} log-uniformly in the case of the mass and uniformly in the case of the position\footnote{Except for $\theta_{\bullet,0}$, which is sampled from $(180^{\circ}/\pi)\arccos(-1+2 \times [0$--$1])$, where [0--1] is a uniform distribution, to ensure that the three-dimensional space is isotropically sampled in spherical coordinates. 
} and velocity. The initial distributions of these six quantities are shown (in gray) in Figures~\ref{fig:SAM} and \ref{fig:SAM_B22} for our $N_{\rm SAM}$ simulations and the $N_{\rm SAM,B22}$ non-barred runs of \citetalias{Bortolas_et_al_2022}, respectively. In Figure~\ref{fig:SAM}, we additionally show the initial distributions of $M_{\rm bar}$ and $a_{\rm bar}$ from Section~\ref{sec:galactic_model}. The noisier initial conditions distributions of \citetalias{Bortolas_et_al_2022} are entirely due to the random sampling. We also note that, while the position and velocity of the MBH obviously vary during the integration, all other quantities (the mass of the MBH and all galactic properties) do not.

\subsection{The machine learning surrogate model}\label{sec:cnf_model}

For each set of initial conditions, given by the properties of the galaxies and of the MBHs discussed in Section~\ref{sec:SAM}, the complete dynamical evolution of the system is obtained running the SAM simulations. The time \td it takes for each secondary MBH to reach 10~pc from the galactic centre is studied at the population level. In order to study the distribution of \td under different conditions, a large number of simulations needs to be carried out, resulting in a large computational and time consumption. In this work, we use a surrogate model based on conditional NFs to learn the probability distribution of \td conditioned on a set of physical parameters describing the system. This architecture was chosen as it accurately captures the stochastic behaviour of the simulations at population level while requiring relatively low computational cost at training. 
Once trained, the model can evaluate the expected distribution of \td for a given population orders of magnitude faster than the actual simulations, with minimal computational effort.

The model is conditioned on the most relevant properties that define the galaxy-MBH systems for this problem. They are fed into the model as a vector of nine features, $\vec{c} = [\log_{10}(M_{\bullet}), r_{\bullet,0}, \phi_{\bullet,0}, \theta_{\bullet,0}, f_{\rm circ,\bullet,0}, \alpha_{\bullet,0},  f_{\rm bar}, a_{\rm bar}, x_{\rm bar}]$, where $x_{\rm bar}$ is a Boolean flag encoding whether the galaxy is barred, which is introduced to allow the model to learn the difference between galaxies with or without a stellar bar. All features are normalized before being fed to the model. A version of the model (without $x_{\rm bar}$, i.e. with eight features) was trained allowing the bar stellar mass fraction and length to take both discrete values -- zero in the case of non-barred galaxies -- and a continuous range of values -- for barred galaxies -- and the performance was observed to degrade with respect to the training with the flag $x_{\rm bar}$, as the model cannot interpret the ambiguity after input feature normalization without said flag. More details about conditional NFs and input feature normalization can be found in Appendix~\ref{App:CNF}.

The architecture used for the surrogate model is a neural spline flow \citep[][]{Durkan_et_al_2019}, implemented using version 3.3.0 of the \textsc{PZFlow} Python package \citep[][]{Crenshaw_et_al_2024}. The chosen model consists of a conditional NF transforming a base one-dimensional normal distribution into the target distribution by means of a bijective transformation parametrized by a neural spline coupling layer, which follows the rational-quadratic spline formulation. Specifically, a transformation with 128 spline bins, six hidden layers, and 64 units per hidden layer is used.\footnote{These values were chosen for the final implementation of the surrogate model with a hyperparameter scan as the combination with the lowest number of learnable parameters that could reproduce the population of both barred and non-barred galaxies. Models with fewer hidden layers and fewer units per hidden layer failed to reproduce both populations at the same time, whereas models with larger dimensions were found to bring negligible improvement in the predictions.} The flow is conditioned on the nine input features, $\vec{c}$, discussed above. 

The model is trained using $N_{\rm SAM}+N_{\rm SAM,B22}$ SAM simulations, comprising galaxies with and without bars. The simulations run up to a maximum given time and, for the instances wherein the secondary MBH has not decayed (to within 10~pc from the centre) within said time, a value of infinity is assigned to $\tau_{\rm d}$. Conditional NFs can struggle with sharp boundaries and discontinuities in the target distribution to be learned. In addition, training the model only with the systems that have formed a binary within the given simulation time limit means that the model would not be able to learn the morphologies that effectively do not form a binary. Thus, in order to avoid the sharp cut-off in the \td distribution from SAM simulations, and to provide the model with the whole phase space simulated, the infinite values of \td are resampled by extrapolating the finite-$\tau_{\rm d}$ distribution tail into larger times than allowed in the simulation. This extrapolation is done by fitting a Gaussian kernel density estimation (KDE) to the \td distribution and employing a Weibull function to extend the fit beyond the simulated regime. The KDE and Weibull functions are matched by requiring that the probability at the boundary of maximum simulated time is identical for both, ensuring continuity of the total function. The systems that did not form a binary in the SAMs are then assigned a random value for \td drawn from the resulting Weibull function. Since the simulation is allowed to run for a significant time ($t_{\rm H}$ or $2 \, t_{\rm H}$, depending on the SAM), this extrapolation avoids noisy behaviour around said boundaries while affecting only an unphysical regime (longer than the age of the Universe) that does not alter the predictions of the model in a significant way. The simulated SAM data are split randomly into a training and a validation dataset, following an $\sim$80-20~per cent ratio. Model parameters are optimized by minimizing the negative log-likelihood (NLL) of the training data, using the \textsc{Adam} \citep[][]{Kingma_et_al_2015} optimizer. The training is carried out in stages with progressively reduced learning rates in order to optimally find the minima of the NLL function. Early stopping with a patience of 20 epochs is applied based on the NLL loss over the validation dataset to mitigate overfitting.

\subsection{The \textsc{tng50} simulation and the retrieval of the galactic quantities}\label{sec:TNG50}

The first application of our emulator aims to estimate the MBH pairing time-scale distribution in a cosmological simulation. \textsc{IllustrisTNG}\footnote{\url{https://www.tng-project.org/}} \citep[][]{Weinberger_et_al_2017,Pillepich_et_al_2018a} is a publicly available suite of cosmological, magnetohydrodynamical simulations, run over different cosmological volumes and with different resolutions with the \textsc{arepo} code \citep[][]{Springel_2010}. We focus here on the \textsc{tng50-1} simulation \citep[hereafter \textsc{tng50};][]{Pillepich_et_al_2019,Nelson_et_al_2019}, run from $z = 127$ to 0 in a volume of (35~cMpc~$h^{-1}$)$^3$ using $2160^3$ DM particles (of mass $4.5 \times 10^5$~M$_{\sun}$) and (initially) $2160^3$ gas cells (of mass $8.5 \times 10^4$~M$_{\sun}$), and assuming the Planck 2015 \citep[][]{Planck_2016} cosmological parameters ($\Omega_{\rm m} = 0.3089$, $\Omega_{\rm b} = 0.0486$, $\Omega_{\Lambda} = 0.6911$, $h = 0.6774$, $\sigma_8 = 0.8159$, and $n_{\rm s} = 0.9667$). The \textsc{tng50} simulation was preferred to its largest-volume counterparts \citep[\textsc{tng100} and \textsc{tng300};][]{Nelson_et_al_2018,Pillepich_et_al_2018b,Springel_et_al_2018,Naiman_et_al_2018,Marinacci_et_al_2018} because of its superior mass and spatial resolution, with a Plummer equivalent gravitational softening for DM and stars of 576~cpc for $z \ge 1$ and 288~pc for $z < 1$, and an adaptive gravitational softening for the gas, whose minimum is 72~cpc $\forall \,z$. The simulation includes metal and primordial gas heating and cooling in the presence of a redshift-dependent ultraviolet background and a radiation field by AGN; stochastic star formation and evolution of stellar populations from a given initial mass function; stellar feedback and chemical enrichment \citep[for more details, see][]{Pillepich_et_al_2018a}; and MBH seeding, growth, and feedback, as described below.

MBHs in \textsc{tng50} are modelled following the prescription by \citeauthor{Weinberger_et_al_2017} (\citeyear{Weinberger_et_al_2017}; see also \citealt{Weinberger_et_al_2018}). They are seeded (with an initial mass of $1.18 \times 10^6$~M$_{\sun}$) at the potential minimum of a DM halo, if such a halo has a DM mass exceeding $7.38 \times 10^{10}$~M$_{\sun}$ and does not already host an MBH. At every global integration time-step, the MBH is repositioned at the potential minimum of its host (with its velocity set to the mean mass-weighted velocity of the potential minimum region), to prevent spurious motion driven by two-body discreteness effects and numerical $N$-body noise. MBHs can accrete gas according to an Eddington-limited (\citealt{Eddington_1916}; assuming a radiative efficiency $\epsilon_{\rm r} = 0.2$), unboosted, Bondi--Hoyle--Lyttleton recipe (\citealt{Hoyle_Lyttleton_1939,Bondi_Hoyle_1944,Bondi_1952}; using a kernel-weighted average over neighbouring cells) and feed back energy to the surrounding medium in the form of thermal or kinetic feedback, depending on the accretion level. The MBHs merge instantaneously when their separation is smaller than the MBH smoothing length, regardless of their relative velocity. As a result, the ensuing merger separations are relatively large (mostly because of the repositioning scheme) and show a bimodal distribution (with peaks at $\sim$0.7 and $\sim$3~kpc), due to the kinetic feedback, which decreases the gas density around the MBH and, as a consequence, increases the size of the nearest-neighbour gas cells of the MBH, i.e. the smoothing length computed for the pre-merger separation (see \citealt{Izquierdo-Villalba_et_al_2026} for details). Indeed, these limitations in MBH modelling, almost inevitable in large cosmological simulations, are amongst the reasons why we decided to pursue a different approach: we ignore the MBHs from the simulation and instead effectively populate the \textsc{tng50} galaxies with MBHs from the SAM.

For each snapshot of interest, we first considered all galaxies \citep[identified as gravitationally bound substructures by \textsc{subfind};][]{Springel_et_al_2001} and selected those with at least $10^4$ stellar particles (thus with a total stellar mass $\gtrsim 5$--$7 \times 10^8$~M$_{\sun}$, depending on the redshift). The measured galaxy features were quantified using a dedicated code, the \textsc{mor}phological \textsc{d}ec\textsc{o}mpose\textsc{r} \citep[\textsc{mordor};][]{Zana_et_al_2022}, which identifies and characterizes the galactic components in an automated way, without requiring any visual inspection. \textsc{mordor} decomposes the stellar component of each galaxy by splitting the stellar particles in up to five regions of the total specific energy-circularity phase space, resulting in up to five constituents: a thin disc (of mass $M_{\rm thin\,disc,\textsc{tng50}}$), a thick disc (of mass $M_{\rm thick\,disc,\textsc{tng50}}$), a central spheroid (hereafter, bulge; of mass $M_{\rm bulge,\textsc{tng50}}$), a pseudo-bulge (of mass $M_{\rm pseudo{\text -}bulge,\textsc{tng50}}$), and a halo (of mass $M_{\rm halo,\textsc{tng50}}$).\footnote{\label{foot:database}The algorithm also finds kinematically decoupled or unbound structures. However, the combined mass of these particles ($M_{\rm unbound,\textsc{tng50}}$) amounts to less than 3~per cent of the total stellar mass in $\sim$97--100~per cent of the galaxies in each snapshot, depending on the redshift. All these quantities are retrievable in the database at \url{https://www.tng-project.org/data/docs/specifications/\#sec5t} and their distributions are shown in Figure~\ref{fig:TNG-mass-histograms}.} It is not straightforward to recover one-to-one relationships between these components and those of the SAM. The pseudo-bulge particles would likely belong to either the SAM disc or the SAM bulge. Also, many thick-disc particles would belong to the SAM disc, depending on the scale-height of the thick disc of each galaxy. Finally, most halo particles would not belong to any of the SAM components. For these reasons, we consider for simplicity only four components and define the central stellar mass $M_{\rm \star,\textsc{tng50}} = M_{\rm thin\,disc,\textsc{tng50}} + M_{\rm thick\,disc,\textsc{tng50}} + M_{\rm pseudo{\text -}bulge,\textsc{tng50}} + M_{\rm bulge,\textsc{tng50}}$, neglecting both the unbound stellar particles and the stellar halo.

For each of these galaxies, \textsc{mordor} additionally determines the presence (or absence) of a stellar bar, via a Fourier decomposition of the stellar surface density field, applying and improving upon the methods described in \citet{Zana_et_al_2018a,Zana_et_al_2019}. A galaxy is deemed to be barred (and thus have $x_{\rm bar} = 1$ in $\vec{c}$\,; see Section~\ref{sec:cnf_model}) if a combination of several quantities meets set thresholds: these include the fraction of stellar kinetic energy in ordered rotation, the local maximum of the radial profile of the ratio between the second and the zeroth term in the Fourier decomposition (i.e. the bar strength), the bar length, the ratio between the vertical and radial velocity dispersions, and the ratio between the masses of two halves of the bar (for details and for the thresholds selected, see \citealt{Zana_et_al_2022}). For the barred galaxies, the algorithm provides, amongst other things, the bar strength and the bar length, $a_{\rm bar,\textsc{tng50}}$. More specifically, the algorithm computes the radial profile of the ratio of the second to the zeroth term in the Fourier decomposition of the face-on stellar surface density field, $A_2(R)$, and its cumulative counterpart, $A_2(<R)$. It then provides an estimate for the bar strength as the local maximum of the former profile, $A_{\rm 2,max}(R)$, and uses both profiles to calculate the bar edges (both inner and outer) by computing when the phase of the Fourier mode ceases to be approximately constant when moving inwards and outwards from the position of the profiles' peaks.

For this work, we also modified and re-ran \textsc{mordor} in order to additionally provide the bar mass, $M_{\rm bar,\textsc{tng50}}$, defined as the mass within a cuboid of length equal to the bar length (aligned with the bar) and the other two dimensions equal to 2.8 times the (physical) Plummer equivalent stellar gravitational softening (i.e. the stellar spline size, beyond which the gravitational force becomes exactly Newtonian; see, e.g. \citealt{Kim_et_al_2016}), and excluding the particles within the inner bar edge, which is usually smaller than the stellar spline size.

We selected a few representative snapshots (listed in Table~\ref{tab:tng50_snapshots}) to encompass different evolutionary phases: when the bar fraction, defined as the fraction of galaxies with a bar strength larger than a given threshold, is relatively low (snapshots $\mathbb{S}$34--$\mathbb{S}$38, at $z = 1.9$--1.6), when the bar fraction is near its peak (snapshots $\mathbb{S}$44--$\mathbb{S}$50, at $z = 1.25$--1), and at late times (snapshots $\mathbb{S}$67, $\mathbb{S}$72, $\mathbb{S}$78, $\mathbb{S}$84, $\mathbb{S}$91, and $\mathbb{S}$99, at $z = 0.5$, 0.4, 0.3, 0.2, 0.1, and 0, respectively; see figure~17 of \citealt{Zana_et_al_2022}).

\begin{figure*}
\centering
\includegraphics[width=0.99
\textwidth]{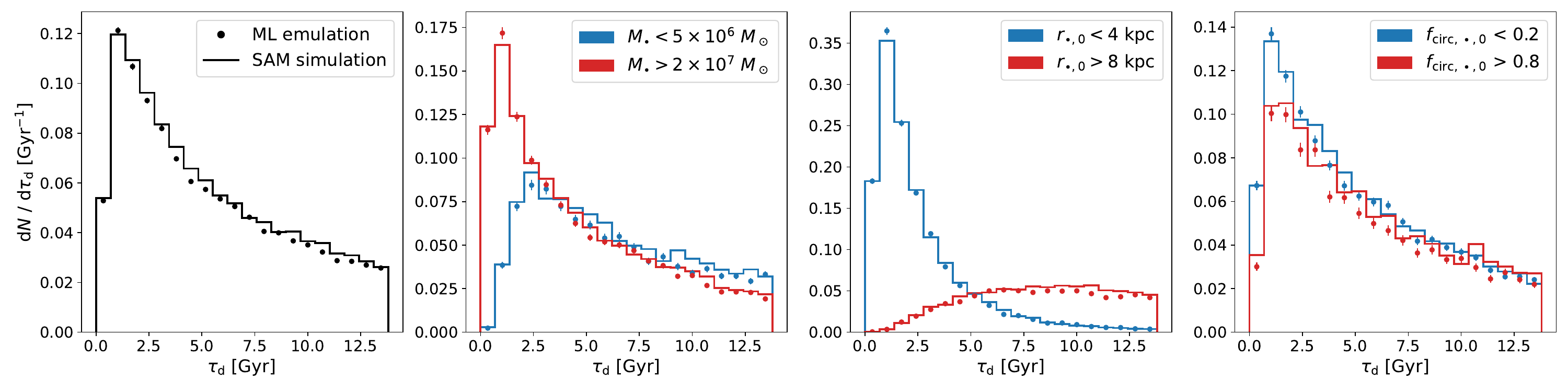}
\caption{Decay time distribution comparison between the SAM simulations (solid lines) and the emulation with the conditional-NF model (dots). The emulation results include statistical error bars that reflect the size of the dataset remaining after each selection. The first panel shows the distribution comparison for the SAM validation set [$\sim$$0.2(N_{\rm SAM}+N_{\rm SAM,B22})$]. The other panels show the same comparison in different selections of the mass, the initial position, and the initial velocity of the MBHs, respectively. All distributions use 20 equally spaced bins between 0 and $t_{\rm H}$ and are normalized to the number of systems considered in each subset.}
\label{fig:td}
\end{figure*}

From each snapshot, we first selected all galaxies with a central stellar mass $M_{\rm \star,\textsc{tng50}} \geq 3 \times 10^9$~M$_{\sun}$ (i.e. at least 100 times more massive than the most massive MBH in the SAM; see the fifth column of Table~\ref{tab:tng50_snapshots}): above this mass threshold, most \textsc{tng50} galaxies are considered ``discy'' \citep[see figure~14 of][]{Zana_et_al_2022}, hence structurally similar to the SAM galaxies. We abstained from imposing also an upper limit on the mass, one that would have avoided considering galaxies much more massive than the SAM galaxy, because the number of galaxies with $M_{\rm \star,\textsc{tng50}} \geq 3 \times 10^{11}$~M$_{\sun}$ is always less than 1 per cent of the total in all the snapshots we analysed. We then divided the sample into non-barred and barred galaxies, according to the \textsc{mordor} algorithm. All $N_{\rm no{\text -}bar}$ non-barred galaxies with $M_{\rm \star,\textsc{tng50}} \geq 3 \times 10^{11}$~M$_{\sun}$ comprise the sample of non-barred galaxies of this work. With regards to the barred galaxies, we imposed three different constraints on the bar strength (keeping all other constraints the same), in order to consider all bars \citep[including the proto-bars; see][]{Zana_et_al_2022}, with $A_{\rm 2,max}(R) \ge 0.1$, the moderate and strong bars, with $A_{\rm 2,max}(R) \ge 0.2$, or the strong bars, with $A_{\rm 2,max}(R) \ge 0.4$ (see the sixth column of Table~\ref{tab:tng50_snapshots}). Finally, from these three sets of massive barred galaxies, we picked those with a bar stellar mass fraction, $f_{\rm bar,\textsc{tng50}} = M_{\rm bar,\textsc{tng50}}/M_{\rm \star,\textsc{tng50}}$,\footnote{We note that the stellar particles deemed by \textsc{mordor} to be part of the bar are also part of one of the other galactic components. Therefore, the bar mass should not be added to the masses of the other structures to obtain the total stellar mass, as opposed to what is done in the SAM, which treats the bar as a separate entity.} and a bar length within the same limits of the SAM (0.0566--0.566 and 1--9~kpc, respectively), obtaining $N_{\rm bar}$ systems (listed in the seventh column of Table~\ref{tab:tng50_snapshots}). Hereafter, when referring to barred galaxies, we exclude the barred galaxies with $f_{\rm bar,\textsc{tng50}}$ and $a_{\rm bar,\textsc{tng50}}$ outside the SAM ranges. The total number of galaxies analyzed in each snapshot is thus $N_{\rm TNG} = N_{\rm no{\text -}bar} + N_{\rm bar}$. The central stellar mass distribution of these galaxies (along with the distributions given by the intermediate steps of this filtering procedure) are shown in Figure~\ref{fig:TNG-central-mass-histograms}.

\subsection{Evaluation of the surrogate model on TNG50 simulations}\label{sec:sampling}

In order to study the distribution and evolution of MBH binary formation time-scales in the \textsc{tng50} simulation, we extrapolate the performance of the conditional-NF model trained on the SAM data described in Section~\ref{sec:SAM}. Both the SAM and \textsc{tng50} datasets are considered simultaneously by means of a cross joint sampling. In practice, this sampling approach is implemented by taking galaxy-MBH systems in the SAM dataset and replacing key features of that galaxy by their equivalent from \textsc{tng50} galaxies. This is done for each element in both datasets, with each galaxy from the \textsc{tng50} simulation sampled as many times as elements in the SAM dataset. This effectively corresponds to injecting the perturber MBHs into the \textsc{tng50} galaxies. The only systems considered from the SAM dataset in the cross joint are the ones from the validation dataset, i.e. the dataset resulting from the initial $\sim$80--20 split of the SAM data prior to the model training and that is only seen by the model for the early-stopping criterion: $N_{\rm MBH} \simeq 0.2 (N_{\rm SAM}+N_{\rm SAM,B22})$. The size of the final set is thus the product of the sizes of the validation SAM runs and \textsc{tng50} simulations: $N_{\rm MBH} \times N_{\rm TNG}$. The conditional-NF model is then evaluated on the resulting cross joint dataset, which now contains all the necessary information from the relevant \textsc{tng50} physics and complemented with the SAM runs for the less relevant features, and with the results being discussed in Section~\ref{sec:individual_snapshots}. This cross joint approach was validated with different runs, for which each \textsc{tng50} galaxy was injected with a different number of SAM MBHs to observe the potential induced biases or correlations. The statistical distributions of \td obtained in these runs were consistent within statistical uncertainties with each other, showing that no such biases are introduced into the resulting dataset.

The only key features that are replaced in the SAM galaxies by \textsc{tng50} information are the bar stellar mass fraction and the bar length, which are expected to capture the dominant dependence of the orbital evolution on the large-scale barred structure of the galaxy. In addition, the model is trained to estimate the distributions of decay times, $\tau_{\rm d^{\prime}}$, for galaxies of a fixed mass, which was chosen to be $M_{\star}=5.3\times 10^{10}M_{\sun}$. The time of MBH binary formation as induced by DF is expected to depend on the total mass of the galaxy and, since the surrogate model is blind to this information, a post-processing correction to the predicted decay time for \textsc{tng50} galaxies is applied. We renormalize the results in the following way: assuming a Maxwellian velocity distribution (and a fixed Coulomb logarithm, $\ln \Lambda$), the DF time is proportional to the product of the mass of the host, $M_{\rm host}$, and its dynamical time, $t_{\rm dyn}$, hence on the square root of the mass of the host \citep[e.g.][]{Binney_Tremaine_2008,Bovy_2026}: $\tau_{\rm DF} \propto M_{\rm host} t_{\rm dyn} / (M_{\rm \bullet} \ln \Lambda) \propto M_{\rm host}^{1/2} / (M_{\rm \bullet} \ln \Lambda)$. This is an exact result in the case of the singular isothermal sphere, but it has been shown to be valid also when assuming other profiles (e.g. the \citeauthor{Hernquist_1990} profile assumed in our SAM; see \citealt{SouzaLima_et_al_2017,Tamburello_et_al_2017}). Therefore, we multiply each $\tau_{\rm d^{\prime}}$ by $(M_{\rm \star,\textsc{tng50}}/M_{\star})^{1/2}$, to finally obtain the decay time, $\tau_{\rm d}$. Consequently, the decay times predicted by the emulator are increased for the \textsc{tng50} galaxies that are more massive than the fiducial SAM galaxy, and reduced for less massive systems. This correction should be regarded as an approximate scaling, which we discuss in Section~\ref{sec:conclusions}.


\section{Results}\label{sec:results}

\begin{figure}
\centering
\includegraphics[width=0.45
\textwidth]{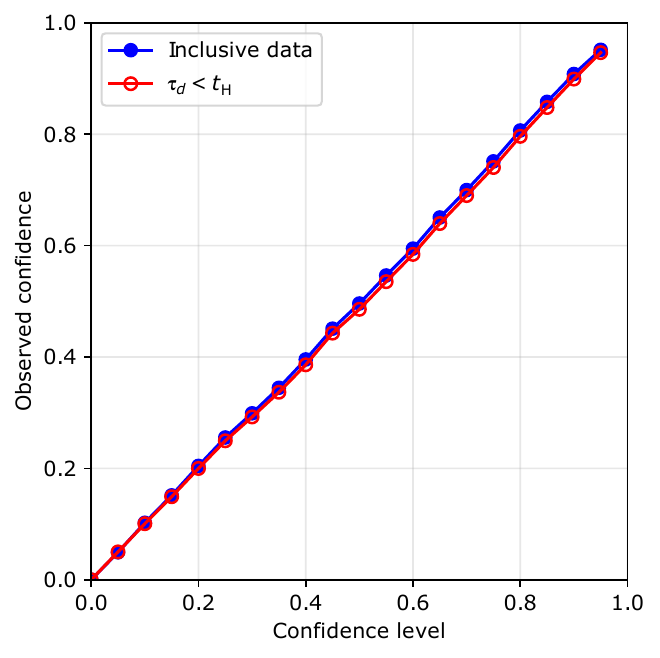}
\caption{Probability-probability distribution for the trained model with the inclusively generated dataset (blue) and for systems with SAM prediction of $\tau_{\rm d} < t_{\rm H}$. The $x$-axis shows reference confidence intervals, whereas the $y$-axis shows the frequency of the simulated $\tau_{\rm d}$ value to be contained within such interval according to the conditional-NF emulator. The closer the drawn line is to the diagonal line, the higher the confidence in the calibration of the model. }
\label{fig:pp}
\end{figure}

\subsection{Semi-analytical results}\label{sec:semi-analytical_results}

In this section, we briefly describe the results of the SAM runs, both for the barred and non-barred case. In Figure~\ref{fig:SAM} and in the top panels of Figure~\ref{fig:SAM_B22}, we show the final distributions (in blue) of the free parameters of our SAM model, i.e. the distribution of the MBH (and bar, in the barred SAM case) quantities yielding an MBH binary formation time $\tau_{\rm d} < t_{\rm H}$. As expected, MBH pairing is promoted for MBHs that are initially relatively more massive (large $M_{\bullet}$), closer to the centre (small $r_{\bullet,0}$), on more radial orbits (small $f_{\rm circ,\bullet,0}$), and on more coplanar orbits ($\theta_{\bullet,0}$ around $90^{\circ}$ and $\alpha_{\bullet,0}$ close to 0 or $180^{\circ}$ -- this last result is not so clear from the \citetalias{Bortolas_et_al_2022} sample, likely due to the increased noise): all these choices increase the efficiency of DF, which is stronger when the perturber is more massive and finds itself in a denser environment \citep[][]{Chandrasekhar_1943}. The effect of the bar properties on the distributions is not massive, with the MBH decay being very slightly hindered with more massive (large $M_{\rm bar}$) and shorter (small $a_{\rm bar}$) bars (i.e. when the bar density is enhanced) and being virtually independent of the initial position of the MBH with respect to the bar ($\phi_{\bullet,0}$). We also note that the presence or absence of the bar itself has a measurable effect, with the fraction of MBHs decaying within a Hubble time being slightly larger in non-barred galaxies (53.3 per cent) than in barred galaxies (45.2 per cent).\footnote{As a comparison, the same fraction in the case of the barred galaxies in the sample of \citetalias{Bortolas_et_al_2022} is 51.1 per cent (see the bottom panels of Figure~\ref{fig:SAM_B22}).}

It is instructive to study the distribution of decay times for the systems with $\tau_{\rm d} \le t_{\rm H}$ (i.e. those depicted by the blue histograms in Figure~\ref{fig:SAM}). This is shown by the solid line in the first panel of Figure~\ref{fig:td}, which displays a clear peak around $\sim$1~Gyr and then a decay towards longer $\tau_{\rm d}$'s. The shape of this distribution can be understood when investigating the second and third panels: heavier/closer MBHs produce a narrow, peaked distribution, whereas the lighter/farther MBHs produce a wider tail.

\subsection{Training and performance of the surrogate model on semi-analytical model data}\label{ML_performance}

The conditional-NF model discussed in Section~\ref{sec:cnf_model} was trained with $\sim$$0.8(N_{\rm SAM}+N_{\rm SAM,B22}) = 8.6 \times 10^4$ 
SAM simulations to learn as target the distribution of times \td that the perturber takes to reach 10~pc from the galactic centre. The model was trained in a workstation equipped with an NVIDIA RTX A5000 GPU with 24 GB of memory. The final model converged after 400 epochs and required approximately 10 minutes of wall-clock training time. Once trained, the model can produce predictions for $10^7$ samples in about 15 seconds, which entails a very small fraction of the time used to run the SAM simulations (Section~\ref{sec:SAM}).

\begin{figure}
\includegraphics[width=0.47\textwidth]{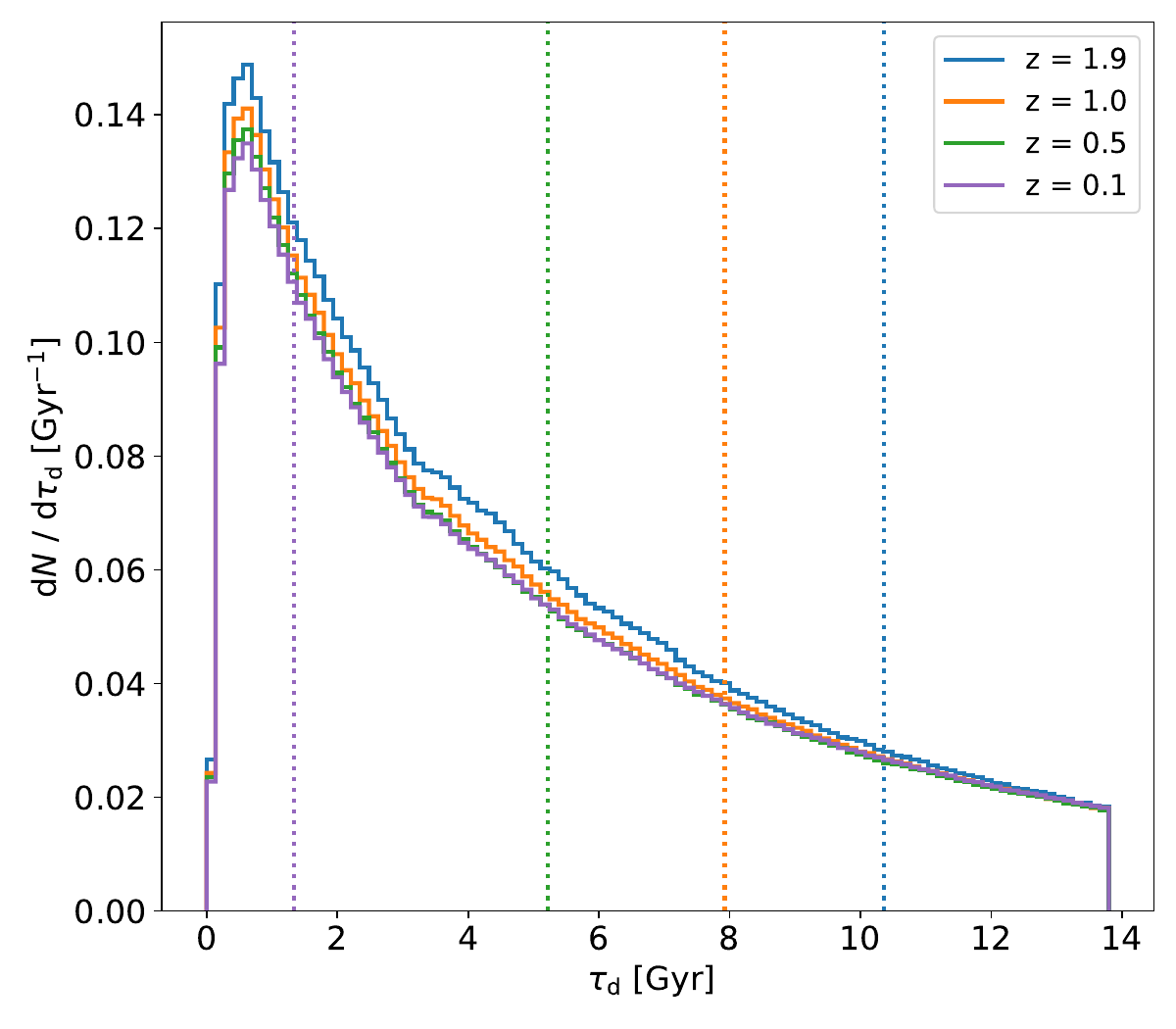}
\caption{Distribution of decay times for four select \textsc{tng50} snapshots ($\mathbb{S}34$, $\mathbb{S}50$, $\mathbb{S}67$, and $\mathbb{S}91$; see Table~\ref{tab:tng50_snapshots}), including all galaxies contemplated in this work (i.e. non-barred galaxies and barred galaxies with a bar stellar mass fraction and a bar length within the SAM ranges, all with a central stellar mass above $3 \times 10^9$~M$_{\sun}$), normalized by the number of galaxies considered in each snapshot. The vertical lines denote the lookback time of each snapshot.}
\label{fig:TNG_some_snapshots_linlin_all_galaxies}
\end{figure}

\begin{figure*}
\includegraphics[width=0.97\textwidth]{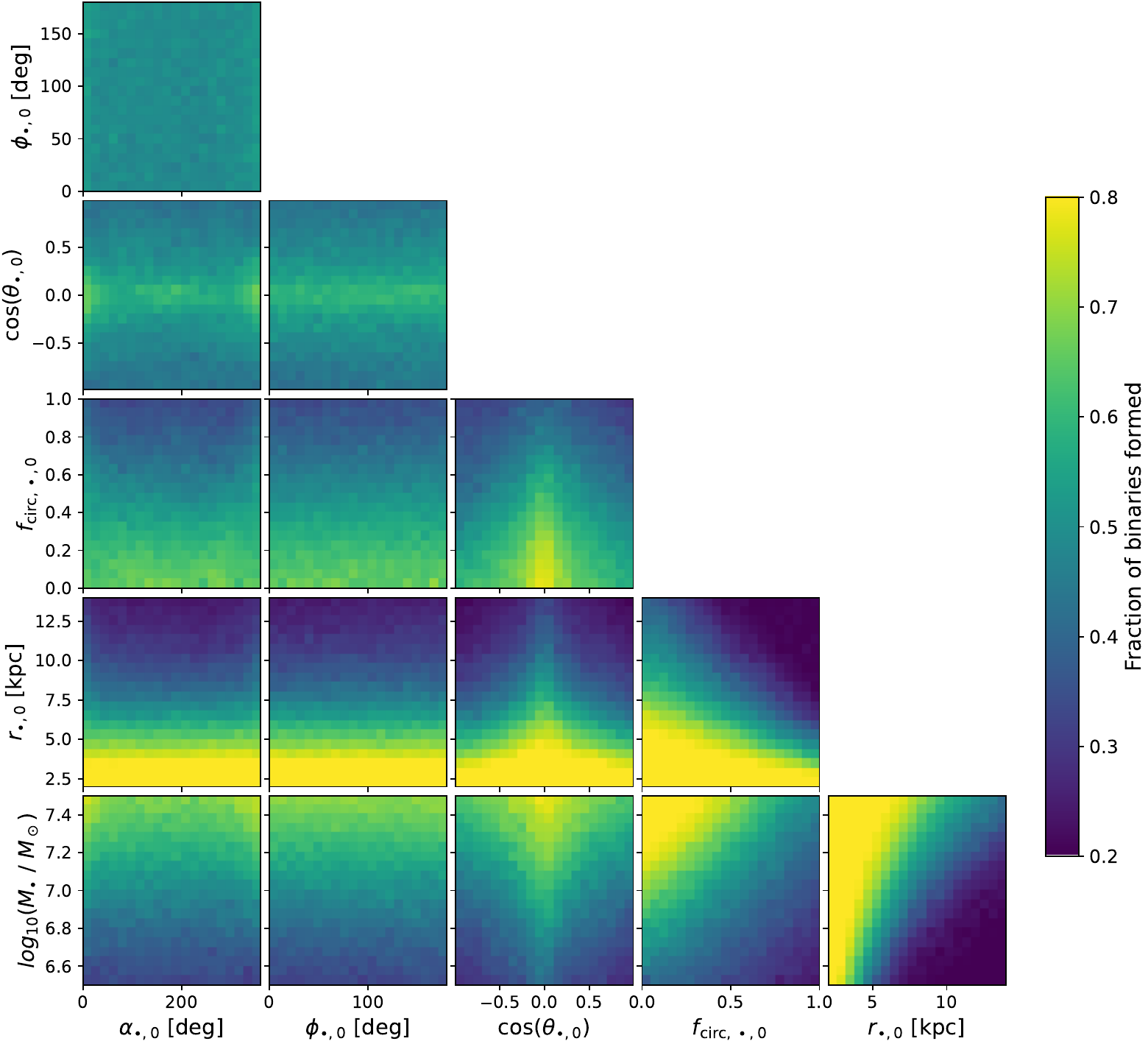}
\caption{Fraction of secondary MBHs for all considered \textsc{tng50} galaxies at $z = 1$ (snapshot~$\mathbb{S}50$) that form an MBH binary with the primary MBH by $z = 0$, as a function of all six MBH model quantities. These fractions are consistent with Figures~\ref{fig:SAM} and \ref{fig:SAM_B22}. We do not show the fraction as a function of the bar model quantities, because of low-number statistics in the \textsc{tng50} simulation.}
\label{fig:heatmap}
\end{figure*}

The performance of the model on SAM data is shown in Figure~\ref{fig:td}. The first panel shows the agreement between the SAM (solid line) and the ML emulator (dots) predictions for the \td distribution, with remarkable agreement. Furthermore, to understand whether the correlations between \td and relevant input features of the conditioning vector $\vec{c}$ were correctly learned, three additional panels are shown with different selections on the MBH mass (second panel), its initial distance from the centre (third panel), and its initial velocity's magnitude (fourth panel). These quantities are found to be amongst the dominant factors governing the orbital decay, and therefore looking at the $\td$ distribution for different values of these variables provides a stringent test of the emulator beyond reproducing the overall population distribution. As shown, the model can accurately describe the \td distribution in all the regimes represented at training time, successfully recovering both the shifts in the typical decay time and the changes in the shape of the distribution in different regions of the parameter space. The close agreement observed in all cases indicates that the conditional NF captures the underlying dependence of the binary formation time \td on the physical properties of the system, rather than merely reproducing the marginal distribution of the training set.

Figure~\ref{fig:pp} shows the probability-probability plot for the trained emulator, comparing the confidence interval of the predictions of the model to the true decay times. The observed coverage is well aligned with the diagonal relation expected for a perfectly calibrated probabilistic model, with no significant deviations observed for either the inclusive SAM dataset (i.e. considering all systems; blue curve) nor the SAM dataset restricted to systems whose binary formation time is smaller than the age of the Universe, $\tau_{\rm d} < t_{\rm H}$ (red curve). Together with Figure~\ref{fig:td}, this demonstrates that the emulator is not only capable of reproducing the decay time distribution, but also provides reliable probabilistic estimates of its associated uncertainty.

\subsection{Analysis on the \textsc{tng50} simulation}\label{sec:individual_snapshots}

We now apply our model on individual snapshots from the \textsc{tng50} simulation. For our analysis, we picked 18 snapshots at very different redshifts (see Section~\ref{sec:TNG50}), to assess the effects of the bar fraction -- which varies significantly as a function of cosmic time -- on the distribution of MBH decay times. Each snapshot contains a few thousand galaxies (i.e. gravitationally bound substructures), of which 36--40~per cent have a central stellar mass greater than $3 \times 10^9$~M$_{\sun}$. Of these massive galaxies, the fraction of non-barred, barred [$A_{\rm 2,max}(R) \ge 0.1$\,], moderately and strongly barred [$A_{\rm 2,max}(R) \ge 0.2$\,], and strongly barred galaxies [$A_{\rm 2,max}(R) \ge 0.4$\,] is 59--73, 27--41, 22--34, and 10--17~per cent, respectively. If we now consider only the barred galaxies with bar stellar mass fractions and lengths within the SAM ranges, the fractions of barred galaxies become 12--35, 12--34, and 10--21~per cent, respectively (see Table~\ref{tab:tng50_snapshots} for the exact numbers). The mass distributions of these subsets of galaxies are shown in Figure~\ref{fig:TNG-central-mass-histograms}, whereas the distributions of bar stellar mass fractions and lengths for the selected barred galaxies are shown in Figures~\ref{fig:TNG-bar-histograms-1} and \ref{fig:TNG-bar-histograms-2}.

In Figure~\ref{fig:TNG_some_snapshots_linlin_all_galaxies}, we show the distribution of MBH decay times (normalized by the number of galaxies considered in each snapshot) for a few select \textsc{tng50} snapshots, wherein we consider all galaxies -- non-barred and barred (with bar stellar mass fractions and lengths within the SAM ranges) -- with $M_{\rm \star,\textsc{tng50}} \ge 3 \times 10^9$~M$_{\sun}$. The distributions are all quite similar to each other, showing a prominent peak between 0 and 1~Gyr and a slow decrease for longer decay times, exactly like those shown in the first panel of Figure~\ref{fig:td}, which were obtained from the SAM simulations and from the ML surrogate model, with no \textsc{tng50} information. This is expected, since the \textsc{tng50} galaxies we considered have SAM-like bar quantities and their masses [\,which have an effect on $\tau_{\rm d}$ via the $(M_{\rm \star,\textsc{tng50}}/M_{\star})^{1/2}$ scaling\,] are not too different from the SAM mass.

We note, however, that the galaxies in the early-time snapshots promote more decays (relative to the number of galaxies) than those in the late-time snapshots (assuming similar seedings of secondary MBHs). The main difference in the galaxy populations of these snapshots is in the fraction of barred galaxies, which is as low as 13.4~per cent at $z = 1.9$ and as high as 31.6~per cent at $z = 0.1$. The observed pattern thus seems to imply that bars tend to slightly demote decay. This is consistent with the trends shown in Figure~\ref{fig:SAM}, wherein the fraction of decays in the SAM decreases for more massive and shorter bars. Decays are also promoted for MBHs that are massive, close to the centre, or in coplanar or radial orbits, as shown in Figure~\ref{fig:heatmap}, which presents the fraction of secondary MBHs that reach 10~pc from the centre by $z = 0$ for a specific \textsc{tng50} snapshot ($\mathbb{S}50$, $z = 1$) as a function of the MBH initial quantities. These results are consistent with what is shown in Figures~\ref{fig:SAM} and \ref{fig:SAM_B22} in the case of the SAMs, and additionally show the interplay between the MBH quantities. The panels of Figure~\ref{fig:heatmap} were constructed by integrating the decay distribution in Figure~\ref{fig:TNG_some_snapshots_linlin_all_galaxies} between 0 and the lookback time of the considered snapshot, for two subsets of initial MBH variables at a time.

\begin{figure}
\includegraphics[width=0.47\textwidth]{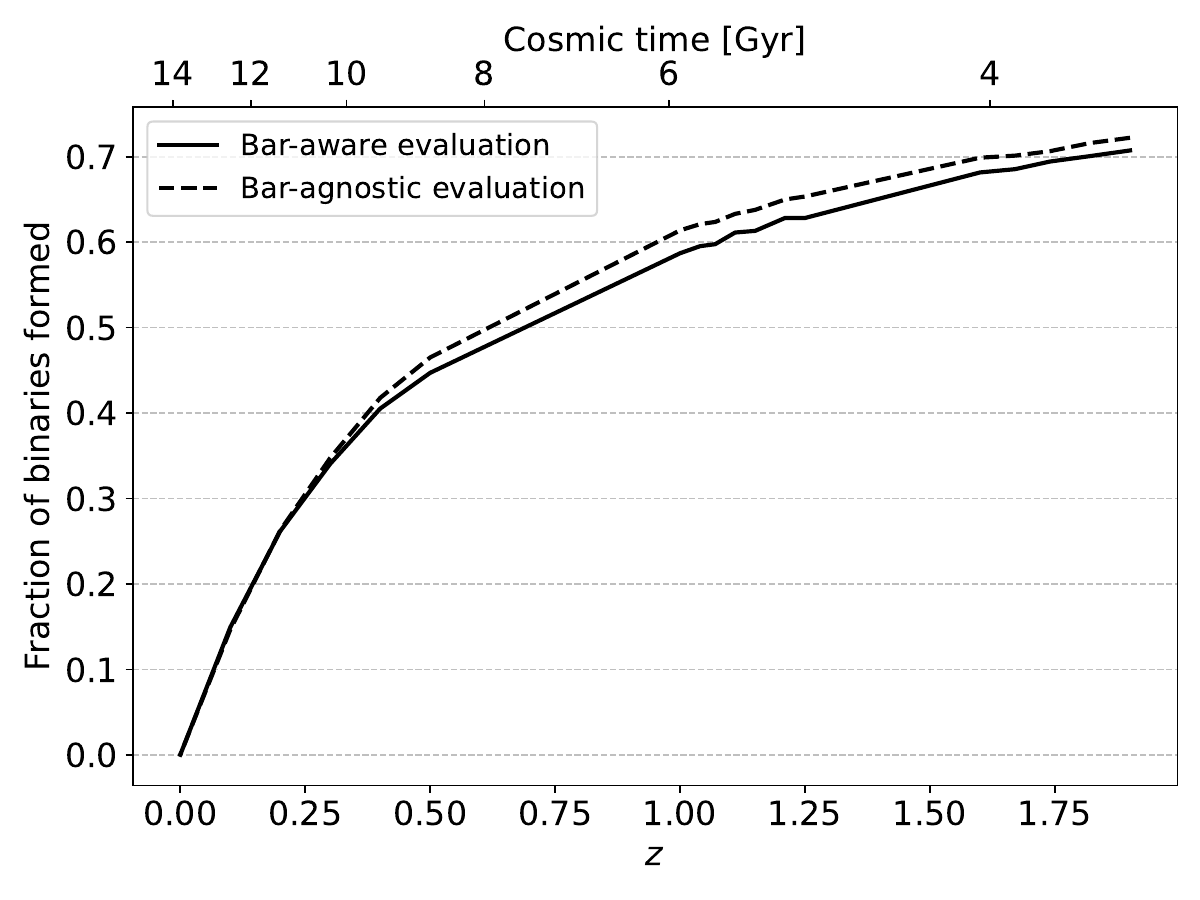}
\caption{Fraction of binary formation events occurred between the time of the considered snapshot and $z = 0$, for each of the 18 \textsc{tng50} snapshots we considered, treating the galaxies according to their own morphology (solid line) and treating all galaxies as non-barred (dashed line). Taking into account the stellar bars tends to produce fewer events, with discrepancies of up to five per cent.}
\label{fig:redshift_dependence}
\end{figure}

We now perform the same exercise but consider instead all possible MBH variables. In other words, we simply integrate the curves in Figure~\ref{fig:TNG_some_snapshots_linlin_all_galaxies} between 0 and the lookback time (this is equivalent to adding each fraction from the pixels in any given panel of Figure~\ref{fig:heatmap} and dividing by the number of pixels, since all pixels are uniformly populated). The results are displayed by the solid curve in Figure~\ref{fig:redshift_dependence}, which shows the cumulative number of binary formation events that has occurred between the redshift of the considered snapshot and $z = 0$, considering all the 18 \textsc{tng50} snapshots analysed in this work and assuming similar seedings of secondary MBHs. Since the lookback time decreases with decreasing redshift and the distributions of Figure~\ref{fig:TNG_some_snapshots_linlin_all_galaxies} are all quite similar to each other, the computed fraction expectedly increases with increasing redshift. It is however difficult to disentangle the effects of cosmic time and those of bars.

\begin{figure}
\includegraphics[width=0.47\textwidth]{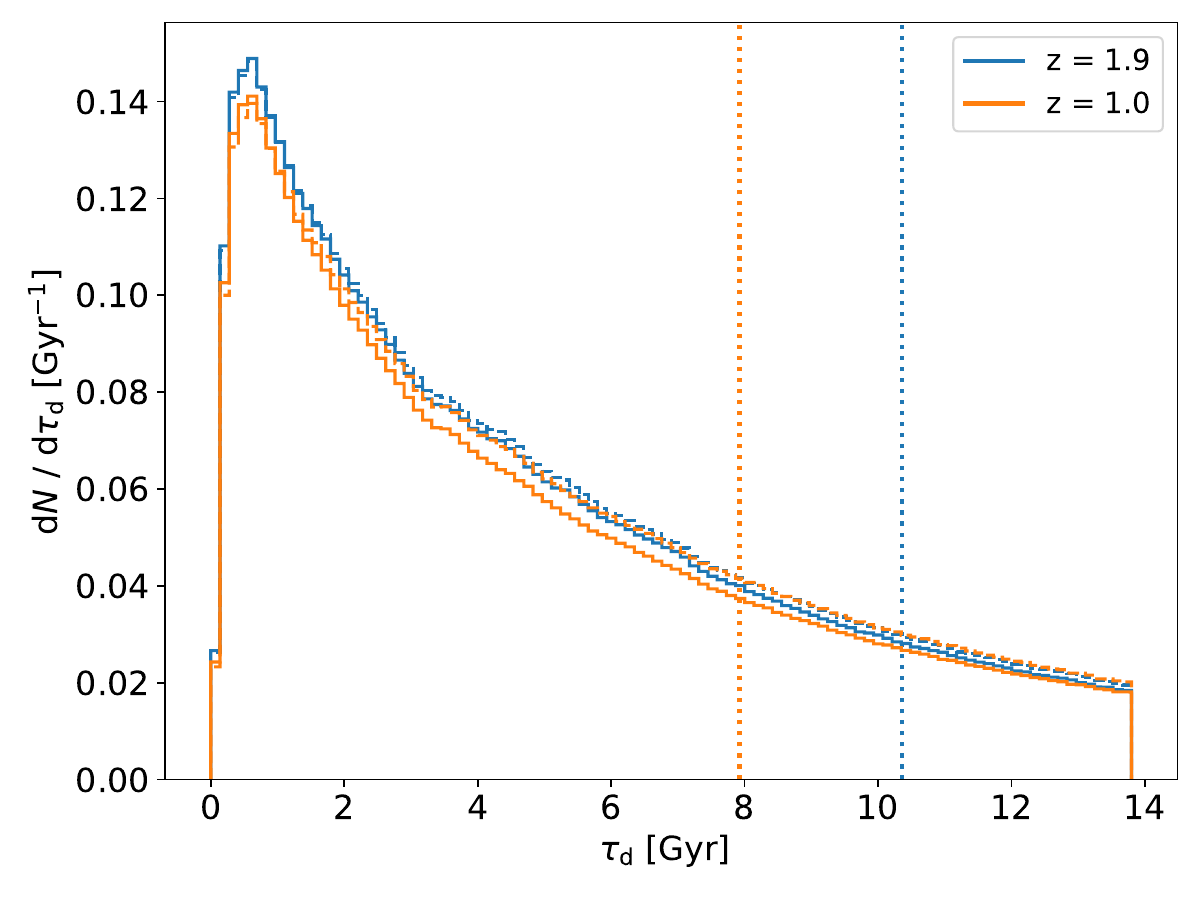}
\includegraphics[width=0.47\textwidth]{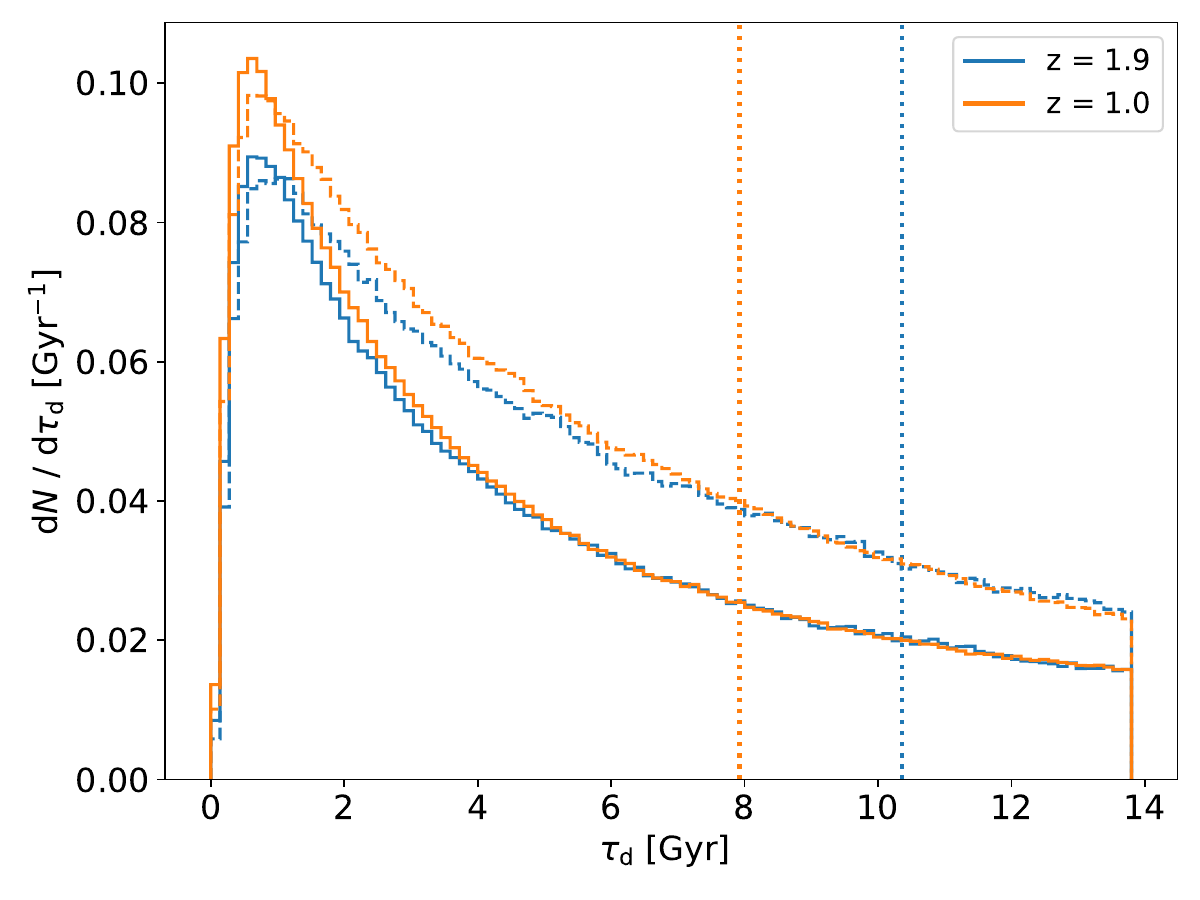}
\includegraphics[width=0.47\textwidth]{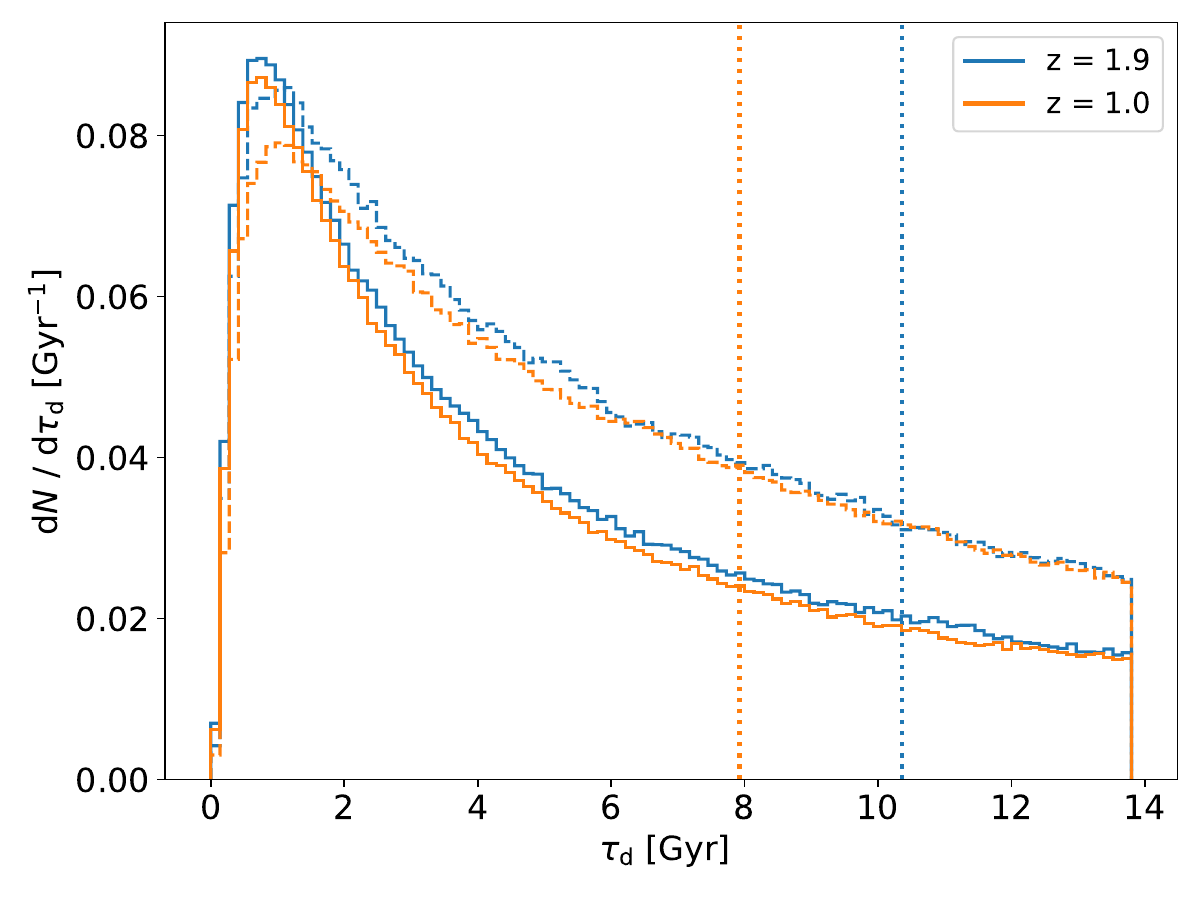}
\caption{Distribution of decay times for two \textsc{tng50} snapshots ($\mathbb{S}34$ and $\mathbb{S}50$, at $z = 1.9$ and 1, respectively), using two different methods. Top panel: comparison between treating galaxies according to their bar morphology (solid lines) and assuming that all galaxies are non-barred (dashed lines). Treating all galaxies as non-barred yields a slightly larger fraction of MBH binaries, more prominent at lower redshift, when the bar fraction is larger. Middle (bottom) panel: same as the top panel, but only for the barred (strongly barred) galaxies. Treating barred (or strongly barred) galaxies as non barred clearly generates a much larger fraction of MBH binaries.}
\label{fig:TNG_comparison_linlin_all_barred_galaxies}
\end{figure}

Another, more informative way to present the effect of bars on the distribution of decay times is to assess how the decay time distributions and MBH binary fractions would change if we assumed all galaxies in each snapshot to be non-barred. This is the approach taken by many past studies, which did not include the complexity of having a time-dependent non-axisymmetric potential. \citet{Li_et_al_2022}, for example, also employed SAMs and cosmological simulations to evaluate the decay time distribution of MBHs in the Universe. In their case, they chose the cosmological, magnetohydrodynamical simulation \textsc{tng50-3}, which is identical in every way to the simulation we used in this work, except for the (initial) numbers of DM and baryonic particles, which are a factor of 64 lower, the (initial) particle masses, which are a factor of 64 larger, and the gravitational softenings, which are a factor of four larger \citep[][]{Pillepich_et_al_2019}. \citet{Li_et_al_2022} started with a population of initially gravitationally unbound MBH pairs, drawn from the \textsc{tng50-3} simulation, with separations of the order of the collisionless gravitational softening at $z = 0$ (1152~pc), and followed their orbital evolution using a SAM developed by \citet{Li_et_al_2020a,Li_et_al_2020b}, adopting a set of parameters obtained by the \textsc{tng50-3} remnant galaxies and assuming four different orbits for each MBH. In the first phase of the evolution, which is what we are interested in here, they take into account DF from a stellar bulge and a gaseous disc and compute the time it takes the MBH to evolve from $\sim$1~kpc down to the influence radius of the MBH binary, whose distribution is shown in the upper-left panel of their figure~1. Comparing our results to theirs is not straightforward, since we use, e.g. somewhat different SAMs, cosmological simulations, and initial separations. Also, whereas we populate each \textsc{tng50} galaxy with a secondary MBH with a position and velocity taken from isotropic distributions, they take the actual position of the MBHs in the \textsc{tng50-3} simulation. Taking all this into account, it is however reassuring that the two distributions show similar shapes (compare their figure~1 with Figure~\ref{fig:TNG_some_snapshots_loglog_all_galaxies}, the logarithmic version of Figure~\ref{fig:TNG_some_snapshots_linlin_all_galaxies}).

To properly compare the two methods, we keep everything the same (including the seeding method of the MBHs) and contrast treating the galaxies according to their own morphology (i.e. our model) to treating them all as non-barred (i.e. setting $x_{\rm bar} = 0$ for all galaxies, regardless of the \textsc{mordor} classification), thus somewhat mimicking the model by \citet{Li_et_al_2022}. In the top panel of Figure~\ref{fig:TNG_comparison_linlin_all_barred_galaxies}, we show the distribution of decay times for two \textsc{tng50} snapshots (at $z = 1.9$ and~1), in which we either treat the galaxies as they should (i.e. barred galaxies as barred, $x_{\rm bar} = 1$, and non-barred galaxies as non-barred, $x_{\rm bar} = 0$; solid lines, which are the same as their counterparts in Figure~\ref{fig:TNG_some_snapshots_linlin_all_galaxies}) or we treat all the galaxies as non-barred ($x_{\rm bar} = 0$; dashed lines). While the difference at $z = 1.9$ is minimal, likely due to the relatively low bar fraction (13.4~per cent), the discrepancy at $z = 1$, when the bar fraction is 28.6~per cent, is more significant: treating all galaxies as non-barred yields a higher fraction of binaries.

The effect is clearly visible when we perform the same exercise only on the barred galaxies of each snapshot (middle panel of Figure~\ref{fig:TNG_comparison_linlin_all_barred_galaxies}). In this case, treating the barred galaxies as non-barred significantly enhances the fraction of binaries, for both the snapshots we considered. The effect is more pronounced for longer binary formation times, whereas the number of binaries formed within $\sim$1~Gyr (`fast binaries'; near the peaks of the distributions) is not much affected by the chosen model. Systems with low values of $\tau_{\rm d}$ are often systems with a short $r_{\rm \bullet,0}$ (see the third panel of Figure~\ref{fig:td}), where the effect of the bar is negligible and the DF process is very efficient. At larger initial separations, DF takes more time and the bar torque starts to have a significant impact on the orbital evolution. Also, we note that the fraction of `fast binaries' is larger at $z = 1$ than at $z = 1.9$, for reasons that become apparent when studying only the strong bars.

When restricting the analysis to strong bars (bottom panel of Figure~\ref{fig:TNG_comparison_linlin_all_barred_galaxies}), the disparity between the two methods is even slightly larger, again consistent with the fact that bars demote decay. Also in this case, we see that the discrepancy is smaller for `fast binaries', but additionally note that the fractions of strong bars in the two redshifts considered are comparable (12 and 16~per cent). This similarity helps explain the disparity of `fast binaries' between snapshots $\mathbb{S}34$ and $\mathbb{S}50$ in the middle panel of Figure~\ref{fig:TNG_comparison_linlin_all_barred_galaxies}, wherein we considered all bars: at $z = 1$, there are relatively more weak and moderate bars than at $z = 1.9$. This is again consistent with bars demoting binary formation.

The discrepancy shown in the top panel of Figure~\ref{fig:TNG_comparison_linlin_all_barred_galaxies} naturally yields different cumulative fractions of MBH binaries, as can be seen when comparing the solid and dashed lines of Figure~\ref{fig:redshift_dependence}: treating all galaxies as non-barred, as it has been done so far in the literature, leads to a higher cumulative fraction of MBH binary formation events, for all (starting) redshifts, with increases of up to five per cent.

\subsection{Binary formation rate estimates}\label{sec:rates}

The studies summarized in this work can be combined to derive estimates of the rate of binary formation events that could eventually lead to LISA detections. On the one hand, we can leverage the enhanced details of the galactic-level physics introduced in the SAM to improve the accuracy of the MBH dynamics and combine it with the redshift-dependent evolution of the galaxy populations encoded in the \textsc{tng50} cosmological simulation (namely, the galaxies' bar stellar mass fraction and length, and their central stellar mass). On the other hand, the use of a surrogate model to speed up the SAM predictions -- while maintaining the probabilistic nature of the problem -- allows us to perform rapid estimates over a large portion of the phase space with very good resolution. Considering the results described for \textsc{tng50} galaxies at different snapshots, we can estimate how many MBH binaries are expected at different redshifts. In order to do so, we consider that at each value of redshift $z$ a number $N_{\rm inj}$ of perturber MBHs are injected in the population of galaxies:

\begin{equation}
\frac{{\rm d}N_{\rm inj}}{{\rm d}z} = \frac{{\rm d}(N_{\rm gal}f_{\rm occ})}{{\rm d}z} = \frac{{\rm d}(n_{\rm gal}f_{\rm occ})}{{\rm d}z}\times V_{\rm sim}\,,
\label{eq:injections}
\end{equation}

\noindent where $N_{\rm gal}$ is the number of galaxies considered and $f_{\rm occ}$ is the MBH occupation fraction of those galaxies, i.e. the fraction of the initial galaxies that receive a secondary MBH. Here we use the \textsc{tng50} simulation as an estimate of the population of galaxies at each redshift, therefore we split the $N_{\rm gal} \equiv N_{\rm TNG}$ term into the product of the comoving number density of galaxies, $n_{\rm gal}$, and the comoving volume of the simulation, $V_{\rm sim}$, and implicitly assume that the \textsc{tng50} volume is representative of the whole Universe. In this context, an injection of a secondary MBH is expected to simulate the result of the merger of two galaxies and, while the merger history of galaxies is tracked in \textsc{tng50}, it is not taken into account in this work.

\begin{figure}
\includegraphics[width=0.47\textwidth]{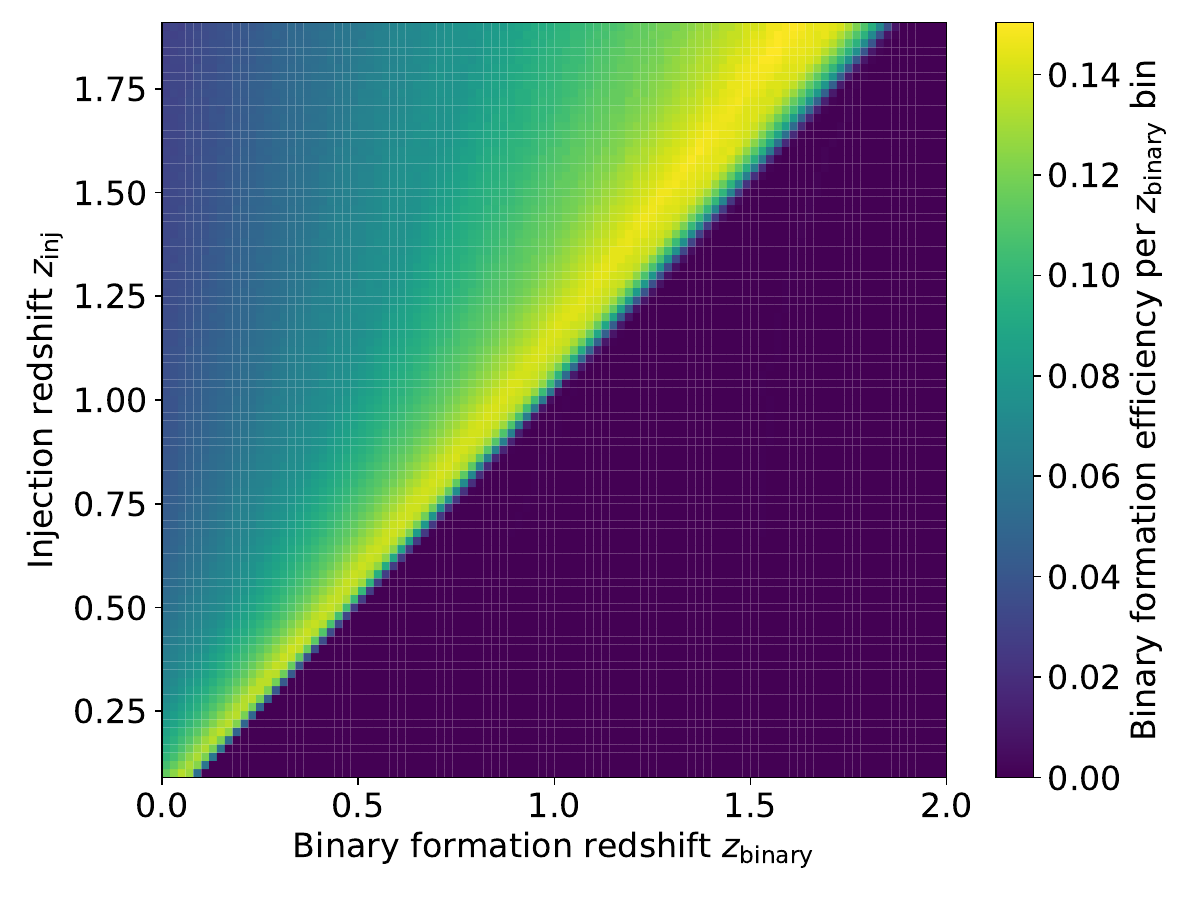}
\caption{Conditional probability kernel of an injected secondary MBH injected at redshift $z_{\rm inj}$ to form a binary with the central MBH at redshift $z_{\rm binary}$.}
\label{fig:prob_map}
\end{figure}

The main assumption made in this calculation is that, once injected, the secondary MBHs will follow the distribution of \td predicted by the model at the time of injection. We note that, in this approximation, the evolution of the galaxies themselves at the cosmological scale is neglected. However, based on the similarity amongst the \td distributions at different redshifts observed in Figure~\ref{fig:TNG_some_snapshots_linlin_all_galaxies}, the difference with respect to a proper evaluation wherein the galactic evolution is taken into account is expected to be small. The probability of a secondary MBH injected at redshift \zinj to form a binary with the central MBH at redshift $z_{\rm binary}$ is shown in Figure~\ref{fig:prob_map}. This 2D map shows a conditional probability kernel, \kzz, whereon each row is the probability density of a secondary MBH to merge at \zbinary conditional on it being injected at \zinj. It is done by stacking together the \td probability distributions obtained from the \textsc{tng50} galaxies at different redshifts. The probabilities shown in the figure are smoothed out in the \zinj dimension by means of quantile interpolation amongst the distributions obtained for the available \textsc{tng50} snapshots. 
This conditional probability kernel is obtained assuming that the properties of the injected MBH are quasi-random variables distributed as discussed in Section~\ref{sec:SAM}.

\begin{figure}
\includegraphics[width=0.47\textwidth]{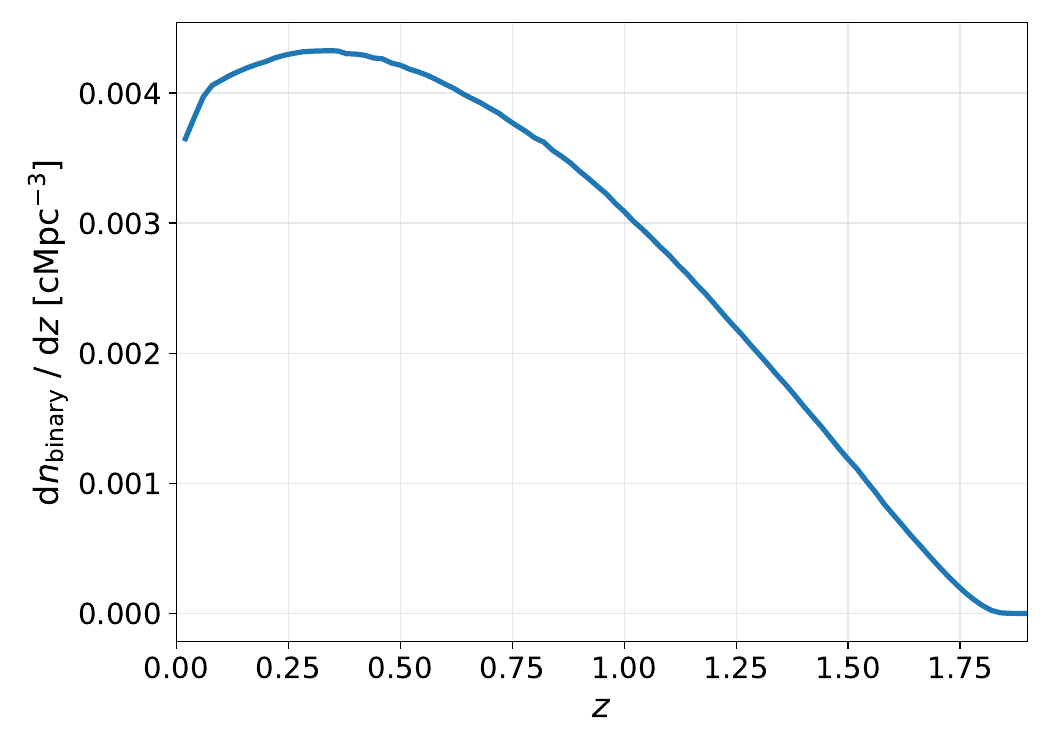}
\caption{Estimated contribution from disc-like galaxies to the global binary formation rates, assuming $f_{\rm occ} = 0.1$ and without tracking galaxy mergers.}
\label{fig:merger_rate}
\end{figure}

The kernel \kzz helps estimate the amount of MBH binary formations expected at each value of \zbinary, as it effectively encodes the fraction of systems that form MBH binaries out of the total injections $N_{\rm inj}$ occurring at each redshift \zinj. The core idea is that the amount of MBH binaries forming at a given redshift will be the accumulation of all systems that were injected at different redshifts \zinj and that form a binary at the same \zbinary. The number of formed binaries per unit comoving volume, $n_{\rm binary}$ at \zbinary, can therefore be obtained by integrating in \zinj the product of the amount of injections and the probability of those injections to form a binary at \zbinary:

\begin{equation}
\left.\frac{{\rm d}n_{\rm binary}}{{\rm d}z}\right|_{\zbinary} = \int_{\zbinary}^\infty {\rm d}z' \lambda(z') K(\zbinary | z')\,,
\end{equation}

\noindent where $\lambda(z')$ contains the information about the amount of injections at each redshift, and depends on the exact modelling of the evolution of galaxy populations at the cosmological scale. In this work, we estimate $\lambda(z)$ based on the galaxy populations of \textsc{tng50} at different redshifts, under the assumption that, once a secondary MBH is injected, it is left to evolve without taking into account the evolution of the galaxy itself. We therefore approximate it, following Equation~\eqref{eq:injections}, as

\begin{equation}
\lambda(z') \approx \left.\frac{ {\rm d}(n_{\rm gal}f_{\rm occ}) }{{\rm d}z}\right|_{z'}\,.
\end{equation}

A more detailed analysis would require evaluating the merger histories of the simulated galaxies in \textsc{tng50} to find a more accurate value of $f_{\rm occ}$ as a function of redshift, which is beyond the scope of this work.

Combining the two factors discussed above, an estimate for the comoving number density of binary formation events per unit redshift is obtained and shown in Figure~\ref{fig:merger_rate}, assuming a constant $f_{\rm occ} = 0.1$. We note that this should not be considered as a global rate of binary formation events, but rather a rate originated in galaxies with a disc-like morphology. The observed increase in the rate of formed binaries as the redshift decreases from $z \sim 2$ to $z \sim 0.5$ is a consequence of the analysis stopping at $z \sim 2$. As the computed rate tracks the contributions from higher redshifts, that hard cut-off leads to an unsaturated behaviour down to redshift $z \sim 0.5$. The subsequent decrease observed at $z \lesssim 0.5$ is due to the fact that the systems that are closer to $z = 0$ will have less time to form a binary, as their lookback time is shorter than that at high redshift. The decision of stopping the analysis at $z \sim 2$ is justified, since the fraction of disc-like galaxies in \textsc{tng50} drops significantly at $z \gtrsim 2$~\citep{Zana_et_al_2022}, and therefore their contribution to the binary formation rate would be drowned by systems with morphologies not covered in this methodology. Meanwhile, the fraction of disc-like galaxies at $z<2$ in the mass range studied in this work is about 80~per cent, implying that the contribution of this type of events to the rate of MBH binary formation at those redshifts is expected to be significant. A global rate estimate would need to include the analysis of different galaxy morphologies that are beyond the scope of this work.


\section{Discussion and conclusions}\label{sec:conclusions}

Previous works, both numerical \citep[][]{Bortolas_et_al_2020} and semi-analytical (\citetalias{Bortolas_et_al_2022}), have shown the highly stochastic behaviour of MBHs in galaxies wherein a prominent stellar bar is present. In this work, we combined three methods to further explore the dynamics of MBHs within the complex structures of the remnants of galaxy mergers: (i) a novel SAM, in which we varied not only the mass, position, and velocity of the orbiting MBH (as it was done in \citetalias{Bortolas_et_al_2022}), but also the mass, length, and orbital frequency of the remnant's stellar bar; (ii) an ML surrogate model, based on conditional NFs, trained on the SAM dataset; and (iii) a hydrodynamical, cosmological simulation (\textsc{tng50}), whose outputs were used to select realistic merger remnants across a wide range of masses, redshifts, and structures.

The new SAM confirmed and built upon the findings of \citetalias{Bortolas_et_al_2022}, showing that MBHs that are initially relatively massive, near the remnant's centre, or in radial or coplanar orbits are more likely to reach the centre within a Hubble time, and that the initial relative distance between the MBH and the bar is inconsequential (see Figure~\ref{fig:SAM}). The dependence of the decay time on the presence itself of the bar and, when present, on its properties showed instead that bars tend to demote the orbital decay of MBHs, with the formation of MBH binaries being more likely when the bar is absent (compare Figures~\ref{fig:SAM} and \ref{fig:SAM_B22}) and, when the bar is present, being less likely when the bar is relatively more massive and shorter (see Figure~\ref{fig:SAM}). For those systems forming an MBH binary within $t_{\rm H}$, the distribution of decay times shows a skewed distribution towards relatively short times ($\sim$1~Gyr), mostly caused by the more massive/central MBHs (see Figure~\ref{fig:td}).

The ML surrogate model chosen is a conditional NF, and we show that it is able to learn the complex correlations between the different seeding conditions of the MBH and the properties of the host galaxy to predict the probabilistic behaviour of the binary formation time (see Figure~\ref{fig:td}), with good coverage of the target probability distribution function (see Figure~\ref{fig:pp}). Once trained, the model can make predictions for large populations in a few seconds with low computing cost, orders of magnitude faster than the SAM simulations.

The hydrodynamical, cosmological \textsc{tng50} simulation provided the framework to test our new model, with thousands of galaxies per snapshot, from which we could retrieve their mass, the absence/presence of a bar, and, when present, the bar stellar mass fraction and length. Applying our ML surrogate model to a few select snapshots across a wide range of redshifts showed that galaxies at late times, when the bar fraction is higher, promote slightly fewer MBH binaries than those at early times (Figure~\ref{fig:TNG_some_snapshots_linlin_all_galaxies}), in agreement with the SAM findings. This trend was confirmed when we compared our model to one wherein barred galaxies are treated as non-barred (Figures~\ref{fig:redshift_dependence} and \ref{fig:TNG_comparison_linlin_all_barred_galaxies}): the number of binaries formed with our bar-aware model decreases significantly, when considering only barred galaxies, and still shows a measurable but modest decrement, when considering all galaxies. Consequently, the number of LISA events may also decrease, although one would need a proper modelling of the binary evolution and LISA data analysis to quantify this.

We note that cosmological simulations, due to their limited resolution and physics approximations, inevitably fail to correctly detect all bar structures, especially at higher redshifts, providing disparate results \citep[see, e.g.][]{Peschken_Lokas_2019,Zhao_et_al_2020,Cavanagh_et_al_2022,RosasGuevara_et_al_2022,Zana_et_al_2022}. In particular, it was shown by \citet{RosasGuevara_et_al_2022} that the fraction of \textsc{tng50} barred galaxies at $z \gtrsim 1$ is lower than that observed \citep[e.g.][]{LeConte_et_al_2024}, when considering only bars with $a_{\rm bar \textsc{tng50}} \ge 2$~kpc, which is close to the length range considered in this work. If we were able to retrieve the correct, presumably higher bar fraction, the observed difference in binaries formed (Figures~\ref{fig:redshift_dependence} and \ref{fig:TNG_comparison_linlin_all_barred_galaxies}) would be even more significant.

We estimate the MBH binary formation rate as a function of redshift by combining the large-scale evolution of a cosmological simulation with the accurate orbital integration allowed by the SAM. The method relies on the accumulation of the expected binary formation times for each \textsc{tng50} snapshot, which is separated in two components. On the one hand, the distribution of seed MBHs and host galaxies can be estimated through cosmological models to provide a proxy for the initial population at each redshift. On the other hand, the probability of each of these MBH-galaxy systems to form a binary is independently estimated with the ML surrogate model (Figure~\ref{fig:prob_map}), and can be computed with high resolution in the parameter space with reasonable computational resources. The results for a toy model wherein 10~per cent of \textsc{tng50} galaxies receive an MBH with uniform seeding conditions is shown in Figure~\ref{fig:merger_rate}. While the toy model is purposely simplistic, assuming a constant, mass- and redshift-independent occupation fraction $f_\text{occ}$, the methodology remains applicable to more complex parametrizations of the injections of MBHs into galaxy populations. 

The results of this work naturally come with caveats associated to the assumptions made in our model. In the SAM calculations, we assumed that all the galactic parameters are constant throughout the orbital evolution of the MBH, which can last for several Gyr. These include the masses and spatial scales of the DM halo, stellar bulge, and stellar disc. More importantly, the bar retains its mass, length, and orbital frequency for the entire duration of the integration. It was shown, however, that stellar bars can become stronger, or weaker, or even temporarily disappear \citep[e.g.][]{Zana_et_al_2018b}. We also assumed that the perturber's mass does not change, both in the case it represents a naked MBH (since we did not model BH accretion) and in the context it is a proxy for an MBH embedded in a stellar structure \citep[since we did not model mass loss; see, e.g.][]{Varisco_et_al_2024}. Another assumption of this study is the value of the occupation fraction of MBHs. Here we have assumed for simplicity that all the galaxies considered are the hosts not only of a primary MBH at their centre (which we did not model) but also of a secondary MBH orbiting in their outskirts (which we modelled). This implies that all these galaxies are surmised to be the remnants of a merger of two galaxies, each with their own central MBH. This is an obvious simplification: $f_{\rm occ}$ is likely not equal to one, and it is a function of galaxy mass and type, and redshift \citep[but see][]{Weller_et_al_2026}. Moreover, it is very difficult to observationally ascertain the real occupation fraction, especially in the case of secondary MBHs at $z \neq 0$, and also numerical simulations/SAMs provide discrepant results at the mass scales considered in this work \citep[see, e.g.][]{Haidar_et_al_2022,Izquierdo-Villalba_et_al_2026}. Finally, we note that the scaling we used to take into account the difference in mass between the \textsc{tng50} galaxies and our SAM galaxy is an approximation based on simplifying assumptions [\,e.g. a Maxwellian velocity distribution and a fixed Coulomb logarithm, even though the latter has a mild dependence on the host mass: $\ln \Lambda \sim K + 2\ln(M_{\rm host})$, with $K$ depending on the initial conditions\,]. We chose this approach for simplicity and to have a scaling that can be used for all systems. Another approach would be to run different sets of SAMs, each set with a different $M_{\rm \star}$ (and appropriately rescaled $M_{\rm vir}$, velocity curve, and size), and use them as an improved training set for the ML surrogate model, so that the host mass dependence is also learnt. We plan on investigating the effects of the simplifying assumptions in the SAM, on $f_{\rm occ}$, and on the mass scaling in future work.

We additionally note that our analysis was performed on individual \textsc{tng50} snapshots, assuming each time that all the considered galaxies in a given snapshot were seeded according to the SAM recipe. This approach implicitly neglects the orbital evolution of the MBHs inside the host galaxies when moving from snapshot to snapshot. For example, MBHs seeded at separations of 2--14~kpc from the centre at a given snapshot are expected, at a later snapshot, due to the orbital evolution, to either find themselves at a different range of separations or to have formed an MBH binary (thus changing the occupation fraction). Taking this into account requires a more sophisticated approach, which will be the basis of a future work.

Finally, we wish to emphasize that what we presented here is a proof of concept, to show the capability and performance of the conditional NFs, properly trained with sophisticated SAMs, to bridge small and large scales in astrophysics. Even though we applied our novel method to a specific simulation (\textsc{tng50}) and question (inferring MBH binary formation time-scales within complex galactic systems), this technique can naturally be applied to a multitude of simulations (and observational surveys) and investigations.

\section*{Acknowledgements}
We thank the anonymous reviewer for the useful comments that greatly improved this work. We are grateful for useful discussions with Tamara Bogdanovi\'{c} and Massimo Dotti. PRC acknowledges support from the Swiss National Science Foundation under the Sinergia Grant CRSII5\_213497 (GW-Learn). MB acknowledges support from the Italian Ministry for Universities and Research (MUR) program ``Dipartimenti di Eccellenza 2023-2027'', within the framework of the activities of the Centro Bicocca di Cosmologia Quantitativa (BiCoQ). TZ acknowledges support from the MUR project FIS-2024-01621 DAWN, the INFN TEONGRAV initiative, and the EU-Recovery Fund PNRR -- National Centre for HPC, Big Data and Quantum Computing.

\section*{Data Availability Statement}
The data underlying this article will be shared on reasonable request to the corresponding author.

\scalefont{0.94}
\setlength{\bibhang}{1.6em}
\setlength\labelwidth{0.0em}
\bibliographystyle{mnras}
\bibliography{GW-learn-bars}

\begin{thebibliography}{}
\makeatletter
\relax
\def\mn@urlcharsother{\let\do\@makeother \do\$\do\&\do\#\do\^\do\_\do\%\do\~}
\def\mn@doi{\begingroup\mn@urlcharsother \@ifnextchar [ {\mn@doi@}
  {\mn@doi@[]}}
\def\mn@doi@[#1]#2{\def\@tempa{#1}\ifx\@tempa\@empty \href
  {http://dx.doi.org/#2} {doi:#2}\else \href {http://dx.doi.org/#2} {#1}\fi
  \endgroup}
\def\mn@eprint#1#2{\mn@eprint@#1:#2::\@nil}
\def\mn@eprint@arXiv#1{\href {http://arxiv.org/abs/#1} {{\tt arXiv:#1}}}
\def\mn@eprint@dblp#1{\href {http://dblp.uni-trier.de/rec/bibtex/#1.xml}
  {dblp:#1}}
\def\mn@eprint@#1:#2:#3:#4\@nil{\def\@tempa {#1}\def\@tempb {#2}\def\@tempc
  {#3}\ifx \@tempc \@empty \let \@tempc \@tempb \let \@tempb \@tempa \fi \ifx
  \@tempb \@empty \def\@tempb {arXiv}\fi \@ifundefined
  {mn@eprint@\@tempb}{\@tempb:\@tempc}{\expandafter \expandafter \csname
  mn@eprint@\@tempb\endcsname \expandafter{\@tempc}}}

\bibitem[\protect\citeauthoryear{{Agazie} et~al.,}{{Agazie}
  et~al.}{2023a}]{Agazie_et_al_2023a}
{Agazie} G.,  et~al., 2023a, \mn@doi [\apjl] {10.3847/2041-8213/acdac6}, \href
  {https://ui.adsabs.harvard.edu/abs/2023ApJ...951L...8A} {951, L8}

\bibitem[\protect\citeauthoryear{{Agazie} et~al.,}{{Agazie}
  et~al.}{2023b}]{Agazie_et_al_2023b}
{Agazie} G.,  et~al., 2023b, \mn@doi [\apjl] {10.3847/2041-8213/ace18b}, \href
  {https://ui.adsabs.harvard.edu/abs/2023ApJ...952L..37A} {952, L37}

\bibitem[\protect\citeauthoryear{{Amaro-Seoane} et~al.,}{{Amaro-Seoane}
  et~al.}{2023}]{AmaroSeoane_et_al_2023}
{Amaro-Seoane} P.,  et~al., 2023, \mn@doi [Living Reviews in Relativity]
  {10.1007/s41114-022-00041-y}, \href
  {https://ui.adsabs.harvard.edu/abs/2023LRR....26....2A} {26, 2}

\bibitem[\protect\citeauthoryear{{Bairagi}, {Wandelt}  \&
  {Villaescusa-Navarro}}{{Bairagi} et~al.}{2025}]{Bairagi_et_al_2025}
{Bairagi} A.,  {Wandelt} B.,   {Villaescusa-Navarro} F.,  2025, \mn@doi [\aap]
  {10.1051/0004-6361/202554602}, \href
  {https://ui.adsabs.harvard.edu/abs/2025A&A...703A.301B} {703, A301}

\bibitem[\protect\citeauthoryear{{Battistini} et~al.,}{{Battistini}
  et~al.}{2026}]{Battistini_et_al_2026}
{Battistini} L.,  et~al., 2026, \mn@doi [\aap] {10.1051/0004-6361/202659543},
  \href {https://ui.adsabs.harvard.edu/abs/2026A&A...710A.232B} {710, A232}

\bibitem[\protect\citeauthoryear{{Begelman}, {Blandford}  \& {Rees}}{{Begelman}
  et~al.}{1980}]{Begelman_et_al_1980}
{Begelman} M.~C.,  {Blandford} R.~D.,   {Rees} M.~J.,  1980, \mn@doi [\nat]
  {10.1038/287307a0}, \href
  {https://ui.adsabs.harvard.edu/abs/1980Natur.287..307B} {287, 307}

\bibitem[\protect\citeauthoryear{{Berczik}, {Merritt}, {Spurzem}  \&
  {Bischof}}{{Berczik} et~al.}{2006}]{Berczik_et_al_2006}
{Berczik} P.,  {Merritt} D.,  {Spurzem} R.,   {Bischof} H.-P.,  2006, \mn@doi
  [\apjl] {10.1086/504426}, \href
  {https://ui.adsabs.harvard.edu/abs/2006ApJ...642L..21B} {642, L21}

\bibitem[\protect\citeauthoryear{{Binney} \& {Tremaine}}{{Binney} \&
  {Tremaine}}{2008}]{Binney_Tremaine_2008}
{Binney} J.,  {Tremaine} S.,  2008, {Galactic Dynamics: Second Edition}.
Princeton University Press, Princeton

\bibitem[\protect\citeauthoryear{{Bird}, {Ni}, {Di Matteo}, {Croft}, {Feng}  \&
  {Chen}}{{Bird} et~al.}{2022}]{Bird_et_al_2022}
{Bird} S.,  {Ni} Y.,  {Di Matteo} T.,  {Croft} R.,  {Feng} Y.,   {Chen} N.,
  2022, \mn@doi [\mnras] {10.1093/mnras/stac648}, \href
  {https://ui.adsabs.harvard.edu/abs/2022MNRAS.512.3703B} {512, 3703}

\bibitem[\protect\citeauthoryear{{Blumenthal} \& {Barnes}}{{Blumenthal} \&
  {Barnes}}{2018}]{Blumenthal_Barnes_2018}
{Blumenthal} K.~A.,  {Barnes} J.~E.,  2018, \mn@doi [\mnras]
  {10.1093/mnras/sty1605}, \href
  {https://ui.adsabs.harvard.edu/abs/2018MNRAS.479.3952B} {479, 3952}

\bibitem[\protect\citeauthoryear{{Blumenthal}, {Faber}, {Primack}  \&
  {Rees}}{{Blumenthal} et~al.}{1984}]{Blumenthal_et_al_1984}
{Blumenthal} G.~R.,  {Faber} S.~M.,  {Primack} J.~R.,   {Rees} M.~J.,  1984,
  \mn@doi [\nat] {10.1038/311517a0}, \href
  {https://ui.adsabs.harvard.edu/abs/1984Natur.311..517B} {311, 517}

\bibitem[\protect\citeauthoryear{{Bogdanovi{\'c}}, {Miller}  \&
  {Blecha}}{{Bogdanovi{\'c}} et~al.}{2022}]{Bogdanovic_et_al_2022}
{Bogdanovi{\'c}} T.,  {Miller} M.~C.,   {Blecha} L.,  2022, \mn@doi [Living
  Reviews in Relativity] {10.1007/s41114-022-00037-8}, \href
  {https://ui.adsabs.harvard.edu/abs/2022LRR....25....3B} {25, 3}

\bibitem[\protect\citeauthoryear{{Bondi}}{{Bondi}}{1952}]{Bondi_1952}
{Bondi} H.,  1952, \mn@doi [\mnras] {10.1093/mnras/112.2.195}, \href
  {https://ui.adsabs.harvard.edu/abs/1952MNRAS.112..195B} {112, 195}

\bibitem[\protect\citeauthoryear{{Bondi} \& {Hoyle}}{{Bondi} \&
  {Hoyle}}{1944}]{Bondi_Hoyle_1944}
{Bondi} H.,  {Hoyle} F.,  1944, \mn@doi [\mnras] {10.1093/mnras/104.5.273},
  \href {https://ui.adsabs.harvard.edu/abs/1944MNRAS.104..273B} {104, 273}

\bibitem[\protect\citeauthoryear{{Bonetti}, {Bortolas}, {Lupi}, {Dotti}  \&
  {Raimundo}}{{Bonetti} et~al.}{2020}]{Bonetti_et_al_2020}
{Bonetti} M.,  {Bortolas} E.,  {Lupi} A.,  {Dotti} M.,   {Raimundo} S.~I.,
  2020, \mn@doi [\mnras] {10.1093/mnras/staa964}, \href
  {https://ui.adsabs.harvard.edu/abs/2020MNRAS.494.3053B} {494, 3053}

\bibitem[\protect\citeauthoryear{{Bonetti}, {Bortolas}, {Lupi}  \&
  {Dotti}}{{Bonetti} et~al.}{2021}]{Bonetti_et_al_2021}
{Bonetti} M.,  {Bortolas} E.,  {Lupi} A.,   {Dotti} M.,  2021, \mn@doi [\mnras]
  {10.1093/mnras/stab222}, \href
  {https://ui.adsabs.harvard.edu/abs/2021MNRAS.502.3554B} {502, 3554}

\bibitem[\protect\citeauthoryear{{Bortolas}, {Capelo}, {Zana}, {Mayer},
  {Bonetti}, {Dotti}, {Davies}  \& {Madau}}{{Bortolas}
  et~al.}{2020}]{Bortolas_et_al_2020}
{Bortolas} E.,  {Capelo} P.~R.,  {Zana} T.,  {Mayer} L.,  {Bonetti} M.,
  {Dotti} M.,  {Davies} M.~B.,   {Madau} P.,  2020, \mn@doi [\mnras]
  {10.1093/mnras/staa2628}, \href
  {https://ui.adsabs.harvard.edu/abs/2020MNRAS.498.3601B} {498, 3601}

\bibitem[\protect\citeauthoryear{{Bortolas}, {Bonetti}, {Dotti}, {Lupi},
  {Capelo}, {Mayer}  \& {Sesana}}{{Bortolas}
  et~al.}{2022}]{Bortolas_et_al_2022}
{Bortolas} E.,  {Bonetti} M.,  {Dotti} M.,  {Lupi} A.,  {Capelo} P.~R.,
  {Mayer} L.,   {Sesana} A.,  2022, \mn@doi [\mnras] {10.1093/mnras/stac645},
  \href {https://ui.adsabs.harvard.edu/abs/2022MNRAS.512.3365B} {512, 3365}

\bibitem[\protect\citeauthoryear{{Bovy}}{{Bovy}}{2026}]{Bovy_2026}
{Bovy} J.,  2026, {Dynamics and Astrophysics of Galaxies}.
Princeton University Press, Princeton

\bibitem[\protect\citeauthoryear{{Buttigieg}, {Sijacki}, {Moore}  \&
  {Bourne}}{{Buttigieg} et~al.}{2025}]{Buttigieg_et_al_2025}
{Buttigieg} S.,  {Sijacki} D.,  {Moore} C.~J.,   {Bourne} M.~A.,  2025, \mn@doi
  [\mnras] {10.1093/mnras/staf1336}, \href
  {https://ui.adsabs.harvard.edu/abs/2025MNRAS.542.2019B} {542, 2019}

\bibitem[\protect\citeauthoryear{{Capelo} \& {Dotti}}{{Capelo} \&
  {Dotti}}{2017}]{Capelo_Dotti_2017}
{Capelo} P.~R.,  {Dotti} M.,  2017, \mn@doi [\mnras] {10.1093/mnras/stw2872},
  \href {https://ui.adsabs.harvard.edu/abs/2017MNRAS.465.2643C} {465, 2643}

\bibitem[\protect\citeauthoryear{{Capelo}, {Volonteri}, {Dotti}, {Bellovary},
  {Mayer}  \& {Governato}}{{Capelo} et~al.}{2015}]{Capelo_et_al_2015}
{Capelo} P.~R.,  {Volonteri} M.,  {Dotti} M.,  {Bellovary} J.~M.,  {Mayer} L.,
   {Governato} F.,  2015, \mn@doi [\mnras] {10.1093/mnras/stu2500}, \href
  {https://ui.adsabs.harvard.edu/abs/2015MNRAS.447.2123C} {447, 2123}

\bibitem[\protect\citeauthoryear{{Capelo}, {Dotti}, {Volonteri}, {Mayer},
  {Bellovary}  \& {Shen}}{{Capelo} et~al.}{2017}]{Capelo_et_al_2017}
{Capelo} P.~R.,  {Dotti} M.,  {Volonteri} M.,  {Mayer} L.,  {Bellovary} J.~M.,
   {Shen} S.,  2017, \mn@doi [\mnras] {10.1093/mnras/stx1067}, \href
  {https://ui.adsabs.harvard.edu/abs/2017MNRAS.469.4437C} {469, 4437}

\bibitem[\protect\citeauthoryear{{Capelo}, {Feruglio}, {Hickox}  \&
  {Tombesi}}{{Capelo} et~al.}{2023}]{Capelo_et_al_2023}
{Capelo} P.~R.,  {Feruglio} C.,  {Hickox} R.~C.,   {Tombesi} F.,  2023, in ,
  Handbook of X-ray and Gamma-ray Astrophysics.
Springer Nature Singapore, p.~126, \mn@doi{10.1007/978-981-16-4544-0_115-1}

\bibitem[\protect\citeauthoryear{{Capelo}, {Mangiagli}, {Zwick}, {Mayer}  \&
  {Volonteri}}{{Capelo} et~al.}{2026}]{Capelo_et_al_2026}
{Capelo} P.~R.,  {Mangiagli} A.,  {Zwick} L.,  {Mayer} L.,   {Volonteri} M.,
  2026, in {Bonaldi} A.,  et~al., eds, , Advancing Astrophysics with the SKA
  II.
p.~189, \mn@doi{10.48550/arXiv.2606.25020}

\bibitem[\protect\citeauthoryear{{Cavanagh}, {Bekki}, {Groves}  \&
  {Pfeffer}}{{Cavanagh} et~al.}{2022}]{Cavanagh_et_al_2022}
{Cavanagh} M.~K.,  {Bekki} K.,  {Groves} B.~A.,   {Pfeffer} J.,  2022, \mn@doi
  [\mnras] {10.1093/mnras/stab3786}, \href
  {https://ui.adsabs.harvard.edu/abs/2022MNRAS.510.5164C} {510, 5164}

\bibitem[\protect\citeauthoryear{{Chakraborty}, {Gallerani}, {Di Mascia},
  {Zana}, {Valentini}, {Carniani}, {Vito}  \& {Bhatt}}{{Chakraborty}
  et~al.}{2025}]{Chakraborty_et_al_2025}
{Chakraborty} S.,  {Gallerani} S.,  {Di Mascia} F.,  {Zana} T.,  {Valentini}
  M.,  {Carniani} S.,  {Vito} F.,   {Bhatt} M.,  2025, \mn@doi [\aap]
  {10.1051/0004-6361/202452926}, \href
  {https://ui.adsabs.harvard.edu/abs/2025A&A...698A.268C} {698, A268}

\bibitem[\protect\citeauthoryear{{Chandrasekhar}}{{Chandrasekhar}}{1943}]{Chandrasekhar_1943}
{Chandrasekhar} S.,  1943, \mn@doi [\apj] {10.1086/144517}, \href
  {https://ui.adsabs.harvard.edu/abs/1943ApJ....97..255C} {97, 255}

\bibitem[\protect\citeauthoryear{{Charnock}, {Lavaux}  \& {Wandelt}}{{Charnock}
  et~al.}{2018}]{Charnock_et_al_2018}
{Charnock} T.,  {Lavaux} G.,   {Wandelt} B.~D.,  2018, \mn@doi [\prd]
  {10.1103/PhysRevD.97.083004}, \href
  {https://ui.adsabs.harvard.edu/abs/2018PhRvD..97h3004C} {97, 083004}

\bibitem[\protect\citeauthoryear{{Chen}, {Ni}, {Tremmel}, {Di Matteo}, {Bird},
  {DeGraf}  \& {Feng}}{{Chen} et~al.}{2022}]{Chen_et_al_2022}
{Chen} N.,  {Ni} Y.,  {Tremmel} M.,  {Di Matteo} T.,  {Bird} S.,  {DeGraf} C.,
   {Feng} Y.,  2022, \mn@doi [\mnras] {10.1093/mnras/stab3411}, \href
  {https://ui.adsabs.harvard.edu/abs/2022MNRAS.510..531C} {510, 531}

\bibitem[\protect\citeauthoryear{{Colpi} et~al.,}{{Colpi}
  et~al.}{2024}]{Colpi_et_al_2024}
{Colpi} M.,  et~al., 2024, \mn@doi [arXiv e-prints]
  {10.48550/arXiv.2402.07571}, \href
  {https://ui.adsabs.harvard.edu/abs/2024arXiv240207571C} {p. arXiv:2402.07571}

\bibitem[\protect\citeauthoryear{{Crenshaw}, {Kalmbach}, {Gagliano}, {Yan},
  {Connolly}, {Malz}, {Schmidt}  \& {The LSST Dark Energy Science
  Collaboration}}{{Crenshaw} et~al.}{2024}]{Crenshaw_et_al_2024}
{Crenshaw} J.~F.,  {Kalmbach} J.~B.,  {Gagliano} A.,  {Yan} Z.,  {Connolly}
  A.~J.,  {Malz} A.~I.,  {Schmidt} S.~J.,   {The LSST Dark Energy Science
  Collaboration} 2024, \mn@doi [\aj] {10.3847/1538-3881/ad54bf}, \href
  {https://ui.adsabs.harvard.edu/abs/2024AJ....168...80C} {168, 80}

\bibitem[\protect\citeauthoryear{{Damiano}, {Valentini}, {Borgani},
  {Tornatore}, {Murante}, {Ragagnin}, {Ragone-Figueroa}  \& {Dolag}}{{Damiano}
  et~al.}{2024}]{Damiano_et_al_2024}
{Damiano} A.,  {Valentini} M.,  {Borgani} S.,  {Tornatore} L.,  {Murante} G.,
  {Ragagnin} A.,  {Ragone-Figueroa} C.,   {Dolag} K.,  2024, \mn@doi [\aap]
  {10.1051/0004-6361/202450021}, \href
  {https://ui.adsabs.harvard.edu/abs/2024A&A...692A..81D} {692, A81}

\bibitem[\protect\citeauthoryear{{Dattathri}, {van den Bosch}, {Banik},
  {Weinberg}, {Natarajan}, {Li}  \& {Dekel}}{{Dattathri}
  et~al.}{2026}]{Dattathri_et_al_2025}
{Dattathri} S.,  {van den Bosch} F.~C.,  {Banik} U.,  {Weinberg} M.,
  {Natarajan} P.,  {Li} Z.,   {Dekel} A.,  2026, \mn@doi [The Open Journal of
  Astrophysics] {10.33232/001c.169864}, \href
  {https://ui.adsabs.harvard.edu/abs/2026OJAp....969864D} {9, 69864}

\bibitem[\protect\citeauthoryear{{Dax}, {Green}, {Gair}, {Macke}, {Buonanno}
  \& {Sch{\"o}lkopf}}{{Dax} et~al.}{2021}]{Dax_et_al_2021}
{Dax} M.,  {Green} S.~R.,  {Gair} J.,  {Macke} J.~H.,  {Buonanno} A.,
  {Sch{\"o}lkopf} B.,  2021, \mn@doi [\prl] {10.1103/PhysRevLett.127.241103},
  \href {https://ui.adsabs.harvard.edu/abs/2021PhRvL.127x1103D} {127, 241103}

\bibitem[\protect\citeauthoryear{{De Rosa} et~al.,}{{De Rosa}
  et~al.}{2019}]{DeRosa_et_al_2019}
{De Rosa} A.,  et~al., 2019, \mn@doi [\nar] {10.1016/j.newar.2020.101525},
  \href {https://ui.adsabs.harvard.edu/abs/2019NewAR..8601525D} {86, 101525}

\bibitem[\protect\citeauthoryear{{DeGraf} \& {Sijacki}}{{DeGraf} \&
  {Sijacki}}{2020}]{DeGraf_Sijacki_2020}
{DeGraf} C.,  {Sijacki} D.,  2020, \mn@doi [\mnras] {10.1093/mnras/stz3309},
  \href {https://ui.adsabs.harvard.edu/abs/2020MNRAS.491.4973D} {491, 4973}

\bibitem[\protect\citeauthoryear{{Debuhr}, {Quataert}  \& {Ma}}{{Debuhr}
  et~al.}{2011}]{Debuhr_et_al_2011}
{Debuhr} J.,  {Quataert} E.,   {Ma} C.-P.,  2011, \mn@doi [\mnras]
  {10.1111/j.1365-2966.2010.17992.x}, \href
  {https://ui.adsabs.harvard.edu/abs/2011MNRAS.412.1341D} {412, 1341}

\bibitem[\protect\citeauthoryear{{Dehnen}}{{Dehnen}}{1993}]{Dehnen_1993}
{Dehnen} W.,  1993, \mn@doi [\mnras] {10.1093/mnras/265.1.250}, \href
  {https://ui.adsabs.harvard.edu/abs/1993MNRAS.265..250D} {265, 250}

\bibitem[\protect\citeauthoryear{{Durkan}, {Bekasov}, {Murray}  \&
  {Papamakarios}}{{Durkan} et~al.}{2019}]{Durkan_et_al_2019}
{Durkan} C.,  {Bekasov} A.,  {Murray} I.,   {Papamakarios} G.,  2019, \mn@doi
  [arXiv e-prints] {10.48550/arXiv.1906.04032}, \href
  {https://ui.adsabs.harvard.edu/abs/2019arXiv190604032D} {p. arXiv:1906.04032}

\bibitem[\protect\citeauthoryear{{Eddington}}{{Eddington}}{1916}]{Eddington_1916}
{Eddington} A.~S.,  1916, \mn@doi [\mnras] {10.1093/mnras/77.1.16}, \href
  {https://ui.adsabs.harvard.edu/abs/1916MNRAS..77...16E} {77, 16}

\bibitem[\protect\citeauthoryear{{Einstein}}{{Einstein}}{1916}]{Einstein_1916}
{Einstein} A.,  1916, Sitzungsberichte der K{\"o}niglich Preussischen Akademie
  der Wissenschaften, \href
  {https://ui.adsabs.harvard.edu/abs/1916SPAW.......688E} {pp 688--696}

\bibitem[\protect\citeauthoryear{{Fakhouri}, {Ma}  \&
  {Boylan-Kolchin}}{{Fakhouri} et~al.}{2010}]{Fakhouri_et_al_2010}
{Fakhouri} O.,  {Ma} C.-P.,   {Boylan-Kolchin} M.,  2010, \mn@doi [\mnras]
  {10.1111/j.1365-2966.2010.16859.x}, \href
  {https://ui.adsabs.harvard.edu/abs/2010MNRAS.406.2267F} {406, 2267}

\bibitem[\protect\citeauthoryear{{Fastidio}, {Bortolas}, {Gualandris},
  {Sesana}, {Read}  \& {Dehnen}}{{Fastidio} et~al.}{2025}]{Fastidio_et_al_2025}
{Fastidio} F.,  {Bortolas} E.,  {Gualandris} A.,  {Sesana} A.,  {Read} J.~I.,
  {Dehnen} W.,  2025, \mn@doi [\aap] {10.1051/0004-6361/202556035}, \href
  {https://ui.adsabs.harvard.edu/abs/2025A&A...703A..86F} {703, A86}

\bibitem[\protect\citeauthoryear{{Fran{\c{c}}ois}, {Gualandris}  \&
  {Dehnen}}{{Fran{\c{c}}ois} et~al.}{2026}]{Francois_et_al_2026}
{Fran{\c{c}}ois} T.~L.,  {Gualandris} A.,   {Dehnen} W.,  2026, \mn@doi
  [\mnras] {10.1093/mnras/stag1298}, \href
  {https://ui.adsabs.harvard.edu/abs/2026MNRAS.551g1298F} {551, stag1298}

\bibitem[\protect\citeauthoryear{{Gualandris}, {Read}, {Dehnen}  \&
  {Bortolas}}{{Gualandris} et~al.}{2017}]{Gualandris_et_al_2017}
{Gualandris} A.,  {Read} J.~I.,  {Dehnen} W.,   {Bortolas} E.,  2017, \mn@doi
  [\mnras] {10.1093/mnras/stw2528}, \href
  {https://ui.adsabs.harvard.edu/abs/2017MNRAS.464.2301G} {464, 2301}

\bibitem[\protect\citeauthoryear{{Gualandris}, {Khan}, {Bortolas}, {Bonetti},
  {Sesana}, {Berczik}  \& {Holley-Bockelmann}}{{Gualandris}
  et~al.}{2022}]{Gualandris_et_al_2022}
{Gualandris} A.,  {Khan} F.~M.,  {Bortolas} E.,  {Bonetti} M.,  {Sesana} A.,
  {Berczik} P.,   {Holley-Bockelmann} K.,  2022, \mn@doi [\mnras]
  {10.1093/mnras/stac241}, \href
  {https://ui.adsabs.harvard.edu/abs/2022MNRAS.511.4753G} {511, 4753}

\bibitem[\protect\citeauthoryear{{Haidar} et~al.,}{{Haidar}
  et~al.}{2022}]{Haidar_et_al_2022}
{Haidar} H.,  et~al., 2022, \mn@doi [\mnras] {10.1093/mnras/stac1659}, \href
  {https://ui.adsabs.harvard.edu/abs/2022MNRAS.514.4912H} {514, 4912}

\bibitem[\protect\citeauthoryear{{Hernquist}}{{Hernquist}}{1990}]{Hernquist_1990}
{Hernquist} L.,  1990, \mn@doi [\apj] {10.1086/168845}, \href
  {https://ui.adsabs.harvard.edu/abs/1990ApJ...356..359H} {356, 359}

\bibitem[\protect\citeauthoryear{{Hopkins} \& {Quataert}}{{Hopkins} \&
  {Quataert}}{2010}]{Hopkins_Quataert_2010}
{Hopkins} P.~F.,  {Quataert} E.,  2010, \mn@doi [\mnras]
  {10.1111/j.1365-2966.2010.17064.x}, \href
  {https://ui.adsabs.harvard.edu/abs/2010MNRAS.407.1529H} {407, 1529}

\bibitem[\protect\citeauthoryear{{Hoyle} \& {Lyttleton}}{{Hoyle} \&
  {Lyttleton}}{1939}]{Hoyle_Lyttleton_1939}
{Hoyle} F.,  {Lyttleton} R.~A.,  1939, \mn@doi [Proceedings of the Cambridge
  Philosophical Society] {10.1017/S0305004100021150}, \href
  {https://ui.adsabs.harvard.edu/abs/1939PCPS...35..405H} {35, 405}

\bibitem[\protect\citeauthoryear{{Izquierdo-Villalba}
  et~al.,}{{Izquierdo-Villalba} et~al.}{2026}]{Izquierdo-Villalba_et_al_2026}
{Izquierdo-Villalba} D.,  et~al., 2026, arXiv e-prints, \href
  {https://ui.adsabs.harvard.edu/abs/2026arXiv260500092I} {p. arXiv:2605.00092}

\bibitem[\protect\citeauthoryear{{Jimenez Rezende} \& {Mohamed}}{{Jimenez
  Rezende} \& {Mohamed}}{2015}]{JimenezRezende_Mohamed_2015}
{Jimenez Rezende} D.,  {Mohamed} S.,  2015, \mn@doi [arXiv e-prints]
  {10.48550/arXiv.1505.05770}, \href
  {https://ui.adsabs.harvard.edu/abs/2015arXiv150505770J} {p. arXiv:1505.05770}

\bibitem[\protect\citeauthoryear{{Kacprzak}, {Fluri}, {Schneider}, {Refregier}
  \& {Stadel}}{{Kacprzak} et~al.}{2023}]{Kacprzak_et_al_2023}
{Kacprzak} T.,  {Fluri} J.,  {Schneider} A.,  {Refregier} A.,   {Stadel} J.,
  2023, \mn@doi [\jcap] {10.1088/1475-7516/2023/02/050}, \href
  {https://ui.adsabs.harvard.edu/abs/2023JCAP...02..050K} {2023, 050}

\bibitem[\protect\citeauthoryear{{Katz}, {Kelley}, {Dosopoulou}, {Berry},
  {Blecha}  \& {Larson}}{{Katz} et~al.}{2020}]{Katz_et_al_2020}
{Katz} M.~L.,  {Kelley} L.~Z.,  {Dosopoulou} F.,  {Berry} S.,  {Blecha} L.,
  {Larson} S.~L.,  2020, \mn@doi [\mnras] {10.1093/mnras/stz3102}, \href
  {https://ui.adsabs.harvard.edu/abs/2020MNRAS.491.2301K} {491, 2301}

\bibitem[\protect\citeauthoryear{{Khan}, {Just}  \& {Merritt}}{{Khan}
  et~al.}{2011}]{Khan_et_al_2011}
{Khan} F.~M.,  {Just} A.,   {Merritt} D.,  2011, \mn@doi [\apj]
  {10.1088/0004-637X/732/2/89}, \href
  {https://ui.adsabs.harvard.edu/abs/2011ApJ...732...89K} {732, 89}

\bibitem[\protect\citeauthoryear{{Khan}, {Capelo}, {Mayer}  \&
  {Berczik}}{{Khan} et~al.}{2018}]{Khan_et_al_2018}
{Khan} F.~M.,  {Capelo} P.~R.,  {Mayer} L.,   {Berczik} P.,  2018, \mn@doi
  [\apj] {10.3847/1538-4357/aae77b}, \href
  {https://ui.adsabs.harvard.edu/abs/2018ApJ...868...97K} {868, 97}

\bibitem[\protect\citeauthoryear{{Kim} et~al.,}{{Kim}
  et~al.}{2016}]{Kim_et_al_2016}
{Kim} J.-h.,  et~al., 2016, \mn@doi [\apj] {10.3847/1538-4357/833/2/202}, \href
  {https://ui.adsabs.harvard.edu/abs/2016ApJ...833..202K} {833, 202}

\bibitem[\protect\citeauthoryear{Kingma \& Ba}{Kingma \&
  Ba}{2015}]{Kingma_et_al_2015}
Kingma D.~P.,  Ba J.,  2015, in Proceedings of the 3rd International Conference
  on Learning Representations.  (\mn@eprint {arXiv} {1412.6980})

\bibitem[\protect\citeauthoryear{{Kormendy} \& {Ho}}{{Kormendy} \&
  {Ho}}{2013}]{Kormendy_Ho_2013}
{Kormendy} J.,  {Ho} L.~C.,  2013, \mn@doi [\araa]
  {10.1146/annurev-astro-082708-101811}, \href
  {https://ui.adsabs.harvard.edu/abs/2013ARA&A..51..511K} {51, 511}

\bibitem[\protect\citeauthoryear{{Koss}, {Mushotzky}, {Treister}, {Veilleux},
  {Vasudevan}  \& {Trippe}}{{Koss} et~al.}{2012}]{Koss_et_al_2012}
{Koss} M.,  {Mushotzky} R.,  {Treister} E.,  {Veilleux} S.,  {Vasudevan} R.,
  {Trippe} M.,  2012, \mn@doi [\apjl] {10.1088/2041-8205/746/2/L22}, \href
  {https://ui.adsabs.harvard.edu/abs/2012ApJ...746L..22K} {746, L22}

\bibitem[\protect\citeauthoryear{{Krause} et~al.,}{{Krause}
  et~al.}{2025}]{Krause_et_al_2025}
{Krause} M. G.~H.,  et~al., 2025, \mn@doi [\pasa] {10.1017/pasa.2025.10120},
  \href {https://ui.adsabs.harvard.edu/abs/2025PASA...42..162K} {42, e162}

\bibitem[\protect\citeauthoryear{{Le Conte} et~al.,}{{Le Conte}
  et~al.}{2024}]{LeConte_et_al_2024}
{Le Conte} Z.~A.,  et~al., 2024, \mn@doi [\mnras] {10.1093/mnras/stae921},
  \href {https://ui.adsabs.harvard.edu/abs/2024MNRAS.530.1984L} {530, 1984}

\bibitem[\protect\citeauthoryear{{Le Conte} et~al.,}{{Le Conte}
  et~al.}{2026}]{LeConte_et_al_2026}
{Le Conte} Z.~A.,  et~al., 2026, \mn@doi [\mnras] {10.1093/mnras/staf2010},
  \href {https://ui.adsabs.harvard.edu/abs/2026MNRAS.545f2010L} {545, staf2010}

\bibitem[\protect\citeauthoryear{{Li}, {Bogdanovi{\'c}}  \& {Ballantyne}}{{Li}
  et~al.}{2020a}]{Li_et_al_2020a}
{Li} K.,  {Bogdanovi{\'c}} T.,   {Ballantyne} D.~R.,  2020a, \mn@doi [\apj]
  {10.3847/1538-4357/ab93c6}, \href
  {https://ui.adsabs.harvard.edu/abs/2020ApJ...896..113L} {896, 113}

\bibitem[\protect\citeauthoryear{{Li}, {Bogdanovi{\'c}}  \& {Ballantyne}}{{Li}
  et~al.}{2020b}]{Li_et_al_2020b}
{Li} K.,  {Bogdanovi{\'c}} T.,   {Ballantyne} D.~R.,  2020b, \mn@doi [\apj]
  {10.3847/1538-4357/abc555}, \href
  {https://ui.adsabs.harvard.edu/abs/2020ApJ...905..123L} {905, 123}

\bibitem[\protect\citeauthoryear{{Li}, {Bogdanovi{\'c}}, {Ballantyne}  \&
  {Bonetti}}{{Li} et~al.}{2022}]{Li_et_al_2022}
{Li} K.,  {Bogdanovi{\'c}} T.,  {Ballantyne} D.~R.,   {Bonetti} M.,  2022,
  \mn@doi [\apj] {10.3847/1538-4357/ac74b5}, \href
  {https://ui.adsabs.harvard.edu/abs/2022ApJ...933..104L} {933, 104}

\bibitem[\protect\citeauthoryear{{Li} et~al.,}{{Li}
  et~al.}{2025}]{Li_et_al_2025}
{Li} E.-K.,  et~al., 2025, \mn@doi [Reports on Progress in Physics]
  {10.1088/1361-6633/adc9be}, \href
  {https://ui.adsabs.harvard.edu/abs/2025RPPh...88e6901L} {88, 056901}

\bibitem[\protect\citeauthoryear{{Long} \& {Murali}}{{Long} \&
  {Murali}}{1992}]{Long_Murali_1992}
{Long} K.,  {Murali} C.,  1992, \mn@doi [\apj] {10.1086/171764}, \href
  {https://ui.adsabs.harvard.edu/abs/1992ApJ...397...44L} {397, 44}

\bibitem[\protect\citeauthoryear{{Luo} et~al.,}{{Luo}
  et~al.}{2016}]{Luo_et_al_2016}
{Luo} J.,  et~al., 2016, \mn@doi [Classical and Quantum Gravity]
  {10.1088/0264-9381/33/3/035010}, \href
  {https://ui.adsabs.harvard.edu/abs/2016CQGra..33c5010L} {33, 035010}

\bibitem[\protect\citeauthoryear{{Luo}, {Wang}, {Wu}, {Hu}  \& {Jin}}{{Luo}
  et~al.}{2021}]{Luo_et_al_2021}
{Luo} Z.,  {Wang} Y.,  {Wu} Y.,  {Hu} W.,   {Jin} G.,  2021, \mn@doi [Progress
  of Theoretical and Experimental Physics] {10.1093/ptep/ptaa083}, \href
  {https://ui.adsabs.harvard.edu/abs/2021PTEP.2021eA108L} {2021, 05A108}

\bibitem[\protect\citeauthoryear{{Mannerkoski}, {Rawlings}, {Johansson},
  {Naab}, {Rantala}, {Springel}, {Irodotou}  \& {Liao}}{{Mannerkoski}
  et~al.}{2023}]{Mannerkoski_et_al_2023}
{Mannerkoski} M.,  {Rawlings} A.,  {Johansson} P.~H.,  {Naab} T.,  {Rantala}
  A.,  {Springel} V.,  {Irodotou} D.,   {Liao} S.,  2023, \mn@doi [\mnras]
  {10.1093/mnras/stad2139}, \href
  {https://ui.adsabs.harvard.edu/abs/2023MNRAS.524.4062M} {524, 4062}

\bibitem[\protect\citeauthoryear{{Marinacci} et~al.,}{{Marinacci}
  et~al.}{2018}]{Marinacci_et_al_2018}
{Marinacci} F.,  et~al., 2018, \mn@doi [\mnras] {10.1093/mnras/sty2206}, \href
  {https://ui.adsabs.harvard.edu/abs/2018MNRAS.480.5113M} {480, 5113}

\bibitem[\protect\citeauthoryear{{Merritt}}{{Merritt}}{2013}]{Merritt_2013}
{Merritt} D.,  2013, {Dynamics and Evolution of Galactic Nuclei}.
Princeton University Press, Princeton

\bibitem[\protect\citeauthoryear{{Merritt}, {Berczik}  \& {Laun}}{{Merritt}
  et~al.}{2007}]{Merritt_et_al_2007}
{Merritt} D.,  {Berczik} P.,   {Laun} F.,  2007, \mn@doi [\aj]
  {10.1086/510294}, \href
  {https://ui.adsabs.harvard.edu/abs/2007AJ....133..553M} {133, 553}

\bibitem[\protect\citeauthoryear{{Milosavljevi{\'c}} \&
  {Merritt}}{{Milosavljevi{\'c}} \&
  {Merritt}}{2001}]{Milosavljevic_Merritt_2001}
{Milosavljevi{\'c}} M.,  {Merritt} D.,  2001, \mn@doi [\apj] {10.1086/323830},
  \href {https://ui.adsabs.harvard.edu/abs/2001ApJ...563...34M} {563, 34}

\bibitem[\protect\citeauthoryear{{Naiman} et~al.,}{{Naiman}
  et~al.}{2018}]{Naiman_et_al_2018}
{Naiman} J.~P.,  et~al., 2018, \mn@doi [\mnras] {10.1093/mnras/sty618}, \href
  {https://ui.adsabs.harvard.edu/abs/2018MNRAS.477.1206N} {477, 1206}

\bibitem[\protect\citeauthoryear{{Navarro}, {Frenk}  \& {White}}{{Navarro}
  et~al.}{1996}]{Navarro_et_al_1996}
{Navarro} J.~F.,  {Frenk} C.~S.,   {White} S. D.~M.,  1996, \mn@doi [\apj]
  {10.1086/177173}, \href
  {https://ui.adsabs.harvard.edu/abs/1996ApJ...462..563N} {462, 563}

\bibitem[\protect\citeauthoryear{{Nelson} et~al.,}{{Nelson}
  et~al.}{2018}]{Nelson_et_al_2018}
{Nelson} D.,  et~al., 2018, \mn@doi [\mnras] {10.1093/mnras/stx3040}, \href
  {https://ui.adsabs.harvard.edu/abs/2018MNRAS.475..624N} {475, 624}

\bibitem[\protect\citeauthoryear{{Nelson} et~al.,}{{Nelson}
  et~al.}{2019}]{Nelson_et_al_2019}
{Nelson} D.,  et~al., 2019, \mn@doi [\mnras] {10.1093/mnras/stz2306}, \href
  {https://ui.adsabs.harvard.edu/abs/2019MNRAS.490.3234N} {490, 3234}

\bibitem[\protect\citeauthoryear{{Ntampaka} et~al.,}{{Ntampaka}
  et~al.}{2019}]{Ntampaka_et_al_2019}
{Ntampaka} M.,  et~al., 2019, \mn@doi [\apj] {10.3847/1538-4357/ab14eb}, \href
  {https://ui.adsabs.harvard.edu/abs/2019ApJ...876...82N} {876, 82}

\bibitem[\protect\citeauthoryear{{Ostriker}}{{Ostriker}}{1999}]{Ostriker_1999}
{Ostriker} E.~C.,  1999, \mn@doi [\apj] {10.1086/306858}, \href
  {https://ui.adsabs.harvard.edu/abs/1999ApJ...513..252O} {513, 252}

\bibitem[\protect\citeauthoryear{{Papamakarios}, {Nalisnick}, {Jimenez
  Rezende}, {Mohamed}  \& {Lakshminarayanan}}{{Papamakarios}
  et~al.}{2021}]{Papamakarios_et_al_2019}
{Papamakarios} G.,  {Nalisnick} E.,  {Jimenez Rezende} D.,  {Mohamed} S.,
  {Lakshminarayanan} B.,  2021, \mn@doi [The Journal of Machine Learning
  Research] {10.48550/arXiv.1912.02762}, \href
  {https://ui.adsabs.harvard.edu/abs/2019arXiv191202762P} {22, 2617}

\bibitem[\protect\citeauthoryear{{Park} \& {Bogdanovi{\'c}}}{{Park} \&
  {Bogdanovi{\'c}}}{2017}]{Park_Bogdanovic_2017}
{Park} K.,  {Bogdanovi{\'c}} T.,  2017, \mn@doi [\apj]
  {10.3847/1538-4357/aa65ce}, \href
  {https://ui.adsabs.harvard.edu/abs/2017ApJ...838..103P} {838, 103}

\bibitem[\protect\citeauthoryear{{Park} \& {Bogdanovi{\'c}}}{{Park} \&
  {Bogdanovi{\'c}}}{2019}]{Park_Bogdanovic_2019}
{Park} K.,  {Bogdanovi{\'c}} T.,  2019, \mn@doi [\apj]
  {10.3847/1538-4357/ab3f30}, \href
  {https://ui.adsabs.harvard.edu/abs/2019ApJ...883..209P} {883, 209}

\bibitem[\protect\citeauthoryear{{Peschken} \& {{\L}okas}}{{Peschken} \&
  {{\L}okas}}{2019}]{Peschken_Lokas_2019}
{Peschken} N.,  {{\L}okas} E.~L.,  2019, \mn@doi [\mnras]
  {10.1093/mnras/sty3277}, \href
  {https://ui.adsabs.harvard.edu/abs/2019MNRAS.483.2721P} {483, 2721}

\bibitem[\protect\citeauthoryear{{Pfeifle}, {Weaver}, {Secrest}, {Rothberg}  \&
  {Patton}}{{Pfeifle} et~al.}{2025}]{Pfeifle_et_al_2025}
{Pfeifle} R.~W.,  {Weaver} K.~A.,  {Secrest} N.~J.,  {Rothberg} B.,   {Patton}
  D.~R.,  2025, \mn@doi [\apjs] {10.3847/1538-4365/adf845}, \href
  {https://ui.adsabs.harvard.edu/abs/2025ApJS..281...25P} {281, 25}

\bibitem[\protect\citeauthoryear{{Pfister}, {Lupi}, {Capelo}, {Volonteri},
  {Bellovary}  \& {Dotti}}{{Pfister} et~al.}{2017}]{Pfister_et_al_2017}
{Pfister} H.,  {Lupi} A.,  {Capelo} P.~R.,  {Volonteri} M.,  {Bellovary} J.~M.,
    {Dotti} M.,  2017, \mn@doi [\mnras] {10.1093/mnras/stx1853}, \href
  {https://ui.adsabs.harvard.edu/abs/2017MNRAS.471.3646P} {471, 3646}

\bibitem[\protect\citeauthoryear{{Pfister}, {Volonteri}, {Dubois}, {Dotti}  \&
  {Colpi}}{{Pfister} et~al.}{2019}]{Pfister_et_al_2019}
{Pfister} H.,  {Volonteri} M.,  {Dubois} Y.,  {Dotti} M.,   {Colpi} M.,  2019,
  \mn@doi [\mnras] {10.1093/mnras/stz822}, \href
  {https://ui.adsabs.harvard.edu/abs/2019MNRAS.486..101P} {486, 101}

\bibitem[\protect\citeauthoryear{{Pillepich} et~al.,}{{Pillepich}
  et~al.}{2018a}]{Pillepich_et_al_2018a}
{Pillepich} A.,  et~al., 2018a, \mn@doi [\mnras] {10.1093/mnras/stx2656}, \href
  {https://ui.adsabs.harvard.edu/abs/2018MNRAS.473.4077P} {473, 4077}

\bibitem[\protect\citeauthoryear{{Pillepich} et~al.,}{{Pillepich}
  et~al.}{2018b}]{Pillepich_et_al_2018b}
{Pillepich} A.,  et~al., 2018b, \mn@doi [\mnras] {10.1093/mnras/stx3112}, \href
  {https://ui.adsabs.harvard.edu/abs/2018MNRAS.475..648P} {475, 648}

\bibitem[\protect\citeauthoryear{{Pillepich} et~al.,}{{Pillepich}
  et~al.}{2019}]{Pillepich_et_al_2019}
{Pillepich} A.,  et~al., 2019, \mn@doi [\mnras] {10.1093/mnras/stz2338}, \href
  {https://ui.adsabs.harvard.edu/abs/2019MNRAS.490.3196P} {490, 3196}

\bibitem[\protect\citeauthoryear{{Planck Collaboration} et~al.,}{{Planck
  Collaboration} et~al.}{2016}]{Planck_2016}
{Planck Collaboration} et~al., 2016, \mn@doi [\aap]
  {10.1051/0004-6361/201525830}, \href
  {https://ui.adsabs.harvard.edu/abs/2016A&A...594A..13P} {594, A13}

\bibitem[\protect\citeauthoryear{{Portail}, {Gerhard}, {Wegg}  \&
  {Ness}}{{Portail} et~al.}{2017}]{Portail_et_al_2017}
{Portail} M.,  {Gerhard} O.,  {Wegg} C.,   {Ness} M.,  2017, \mn@doi [\mnras]
  {10.1093/mnras/stw2819}, \href
  {https://ui.adsabs.harvard.edu/abs/2017MNRAS.465.1621P} {465, 1621}

\bibitem[\protect\citeauthoryear{{Rosas-Guevara} et~al.,}{{Rosas-Guevara}
  et~al.}{2022}]{RosasGuevara_et_al_2022}
{Rosas-Guevara} Y.,  et~al., 2022, \mn@doi [\mnras] {10.1093/mnras/stac816},
  \href {https://ui.adsabs.harvard.edu/abs/2022MNRAS.512.5339R} {512, 5339}

\bibitem[\protect\citeauthoryear{{Ruan}, {Guo}, {Cai}  \& {Zhang}}{{Ruan}
  et~al.}{2020}]{Ruan_et_al_2020}
{Ruan} W.-H.,  {Guo} Z.-K.,  {Cai} R.-G.,   {Zhang} Y.-Z.,  2020, \mn@doi
  [International Journal of Modern Physics A] {10.1142/S0217751X2050075X},
  \href {https://ui.adsabs.harvard.edu/abs/2020IJMPA..3550075R} {35, 2050075}

\bibitem[\protect\citeauthoryear{{Salcido}, {Bower}, {Theuns}, {McAlpine},
  {Schaller}, {Crain}, {Schaye}  \& {Regan}}{{Salcido}
  et~al.}{2016}]{Salcido_et_al_2016}
{Salcido} J.,  {Bower} R.~G.,  {Theuns} T.,  {McAlpine} S.,  {Schaller} M.,
  {Crain} R.~A.,  {Schaye} J.,   {Regan} J.,  2016, \mn@doi [\mnras]
  {10.1093/mnras/stw2048}, \href
  {https://ui.adsabs.harvard.edu/abs/2016MNRAS.463..870S} {463, 870}

\bibitem[\protect\citeauthoryear{{Sato-Polito}, {Zaldarriaga}  \&
  {Quataert}}{{Sato-Polito} et~al.}{2025}]{Sato-Polito_et_al_2025}
{Sato-Polito} G.,  {Zaldarriaga} M.,   {Quataert} E.,  2025, \mn@doi [\prd]
  {10.1103/1br7-s1rc}, \href
  {https://ui.adsabs.harvard.edu/abs/2025PhRvD.112l3018S} {112, 123018}

\bibitem[\protect\citeauthoryear{{Sellwood} \& {Wilkinson}}{{Sellwood} \&
  {Wilkinson}}{1993}]{Sellwood_Wilkinson_1993}
{Sellwood} J.~A.,  {Wilkinson} A.,  1993, \mn@doi [Reports on Progress in
  Physics] {10.1088/0034-4885/56/2/001}, \href
  {https://ui.adsabs.harvard.edu/abs/1993RPPh...56..173S} {56, 173}

\bibitem[\protect\citeauthoryear{{Shallue} \& {Vanderburg}}{{Shallue} \&
  {Vanderburg}}{2018}]{Shallue_et_al_2018}
{Shallue} C.~J.,  {Vanderburg} A.,  2018, \mn@doi [\aj]
  {10.3847/1538-3881/aa9e09}, \href
  {https://ui.adsabs.harvard.edu/abs/2018AJ....155...94S} {155, 94}

\bibitem[\protect\citeauthoryear{{Shannon} et~al.,}{{Shannon}
  et~al.}{2025}]{Shannon_et_al_2025}
{Shannon} R.~M.,  et~al., 2025, \mn@doi [The Open Journal of Astrophysics]
  {10.33232/001c.154243}, \href
  {https://ui.adsabs.harvard.edu/abs/2025OJAp....854243S} {8, 54243}

\bibitem[\protect\citeauthoryear{{Shibuya}, {Ouchi}, {Kubo}  \&
  {Harikane}}{{Shibuya} et~al.}{2016}]{Shibuya_et_al_2016}
{Shibuya} T.,  {Ouchi} M.,  {Kubo} M.,   {Harikane} Y.,  2016, \mn@doi [\apj]
  {10.3847/0004-637X/821/2/72}, \href
  {https://ui.adsabs.harvard.edu/abs/2016ApJ...821...72S} {821, 72}

\bibitem[\protect\citeauthoryear{{Sijacki}, {Vogelsberger}, {Genel},
  {Springel}, {Torrey}, {Snyder}, {Nelson}  \& {Hernquist}}{{Sijacki}
  et~al.}{2015}]{Sijacki_et_al_2015}
{Sijacki} D.,  {Vogelsberger} M.,  {Genel} S.,  {Springel} V.,  {Torrey} P.,
  {Snyder} G.~F.,  {Nelson} D.,   {Hernquist} L.,  2015, \mn@doi [\mnras]
  {10.1093/mnras/stv1340}, \href
  {https://ui.adsabs.harvard.edu/abs/2015MNRAS.452..575S} {452, 575}

\bibitem[\protect\citeauthoryear{{Sobol'}}{{Sobol'}}{1967}]{Sobol_1967}
{Sobol'} I.~M.,  1967, \mn@doi [USSR Computational Mathematics and Mathematical
  Physics] {10.1016/0041-5553(67)90144-9}, 7, 86

\bibitem[\protect\citeauthoryear{{Souza Lima}, {Mayer}, {Capelo}  \&
  {Bellovary}}{{Souza Lima} et~al.}{2017}]{SouzaLima_et_al_2017}
{Souza Lima} R.,  {Mayer} L.,  {Capelo} P.~R.,   {Bellovary} J.~M.,  2017,
  \mn@doi [\apj] {10.3847/1538-4357/aa5d19}, \href
  {https://ui.adsabs.harvard.edu/abs/2017ApJ...838...13S} {838, 13}

\bibitem[\protect\citeauthoryear{{Spitzer}}{{Spitzer}}{1942}]{Spitzer_1942}
{Spitzer} Lyman J.,  1942, \mn@doi [\apj] {10.1086/144407}, \href
  {https://ui.adsabs.harvard.edu/abs/1942ApJ....95..329S} {95, 329}

\bibitem[\protect\citeauthoryear{{Springel}}{{Springel}}{2010}]{Springel_2010}
{Springel} V.,  2010, \mn@doi [\mnras] {10.1111/j.1365-2966.2009.15715.x},
  \href {https://ui.adsabs.harvard.edu/abs/2010MNRAS.401..791S} {401, 791}

\bibitem[\protect\citeauthoryear{{Springel}, {White}, {Tormen}  \&
  {Kauffmann}}{{Springel} et~al.}{2001}]{Springel_et_al_2001}
{Springel} V.,  {White} S. D.~M.,  {Tormen} G.,   {Kauffmann} G.,  2001,
  \mn@doi [\mnras] {10.1046/j.1365-8711.2001.04912.x}, \href
  {https://ui.adsabs.harvard.edu/abs/2001MNRAS.328..726S} {328, 726}

\bibitem[\protect\citeauthoryear{{Springel} et~al.,}{{Springel}
  et~al.}{2018}]{Springel_et_al_2018}
{Springel} V.,  et~al., 2018, \mn@doi [\mnras] {10.1093/mnras/stx3304}, \href
  {https://ui.adsabs.harvard.edu/abs/2018MNRAS.475..676S} {475, 676}

\bibitem[\protect\citeauthoryear{{Tamburello}, {Capelo}, {Mayer}, {Bellovary}
  \& {Wadsley}}{{Tamburello} et~al.}{2017}]{Tamburello_et_al_2017}
{Tamburello} V.,  {Capelo} P.~R.,  {Mayer} L.,  {Bellovary} J.~M.,   {Wadsley}
  J.~W.,  2017, \mn@doi [\mnras] {10.1093/mnras/stw2561}, \href
  {https://ui.adsabs.harvard.edu/abs/2017MNRAS.464.2952T} {464, 2952}

\bibitem[\protect\citeauthoryear{{Tamfal}, {Capelo}, {Kazantzidis}, {Mayer},
  {Potter}, {Stadel}  \& {Widrow}}{{Tamfal} et~al.}{2018}]{Tamfal_et_al_2018}
{Tamfal} T.,  {Capelo} P.~R.,  {Kazantzidis} S.,  {Mayer} L.,  {Potter} D.,
  {Stadel} J.,   {Widrow} L.~M.,  2018, \mn@doi [\apjl]
  {10.3847/2041-8213/aada4b}, \href
  {https://ui.adsabs.harvard.edu/abs/2018ApJ...864L..19T} {864, L19}

\bibitem[\protect\citeauthoryear{{Tamfal}, {Mayer}, {Quinn}, {Capelo},
  {Kazantzidis}, {Babul}  \& {Potter}}{{Tamfal}
  et~al.}{2021}]{Tamfal_et_al_2021}
{Tamfal} T.,  {Mayer} L.,  {Quinn} T.~R.,  {Capelo} P.~R.,  {Kazantzidis} S.,
  {Babul} A.,   {Potter} D.,  2021, \mn@doi [\apj] {10.3847/1538-4357/ac0627},
  \href {https://ui.adsabs.harvard.edu/abs/2021ApJ...916...55T} {916, 55}

\bibitem[\protect\citeauthoryear{{Tremaine} \& {Weinberg}}{{Tremaine} \&
  {Weinberg}}{1984}]{Tremaine_Weinberg_1984}
{Tremaine} S.,  {Weinberg} M.~D.,  1984, \mn@doi [\mnras]
  {10.1093/mnras/209.4.729}, \href
  {https://ui.adsabs.harvard.edu/abs/1984MNRAS.209..729T} {209, 729}

\bibitem[\protect\citeauthoryear{{Tremaine}, {Richstone}, {Byun}, {Dressler},
  {Faber}, {Grillmair}, {Kormendy}  \& {Lauer}}{{Tremaine}
  et~al.}{1994}]{Tremaine_et_al_1994}
{Tremaine} S.,  {Richstone} D.~O.,  {Byun} Y.-I.,  {Dressler} A.,  {Faber}
  S.~M.,  {Grillmair} C.,  {Kormendy} J.,   {Lauer} T.~R.,  1994, \mn@doi [\aj]
  {10.1086/116883}, \href
  {https://ui.adsabs.harvard.edu/abs/1994AJ....107..634T} {107, 634}

\bibitem[\protect\citeauthoryear{{Tremmel}, {Governato}, {Volonteri}  \&
  {Quinn}}{{Tremmel} et~al.}{2015}]{Tremmel_et_al_2015}
{Tremmel} M.,  {Governato} F.,  {Volonteri} M.,   {Quinn} T.~R.,  2015, \mn@doi
  [\mnras] {10.1093/mnras/stv1060}, \href
  {https://ui.adsabs.harvard.edu/abs/2015MNRAS.451.1868T} {451, 1868}

\bibitem[\protect\citeauthoryear{{Truant}, {Izquierdo-Villalba}, {Sesana},
  {Mohiuddin Shaifullah}, {Bonetti}, {Spinoso}  \& {Bonoli}}{{Truant}
  et~al.}{2026}]{Truant_et_al_2026}
{Truant} R.~J.,  {Izquierdo-Villalba} D.,  {Sesana} A.,  {Mohiuddin Shaifullah}
  G.,  {Bonetti} M.,  {Spinoso} D.,   {Bonoli} S.,  2026, \mn@doi [\aap]
  {10.1051/0004-6361/202554846}, \href
  {https://ui.adsabs.harvard.edu/abs/2026A&A...706A.115T} {706, A115}

\bibitem[\protect\citeauthoryear{{Van Wassenhove}, {Capelo}, {Volonteri},
  {Dotti}, {Bellovary}, {Mayer}  \& {Governato}}{{Van Wassenhove}
  et~al.}{2014}]{VanWassenhove_et_al_2014}
{Van Wassenhove} S.,  {Capelo} P.~R.,  {Volonteri} M.,  {Dotti} M.,
  {Bellovary} J.~M.,  {Mayer} L.,   {Governato} F.,  2014, \mn@doi [\mnras]
  {10.1093/mnras/stu024}, \href
  {https://ui.adsabs.harvard.edu/abs/2014MNRAS.439..474V} {439, 474}

\bibitem[\protect\citeauthoryear{{Varisco}, {Dotti}, {Bonetti}, {Bortolas}  \&
  {Lupi}}{{Varisco} et~al.}{2024}]{Varisco_et_al_2024}
{Varisco} L.,  {Dotti} M.,  {Bonetti} M.,  {Bortolas} E.,   {Lupi} A.,  2024,
  \mn@doi [\aap] {10.1051/0004-6361/202449700}, \href
  {https://ui.adsabs.harvard.edu/abs/2024A&A...689A.279V} {689, A279}

\bibitem[\protect\citeauthoryear{{Vasiliev}, {Antonini}  \&
  {Merritt}}{{Vasiliev} et~al.}{2015}]{Vasiliev_et_al_2015}
{Vasiliev} E.,  {Antonini} F.,   {Merritt} D.,  2015, \mn@doi [\apj]
  {10.1088/0004-637X/810/1/49}, \href
  {https://ui.adsabs.harvard.edu/abs/2015ApJ...810...49V} {810, 49}

\bibitem[\protect\citeauthoryear{{Villaescusa-Navarro}
  et~al.,}{{Villaescusa-Navarro} et~al.}{2022}]{Villaescusa_Navarro_et_al_2022}
{Villaescusa-Navarro} F.,  et~al., 2022, \mn@doi [\apjs]
  {10.3847/1538-4365/ac5ab0}, \href
  {https://ui.adsabs.harvard.edu/abs/2022ApJS..259...61V} {259, 61}

\bibitem[\protect\citeauthoryear{{Volonteri} et~al.,}{{Volonteri}
  et~al.}{2020}]{Volonteri_et_al_2020}
{Volonteri} M.,  et~al., 2020, \mn@doi [\mnras] {10.1093/mnras/staa2384}, \href
  {https://ui.adsabs.harvard.edu/abs/2020MNRAS.498.2219V} {498, 2219}

\bibitem[\protect\citeauthoryear{{Weinberg}}{{Weinberg}}{1986}]{Weinberg_1986}
{Weinberg} M.~D.,  1986, \mn@doi [\apj] {10.1086/163785}, \href
  {https://ui.adsabs.harvard.edu/abs/1986ApJ...300...93W} {300, 93}

\bibitem[\protect\citeauthoryear{{Weinberg}}{{Weinberg}}{1989}]{Weinberg_1989}
{Weinberg} M.~D.,  1989, \mn@doi [\mnras] {10.1093/mnras/239.2.549}, \href
  {https://ui.adsabs.harvard.edu/abs/1989MNRAS.239..549W} {239, 549}

\bibitem[\protect\citeauthoryear{{Weinberger} et~al.,}{{Weinberger}
  et~al.}{2017}]{Weinberger_et_al_2017}
{Weinberger} R.,  et~al., 2017, \mn@doi [\mnras] {10.1093/mnras/stw2944}, \href
  {https://ui.adsabs.harvard.edu/abs/2017MNRAS.465.3291W} {465, 3291}

\bibitem[\protect\citeauthoryear{{Weinberger} et~al.,}{{Weinberger}
  et~al.}{2018}]{Weinberger_et_al_2018}
{Weinberger} R.,  et~al., 2018, \mn@doi [\mnras] {10.1093/mnras/sty1733}, \href
  {https://ui.adsabs.harvard.edu/abs/2018MNRAS.479.4056W} {479, 4056}

\bibitem[\protect\citeauthoryear{{Weller}, {Natarajan}  \& {Burke}}{{Weller}
  et~al.}{2026}]{Weller_et_al_2026}
{Weller} E.~J.,  {Natarajan} P.,   {Burke} C.~J.,  2026, \mn@doi [\apjl]
  {10.3847/2041-8213/ae9c38}, \href
  {https://ui.adsabs.harvard.edu/abs/2026ApJ..1008L..54W} {1008, L54}

\bibitem[\protect\citeauthoryear{{White}}{{White}}{1983}]{White_1983}
{White} S.~D.~M.,  1983, \mn@doi [\apj] {10.1086/161425}, \href
  {https://ui.adsabs.harvard.edu/abs/1983ApJ...274...53W} {274, 53}

\bibitem[\protect\citeauthoryear{{Zana}, {Dotti}, {Capelo}, {Bonoli}, {Haardt},
  {Mayer}  \& {Spinoso}}{{Zana} et~al.}{2018a}]{Zana_et_al_2018a}
{Zana} T.,  {Dotti} M.,  {Capelo} P.~R.,  {Bonoli} S.,  {Haardt} F.,  {Mayer}
  L.,   {Spinoso} D.,  2018a, \mn@doi [\mnras] {10.1093/mnras/stx2503}, \href
  {https://ui.adsabs.harvard.edu/abs/2018MNRAS.473.2608Z} {473, 2608}

\bibitem[\protect\citeauthoryear{{Zana}, {Dotti}, {Capelo}, {Mayer}, {Haardt},
  {Shen}  \& {Bonoli}}{{Zana} et~al.}{2018b}]{Zana_et_al_2018b}
{Zana} T.,  {Dotti} M.,  {Capelo} P.~R.,  {Mayer} L.,  {Haardt} F.,  {Shen} S.,
    {Bonoli} S.,  2018b, \mn@doi [\mnras] {10.1093/mnras/sty1850}, \href
  {https://ui.adsabs.harvard.edu/abs/2018MNRAS.479.5214Z} {479, 5214}

\bibitem[\protect\citeauthoryear{{Zana}, {Capelo}, {Dotti}, {Mayer}, {Lupi},
  {Haardt}, {Bonoli}  \& {Shen}}{{Zana} et~al.}{2019}]{Zana_et_al_2019}
{Zana} T.,  {Capelo} P.~R.,  {Dotti} M.,  {Mayer} L.,  {Lupi} A.,  {Haardt} F.,
   {Bonoli} S.,   {Shen} S.,  2019, \mn@doi [\mnras] {10.1093/mnras/stz1834},
  \href {https://ui.adsabs.harvard.edu/abs/2019MNRAS.488.1864Z} {488, 1864}

\bibitem[\protect\citeauthoryear{{Zana} et~al.,}{{Zana}
  et~al.}{2022}]{Zana_et_al_2022}
{Zana} T.,  et~al., 2022, \mn@doi [\mnras] {10.1093/mnras/stac1708}, \href
  {https://ui.adsabs.harvard.edu/abs/2022MNRAS.515.1524Z} {515, 1524}

\bibitem[\protect\citeauthoryear{{Zhang} et~al.,}{{Zhang}
  et~al.}{2023}]{Zhang_et_al_2023}
{Zhang} Y.,  et~al., 2023, \mn@doi [\apj] {10.3847/1538-4357/acc2c2}, \href
  {https://ui.adsabs.harvard.edu/abs/2023ApJ...948..103Z} {948, 103}

\bibitem[\protect\citeauthoryear{{Zhao}, {Du}, {Ho}, {Debattista}  \&
  {Shi}}{{Zhao} et~al.}{2020}]{Zhao_et_al_2020}
{Zhao} D.,  {Du} M.,  {Ho} L.~C.,  {Debattista} V.~P.,   {Shi} J.,  2020,
  \mn@doi [\apj] {10.3847/1538-4357/abbe1b}, \href
  {https://ui.adsabs.harvard.edu/abs/2020ApJ...904..170Z} {904, 170}

\bibitem[\protect\citeauthoryear{{Zwick}, {Derdzinski}, {Garg}, {Capelo}  \&
  {Mayer}}{{Zwick} et~al.}{2022}]{Zwick_et_al_2022}
{Zwick} L.,  {Derdzinski} A.,  {Garg} M.,  {Capelo} P.~R.,   {Mayer} L.,  2022,
  \mn@doi [\mnras] {10.1093/mnras/stac299}, \href
  {https://ui.adsabs.harvard.edu/abs/2022MNRAS.511.6143Z} {511, 6143}

\makeatother
\end{thebibliography}
\normalsize

\appendix

\section{Conditional normalizing flows and data normalization}\label{App:CNF}

\subsection{Overview of normalizing flows}

NFs are a class of generative models that construct complex probability distributions by applying a sequence of invertible and differentiable transformations to a simple base distribution \citep[e.g.][]{JimenezRezende_Mohamed_2015,Papamakarios_et_al_2019}. Let $z$ be a latent variable drawn from a simple prior distribution $p_{\rm Z}(z)$, typically a standard normal $\mathcal{N}(0, 1)$. A bijective mapping $f_\theta : \mathbb{R} \to \mathbb{R}$ parametrized by $\theta$ transforms $z$ into the observed variable $x = f_\theta(z)$. By the change-of-variables formula, the probability density of $x$ is given by

\begin{equation}
p_{\rm X}(x) = p_{\rm Z}(f_\theta^{-1}(x))
\left| \det \frac{\partial f_\theta^{-1}}{\partial x} \right|\,.
\label{eq:nf_change_of_variables}
\end{equation}

Since both $f_\theta$ and its inverse are explicitly defined, this formulation allows for the exact computation of the likelihood $p_{\rm X}(x)$ as well as an efficient sampling from the learned distribution. By composing multiple such transformations, one obtains a highly expressive model capable of representing complex, non-Gaussian data while maintaining tractable likelihood evaluation.

\subsection{Conditional normalizing flows}

To model the conditional density, we employ a conditional extension of the NF. In a conditional NF, the transformations depend explicitly on the conditioning variables $\vec{c}$, so that both the mapping and its inverse take the form

\[
\tau_{\mathrm{d}} = f_\theta(z; \vec{c}\,)\,, \qquad
z = f_\theta^{-1}(\tau_{\rm d}; \vec{c}\,)\,,
\]

\noindent where $z \sim \mathcal{N}(0, 1)$ is a latent variable. The resulting conditional density can be written as

\begin{equation}
p(\tau_{\rm d} \,|\, \vec{c}\,)
= p_{\rm Z}(f_\theta^{-1}(\tau_{\rm d}; \vec{c}\,))
\left| \det \frac{\partial f_\theta^{-1}(\tau_{\rm d}; \vec{c}\,)}{\partial \tau_{\rm d}} \right|\,.
\end{equation}

This structure enables the model to learn complex dependencies between the target variable and the conditioning inputs, while retaining exact likelihood computation.

In practice, we implement $f_\theta$ as a composition of neural spline coupling transformations \citep[][]{Durkan_et_al_2019}, which provide smooth, invertible mappings parametrized by neural networks. Each coupling transformation operates on the target variable $\tau_{\rm d}$ conditioned on the nine input features $\mathbf{c}$, allowing the model to flexibly capture non-linear relationships and heteroskedastic uncertainties in the conditional distribution.

\subsection{Data preprocessing and normalization}

Both the target variable $\tau_{\rm d}$ and the conditioning features were standardized according to

\begin{equation}
x' = \frac{x - \mu_{\rm X}}{\sigma_{\rm X}}, \qquad
y' = \frac{y - \mu_{\rm Y}}{\sigma_{\rm Y}}\,,
\end{equation}

\noindent where $\mu$ and $\sigma$ denote the mean and standard deviation estimated from the training set. The same normalization parameters were subsequently applied to the test data to avoid information leakage.

This normalization procedure ensures that all input variables have zero mean and unit variance, which stabilizes training and improves optimization efficiency. Combined with the conditional-NF architecture, this enables the model to learn an accurate and fully probabilistic representation of the conditional distribution $p(\tau_{\rm d} \,|\, \vec{c}\,)$.

\section{Additional information on the \textsc{tng50} galaxies}\label{sec:TNG_additional_information}

In this section, we provide useful information on the \textsc{tng50} galaxies we considered in this work.

In Table~\ref{tab:tng50_snapshots}, we list the relevant data for each of the 18 \textsc{tng50} snapshots employed in our analysis, including the numbers of non-barred and barred galaxies (additionally split by bar strength) used for our predictions.

In Figure~\ref{fig:TNG-central-mass-histograms}, we show the distributions of central stellar masses for all the subsets of galaxies considered in this paper, for each of the 18 \textsc{tng50} snapshots used here.

Finally, in Figures~\ref{fig:TNG-bar-histograms-1} and \ref{fig:TNG-bar-histograms-2}, we present, for each of the 18 \textsc{tng50} snapshots employed in the analysis, the distributions of bar stellar mass fractions and lengths for all the barred galaxies considered in our work, split by bar strength.

\begin{table*}
  \centering
  \caption{Galaxy numbers in the 18 \textsc{tng50} snapshots we considered. Columns~1--3: snapshot number, redshift, and age (taken from \url{https://www.tng-project.org/data/downloads/TNG50-1/}). Column~4: total number of galaxies (with more than $10^4$ baryon particles, i.e. with a total stellar mass equal to or greater than $\sim$$5$--$7 \times 10^8$~M$_{\sun}$, depending on the redshift) in each snapshot. Column~5: number of galaxies from the subset of Column~4 (and fraction with respect to Column~4) with a central stellar mass larger than $3 \times 10^9$~M$_{\sun}$. Column~6: number of barred galaxies from the subset of Column~5 (and fraction with respect to Column~5), using three different thresholds for the bar strength (with all other constraints being fixed; see text for more details): $A_{\rm 2,max}(R) \ge 0.1$, 0.2, and 0.4. The difference between Column~5 and Column~6 [when assuming $A_{\rm 2,max}(R) \ge 0.1$] yields the number of non-barred galaxies, $N_{\rm no\text{-}bar}$. Column~7: $N_{\rm bar}$, number of galaxies from the subsets of Column~6 with a bar stellar mass fraction between 0.0566 and 0.566 and with a bar length between 1 and 9~kpc [and fraction with respect to the total number of galaxies considered: $N_{\rm bar}/(N_{\rm no\text{-}bar}+N_{\rm bar})$].
  }
  \vspace{-0.2cm}
  \label{tab:tng50_snapshots}
  \begin{center}
    \begin{tabular}{ccccccc}
    \hline
    \hline
    Snapshot & $z$ & $t_{\rm age}$/Gyr & Total & \begin{tabular}{@{}c@{}} $M_{\star,\textsc{tng50}} \ge$ \\ $3 \times 10^9$~M$_{\sun}$ \end{tabular} & \begin{tabular}{@{}c@{}} $A_{\rm 2,max}(R) \ge 0.1$ \\ $A_{\rm 2,max}(R) \ge 0.2$ \\ $A_{\rm 2,max}(R) \ge 0.4$ \end{tabular} & \begin{tabular}{@{}c@{}} $a_{\rm bar,\textsc{tng50}} \in 1$--9~kpc AND \\ $f_{\rm bar,\textsc{tng50}} \in 0.0566$--0.566 \end{tabular} \\
    \hline
    \hline
    $\mathbb{S}$34 & 1.90 & 3.447 & 2120 & 762 (35.9\%) & \begin{tabular}[c]{@{}c@{}}206 (27.0\%)\\176 (23.1\%)\\91 (11.9\%)\end{tabular} & \begin{tabular}[c]{@{}c@{}}86 (13.4\%)\\86 (13.4\%)\\75 (11.9\%)\end{tabular} \\
    \hline
    $\mathbb{S}$35 & 1.82 & 3.593 & 2205 & 800 (36.3\%) & \begin{tabular}[c]{@{}c@{}}230 (28.8\%)\\191 (23.9\%)\\95 (11.9\%)\end{tabular} & \begin{tabular}[c]{@{}c@{}}97 (14.5\%)\\96 (14.4\%)\\73 (11.4\%)\end{tabular} \\
    \hline
    $\mathbb{S}$36 & 1.74 & 3.744 & 2295 & 832 (36.3\%) & \begin{tabular}[c]{@{}c@{}}232 (27.9\%)\\181 (21.8\%)\\86 (10.3\%)\end{tabular} & \begin{tabular}[c]{@{}c@{}}84 (12.3\%)\\84 (12.3\%)\\65 (9.8\%)\end{tabular} \\
    \hline
    $\mathbb{S}$37 & 1.67 & 3.902 & 2380 & 872 (36.6\%) & \begin{tabular}[c]{@{}c@{}}254 (29.1\%)\\218 (25.0\%)\\96 (11.0\%)\end{tabular} & \begin{tabular}[c]{@{}c@{}}104 (14.4\%)\\102 (14.2\%)\\76 (11.0\%)\end{tabular} \\
    \hline
    $\mathbb{S}$38 & 1.60 & 4.038 & 2428 & 910 (37.5\%) & \begin{tabular}[c]{@{}c@{}}278 (30.5\%)\\229 (25.2\%)\\105 (11.5\%)\end{tabular} & \begin{tabular}[c]{@{}c@{}}120 (16.0\%)\\117 (15.6\%)\\81 (11.4\%)\end{tabular} \\
    \hline
    $\mathbb{S}$44 & 1.25 & 4.980 & 2805 & 1086 (38.7\%) & \begin{tabular}[c]{@{}c@{}}375 (34.5\%)\\309 (28.5\%)\\138 (12.7\%)\end{tabular} & \begin{tabular}[c]{@{}c@{}}228 (24.3\%)\\224 (24.0\%)\\131 (15.6\%)\end{tabular} \\
    \hline
    $\mathbb{S}$45 & 1.21 & 5.115 & 2832 & 1107 (39.1\%) & \begin{tabular}[c]{@{}c@{}}366 (33.1\%)\\310 (28.0\%)\\134 (12.1\%)\end{tabular} & \begin{tabular}[c]{@{}c@{}}216 (22.6\%)\\213 (22.3\%)\\124 (14.3\%)\end{tabular} \\
    \hline
    $\mathbb{S}$46 & 1.15 & 5.289 & 2868 & 1129 (39.4\%) & \begin{tabular}[c]{@{}c@{}}395 (35.0\%)\\328 (29.1\%)\\148 (13.1\%)\end{tabular} & \begin{tabular}[c]{@{}c@{}}245 (25.0\%)\\242 (24.8\%)\\141 (16.1\%)\end{tabular} \\
    \hline
    $\mathbb{S}$47 & 1.11 & 5.431 & 2932 & 1141 (38.9\%) & \begin{tabular}[c]{@{}c@{}}375 (32.9\%)\\324 (28.4\%)\\154 (13.5\%)\end{tabular} & \begin{tabular}[c]{@{}c@{}}238 (23.7\%)\\234 (23.4\%)\\142 (15.6\%)\end{tabular} \\
    \hline
    $\mathbb{S}$48 & 1.07 & 5.577 & 2966 & 1160 (39.1\%) & \begin{tabular}[c]{@{}c@{}}403 (34.7\%)\\351 (30.3\%)\\149 (12.8\%)\end{tabular} & \begin{tabular}[c]{@{}c@{}}278 (26.9\%)\\269 (26.2\%)\\142 (15.8\%)\end{tabular} \\
    \hline
    $\mathbb{S}$49 & 1.04 & 5.726 & 3002 & 1191 (39.7\%) & \begin{tabular}[c]{@{}c@{}}418 (35.1\%)\\351 (29.5\%)\\152 (12.8\%)\end{tabular} & \begin{tabular}[c]{@{}c@{}}287 (27.1\%)\\282 (26.7\%)\\144 (15.7\%)\end{tabular} \\
    \hline
    $\mathbb{S}$50 & 1.00 & 5.878 & 3047 & 1204 (39.5\%) & \begin{tabular}[c]{@{}c@{}}447 (37.1\%)\\374 (31.1\%)\\154 (12.8\%)\end{tabular} & \begin{tabular}[c]{@{}c@{}}303 (28.6\%)\\294 (28.0\%)\\146 (16.2\%)\end{tabular} \\
    \hline
    $\mathbb{S}$67 & 0.50 & 8.587 & 3544 & 1401 (39.5\%) & \begin{tabular}[c]{@{}c@{}}581 (41.5\%)\\479 (34.2\%)\\226 (16.1\%)\end{tabular} & \begin{tabular}[c]{@{}c@{}}440 (34.9\%)\\421 (33.9\%)\\220 (21.2\%)\end{tabular} \\
    \hline
    $\mathbb{S}$72 & 0.40 & 9.389 & 3684 & 1442 (39.1\%) & \begin{tabular}[c]{@{}c@{}}570 (39.5\%)\\475 (32.9\%)\\241 (16.7\%)\end{tabular} & \begin{tabular}[c]{@{}c@{}}435 (33.3\%)\\418 (32.4\%)\\234 (21.2\%)\end{tabular} \\
    \hline
    $\mathbb{S}$78 & 0.30 & 10.299 & 3758 & 1447 (38.5\%) & \begin{tabular}[c]{@{}c@{}}573 (39.6\%)\\467 (32.3\%)\\221 (15.3\%)\end{tabular} & \begin{tabular}[c]{@{}c@{}}429 (32.9\%)\\414 (32.1\%)\\210 (19.4\%)\end{tabular} \\
    \hline
    $\mathbb{S}$84 & 0.20 & 11.323 & 3888 & 1457 (37.5\%) & \begin{tabular}[c]{@{}c@{}}569 (39.1\%)\\472 (32.4\%)\\209 (14.3\%)\end{tabular} & \begin{tabular}[c]{@{}c@{}}437 (33.0\%)\\428 (32.5\%)\\202 (18.5\%)\end{tabular} \\
    \hline
    $\mathbb{S}$91 & 0.10 & 12.467 & 4003 & 1466 (36.6\%) & \begin{tabular}[c]{@{}c@{}}563 (38.4\%)\\440 (30.0\%)\\201 (13.7\%)\end{tabular} & \begin{tabular}[c]{@{}c@{}}417 (31.6\%)\\394 (30.4\%)\\190 (17.4\%)\end{tabular} \\
    \hline
    $\mathbb{S}$99 & 0.00 & 13.803 & 4091 & 1497 (36.6\%) & \begin{tabular}[c]{@{}c@{}}571 (38.1\%)\\441 (29.5\%)\\200 (13.4\%)\end{tabular} & \begin{tabular}[c]{@{}c@{}}421 (31.3\%)\\392 (29.7\%)\\190 (17.0\%)\end{tabular} \\
    \hline
    \hline
    \end{tabular}
  \end{center}
\end{table*}

\begin{figure*}
\includegraphics[width=1.00\textwidth]{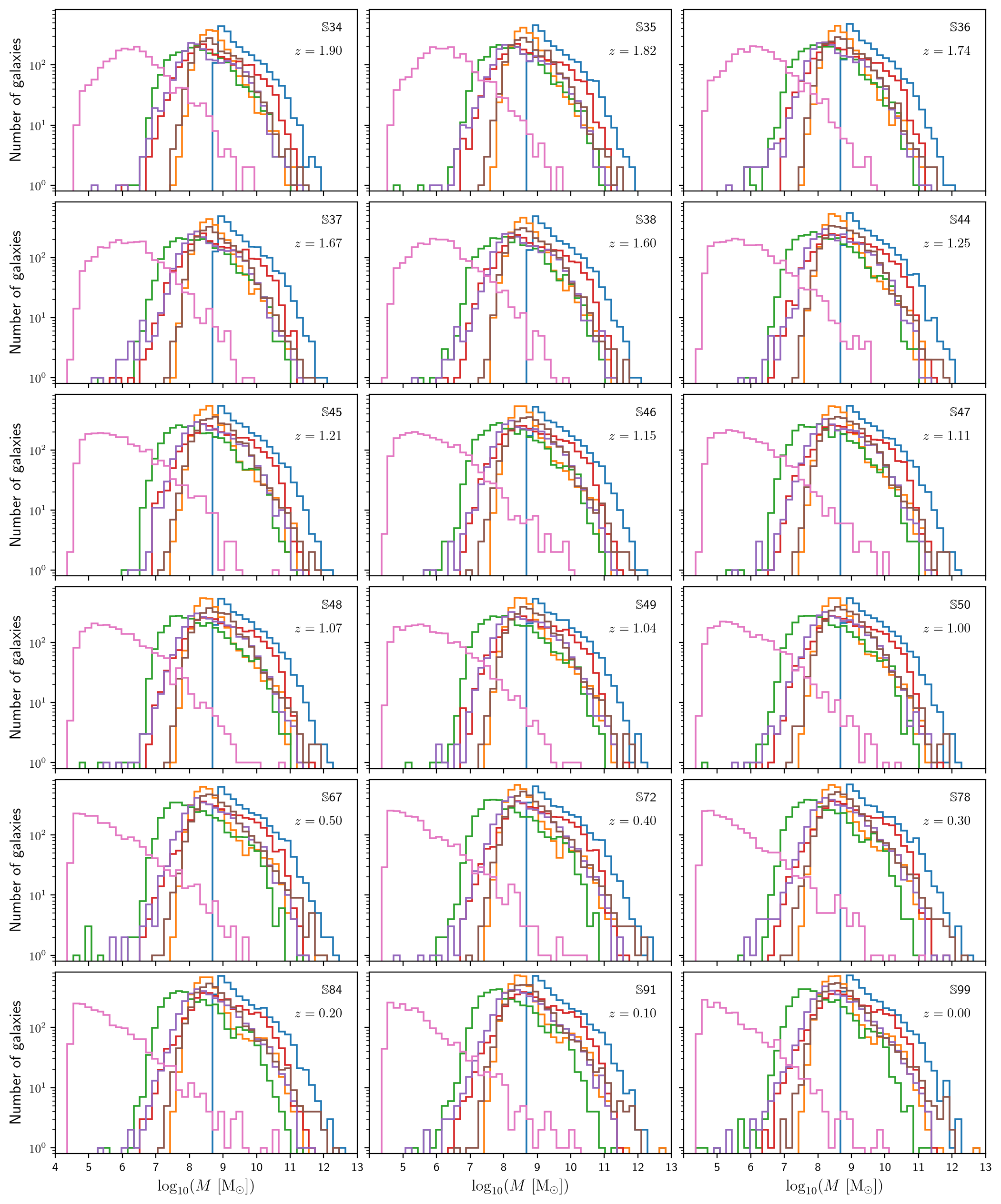}
\caption{For the 18 snapshots listed in Table~\ref{tab:tng50_snapshots}, we show the distribution of stellar masses, for all components of the galaxy separately (bulge in orange, pseudo-bulge in green, thin disc in red, thick disc in purple, halo in brown, and unbound particles in pink) and for the galaxy as a whole (in blue). The sharp cut-off of the blue curve at $\sim$$5$--$7 \times 10^8$~M$_{\sun}$ is due to the minimum number of stellar particles ($10^4$) used by the \textsc{mordor} algorithm. All distributions utilize 50 equally spaced bins in the shown range.}
\label{fig:TNG-mass-histograms}
\end{figure*}

\begin{figure*}
\includegraphics[width=1.00\textwidth]{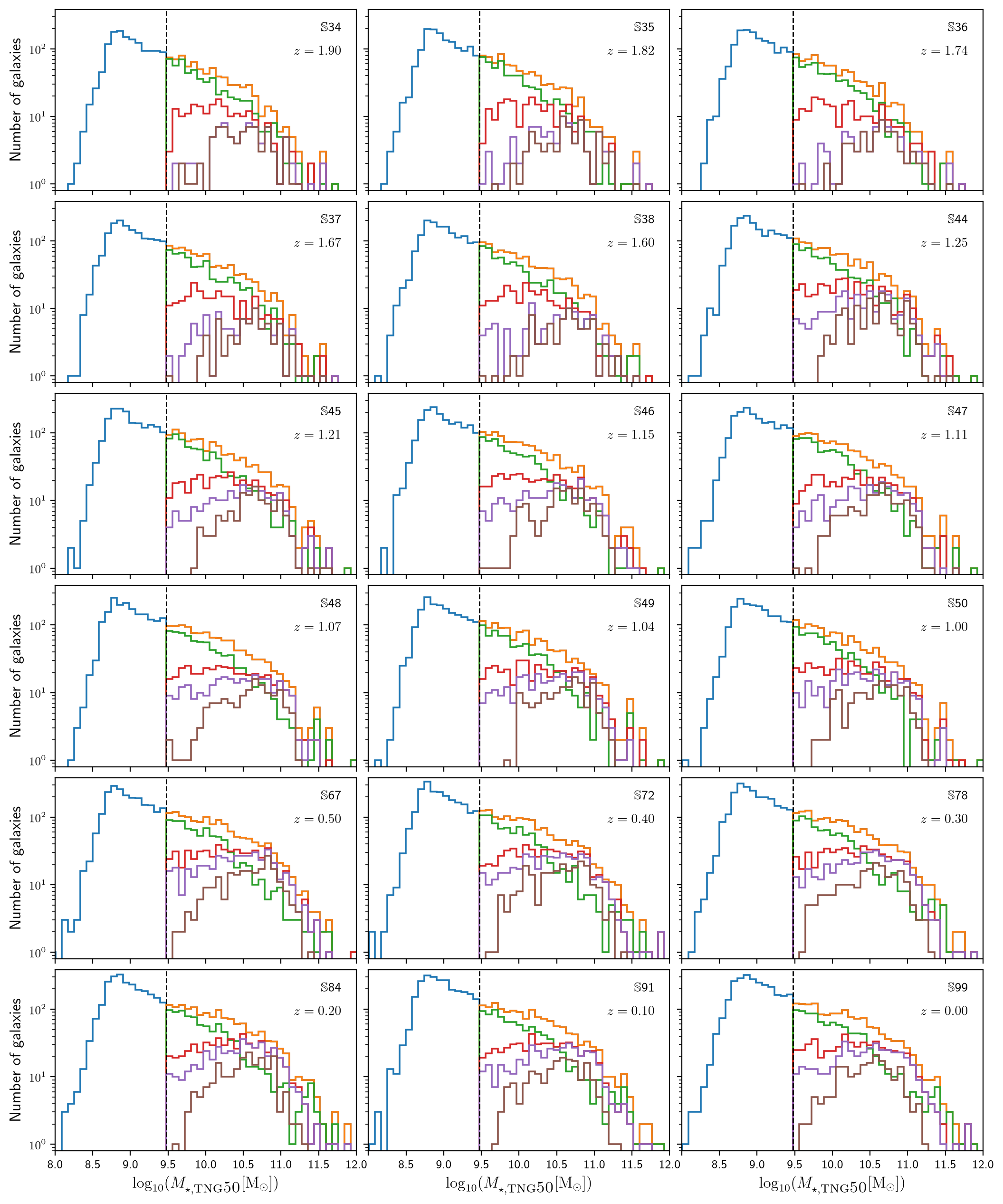}
\caption{For the 18 snapshots listed in Table~\ref{tab:tng50_snapshots}, we show the distribution of central stellar masses for all galaxies (blue; from Column~4) and for all galaxies with a central stellar mass $M_{\rm \star,\textsc{tng50}} \geq 3 \times 10^9$~M$_{\sun}$ (orange; from Column~5). We then split the latter distribution in barred (red; Column~6: here we are showing only the weak-bar constraint of $A_{\rm 2,max}(R) \ge 0.1$) and non-barred (green) galaxies. For the barred galaxies, we further apply the filters of Column~7, considering all bars (purple) and only strong bars (brown; $A_{\rm 2,max}(R) \ge 0.4$). All distributions utilize 50 equally spaced bins in the range $\log_{10}(M_{\rm \star,\textsc{tng50}}/{\rm M}_{\sun}) \simeq 8.0$--12.0, such that the vertical, black, dashed line, showing the central stellar mass threshold applied, is precisely at one of the bin edges.}
\label{fig:TNG-central-mass-histograms}
\end{figure*}

\begin{figure*}
\includegraphics[width=0.49\textwidth]{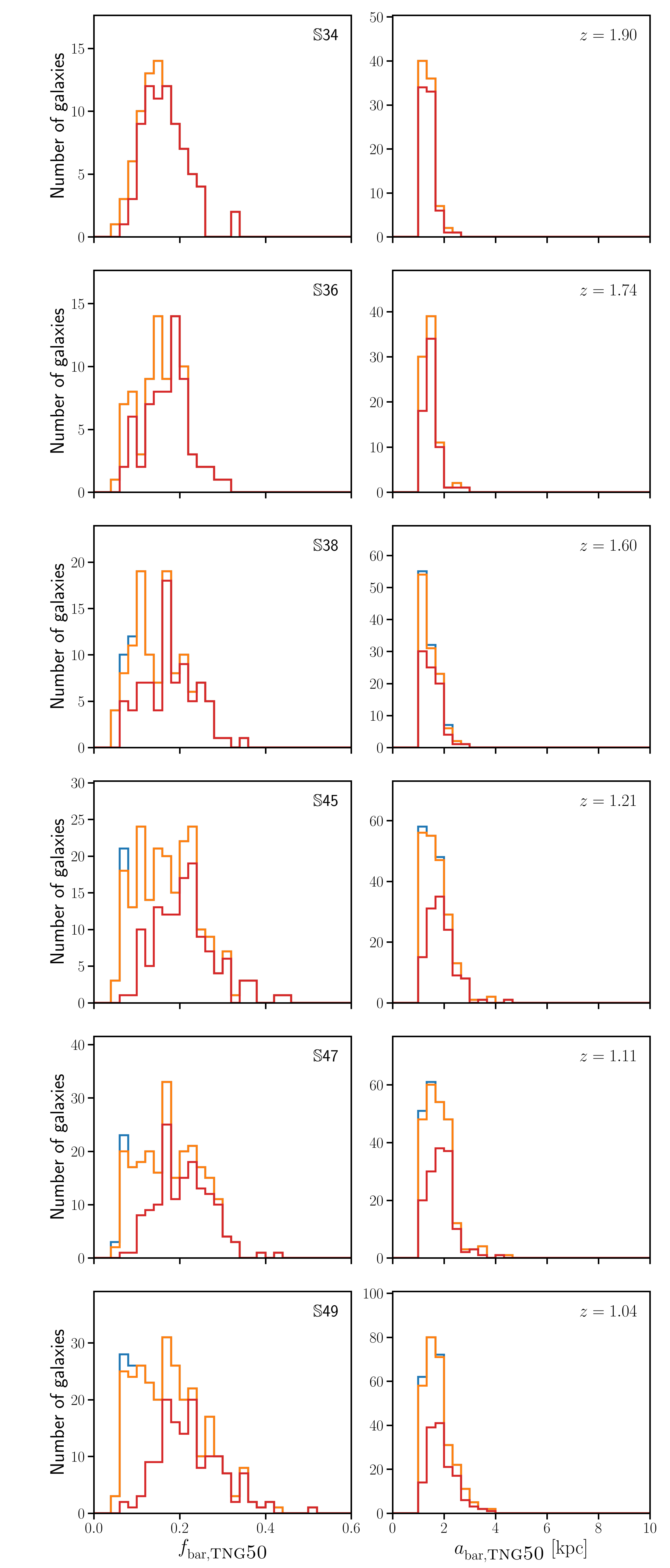}
\includegraphics[width=0.49\textwidth]{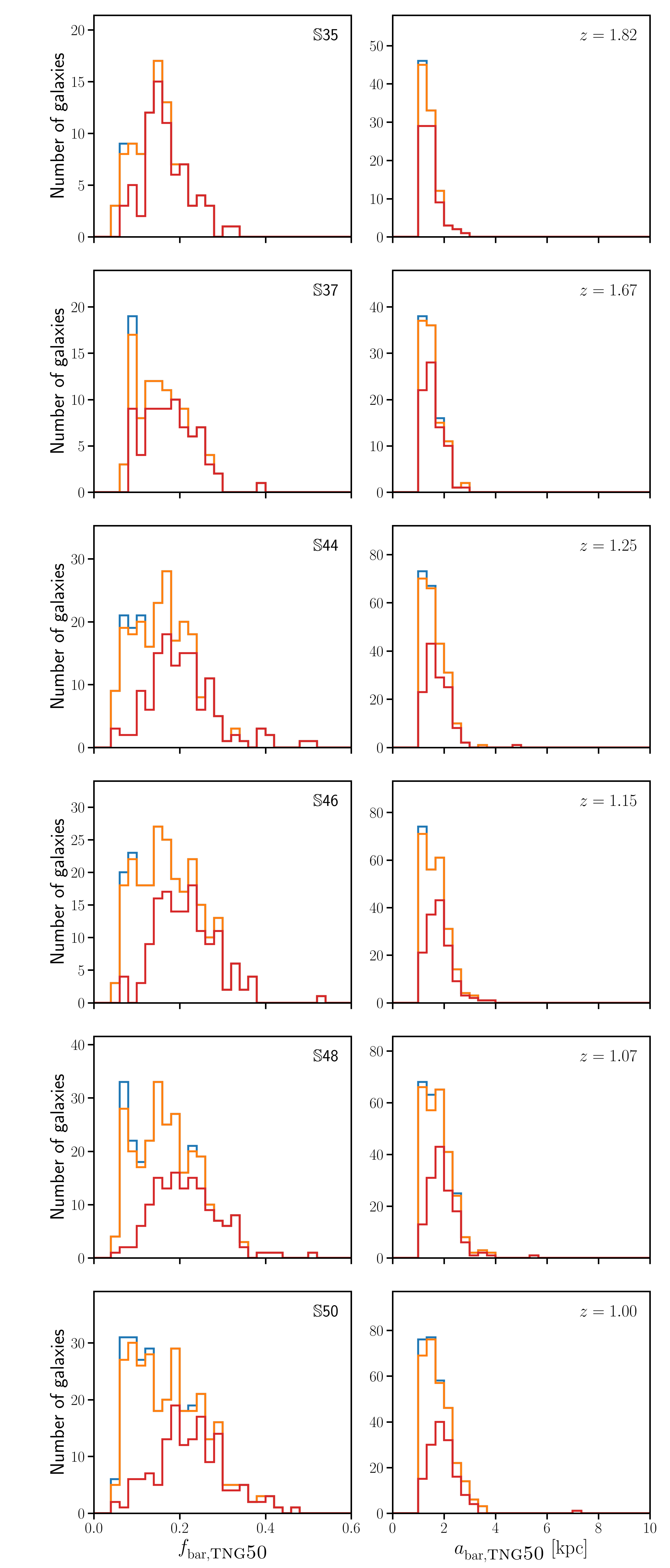}
\caption{For the top 12 snapshots listed in Table~\ref{tab:tng50_snapshots}, we show the distribution of bar stellar mass fractions (left-hand panel) and bar lengths (right-hand panel) for the galaxies with a central stellar mass $M_{\rm \star,\textsc{tng50}} \ge 3 \times 10^9$~M$_{\sun}$, defined as barred galaxies, and with $f_{\rm bar,\textsc{tng50}} \in [0.0566$--0.566$]$ and $a_{\rm bar,\textsc{tng50}} \in [1$--$9]$. We adopt three different constraints on the bar strength to define a barred galaxy (with all other constraints being fixed; see text for more details): $A_{\rm 2,max}(R) \ge 0.1$ (all bars; blue), 0.2 (only moderate and strong bars; orange), and 0.4 (only strong bars; red). All distributions utilize 30 equally spaced bins in the shown ranges.}
\label{fig:TNG-bar-histograms-1}
\end{figure*}

\begin{figure*}
\includegraphics[width=0.49\textwidth]{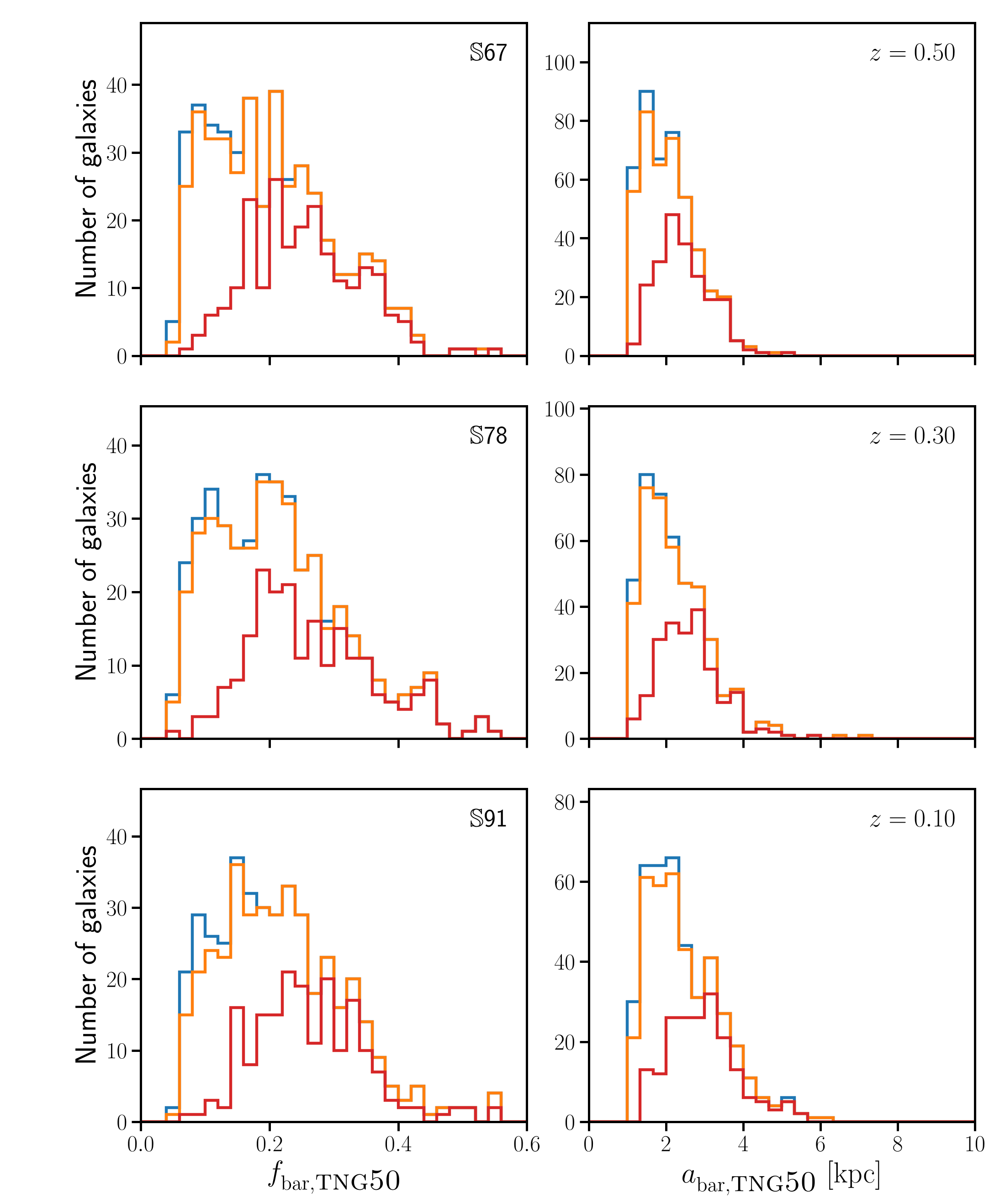}
\includegraphics[width=0.49\textwidth]{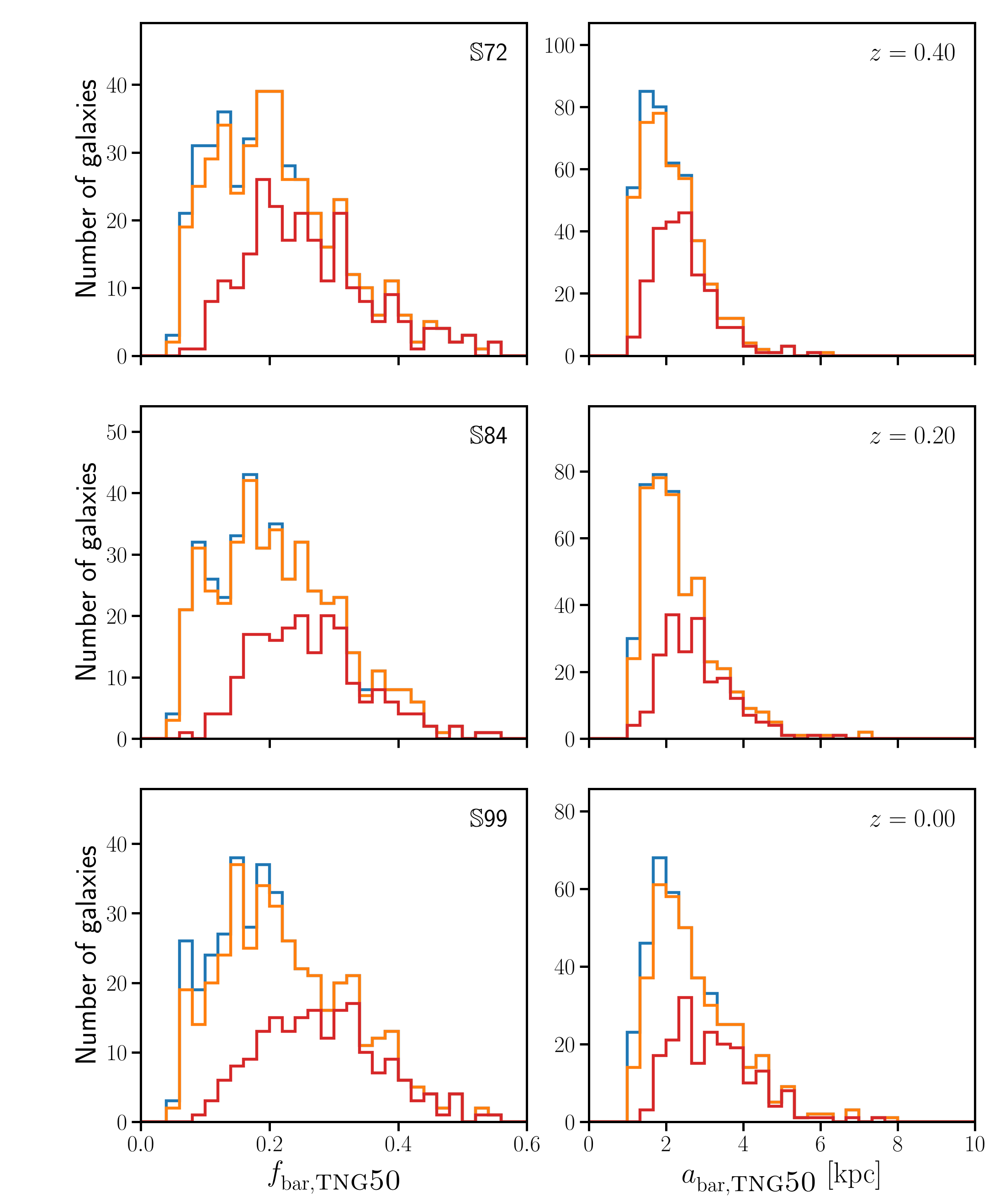}
\caption{Same as Figure~\ref{fig:TNG-bar-histograms-1}, but for the bottom six snapshots listed in Table~\ref{tab:tng50_snapshots}.}
\label{fig:TNG-bar-histograms-2}
\end{figure*}

\newpage

\section{Log-log version of Figure~5}

\begin{figure}
\includegraphics[width=0.47\textwidth]{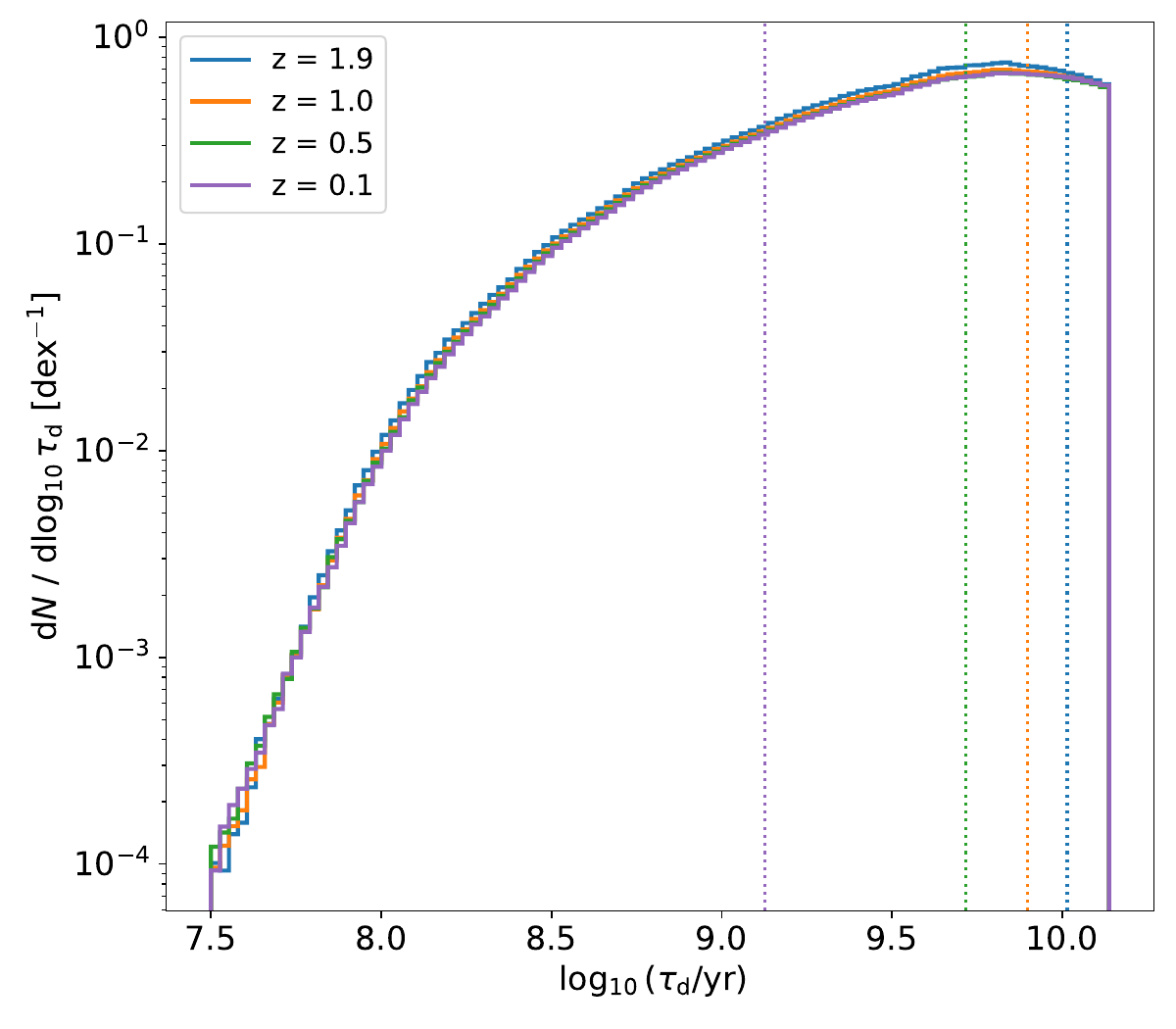}
\caption{Same as Figure~\ref{fig:TNG_some_snapshots_linlin_all_galaxies}, but in log-log.}
\label{fig:TNG_some_snapshots_loglog_all_galaxies}
\end{figure}

For an easier comparison with \citet{Li_et_al_2022}, in Figure~\ref{fig:TNG_some_snapshots_loglog_all_galaxies} we present the log-log version of Figure~\ref{fig:TNG_some_snapshots_linlin_all_galaxies}.

\bsp 
\label{lastpage}
\end{document}